\documentclass[draft]{agujournal2018}
\draftfalse

\usepackage{booktabs}
\usepackage{url}

\usepackage[bookmarks=false,  pdfnewwindow=true, colorlinks=true,  linkcolor=magenta, citecolor=blue, filecolor=cyan, urlcolor=purple, final=true]{hyperref}

\usepackage{journal_macros}

\journalname{JGR-Space Physics}

\graphicspath{{./}{figures/}}

\begin{document}

\sloppy

\title{Magnetic Structure and Propagation of Two Interacting CMEs from the Sun to Saturn}

\authors{Erika~Palmerio\affil{1,2}, Teresa~Nieves-Chinchilla\affil{3}, Emilia~K.~J.~Kilpua\affil{4}, David~Barnes\affil{5}, Andrei~N.~Zhukov\affil{6,7}, Lan~K.~Jian\affil{3}, Olivier~Witasse\affil{8}, Gabrielle~Provan\affil{9}, Chihiro~Tao\affil{10}, Laurent~Lamy\affil{11,12}, Thomas~J.~Bradley\affil{9}, M.~Leila~Mays\affil{3}, Christian~{M\"o}stl\affil{13,14}, Elias~Roussos\affil{15}, Yoshifumi~Futaana\affil{16}, Adam~Masters\affil{17}, and Beatriz~S{\'a}nchez-Cano\affil{9}}

\affiliation{1}{Space Sciences Laboratory, University of California--Berkeley, Berkeley, CA, USA}
\affiliation{2}{CPAESS, University Corporation for Atmospheric Research, Boulder, CO, USA}
\affiliation{3}{Heliophysics Science Division, NASA Goddard Space Flight Center, Greenbelt, MD, USA}
\affiliation{4}{Department of Physics, University of Helsinki, Helsinki, Finland}
\affiliation{5}{STFC RAL Space, Rutherford Appleton Laboratory, Harwell Campus, Oxfordshire, UK}
\affiliation{6}{Solar--Terrestrial Centre of Excellence---SIDC, Royal Observatory of Belgium, Brussels, Belgium}
\affiliation{7}{Skobeltsyn Institute of Nuclear Physics, Moscow State University, Moscow, Russia}
\affiliation{8}{ESTEC, European Space Agency, Noordwijk, Netherlands}
\affiliation{9}{School of Physics and Astronomy, University of Leicester, Leicester, UK}
\affiliation{10}{National Institute of Information and Communications Technology (NICT), Koganei, Japan}
\affiliation{11}{LESIA, Observatoire de Paris, PSL, CNRS, UPMC, Universit{\'e} Paris Diderot, Meudon, France}
\affiliation{12}{LAM, Pyth{\'e}as, Aix Marseille Universit{\'e}, CNRS, CNES, Marseille, France}
\affiliation{13}{Space Research Institute, Austrian Academy of Sciences, Graz, Austria}
\affiliation{14}{Institute of Geodesy, Graz University of Technology, Graz, Austria}
\affiliation{15}{Max Planck Institute for Solar System Research, G{\"o}ttingen, Germany}
\affiliation{16}{Swedish Institute of Space Physics, Kiruna, Sweden}
\affiliation{17}{The Blackett Laboratory, Imperial College London, London, UK}

\correspondingauthor{Erika Palmerio}{epalmerio@berkeley.edu}

\begin{keypoints}
\item We analyse the eruption of two coronal mass ejections on 28~April~2012 and their evolution up to Saturn
\item We study and compare the flux rope magnetic structure at the Sun, at Venus, at Earth, and at Saturn
\item We find a single flux rope structure at all planets, suggesting interaction of the two eruptions in the inner heliosphere
\end{keypoints}


\begin{abstract}
One of the grand challenges in heliophysics is the characterisation of coronal mass ejection (CME) magnetic structure and evolution from eruption at the Sun through heliospheric propagation. At present, the main difficulties are related to the lack of direct measurements of the coronal magnetic fields and the lack of 3D in-situ measurements of the CME body in interplanetary space. Nevertheless, the evolution of a CME magnetic structure can be followed using a combination of multi-point remote-sensing observations and multi-spacecraft in-situ measurements as well as modelling. Accordingly, we present in this work the analysis of two CMEs that erupted from the Sun on 28~April~2012. We follow their eruption and early evolution using remote-sensing data, finding indications of CME--CME interaction, and then analyse their interplanetary counterpart(s) using in-situ measurements at Venus, Earth, and Saturn. We observe a seemingly single flux rope at all locations, but find possible signatures of interaction at Earth, where high-cadence plasma data are available. Reconstructions of the in-situ flux ropes provide almost identical results at Venus and Earth but show greater discrepancies at Saturn, suggesting that the CME was highly distorted and/or that further interaction with nearby solar wind structures took place before 10~AU. This work highlights the difficulties in connecting structures from the Sun to the outer heliosphere and demonstrates the importance of multi-spacecraft studies to achieve a deeper understanding of the magnetic configuration of CMEs.
\end{abstract}


\section{Introduction} \label{sec:intro}

Coronal mass ejections \citep[CMEs; e.g.,][]{webb2012} are spectacular eruptions of magnetic fields and plasma that are regularly launched from the Sun throughout the heliosphere. Their magnetic structure, when they leave the solar atmosphere, is that of a flux rope \citep[e.g.,][]{chen2011,forbes2000,green2018,klimchuk2001}, i.e.\ consisting of a bundle of magnetic fields wrapped about a central axis. After erupting, CMEs usually undergo rapid acceleration and expansion in the low corona \citep[e.g.,][]{patsourakos2010a,patsourakos2010b,temmer2008,temmer2010,veronig2018} as a result of the large energy release and the high internal pressure compared to that of the ambient solar wind \citep{demoulin2009a}. After the initial, impulsive phase, i.e.\ from the outer corona outwards, CMEs generally expand in a self-similar fashion \citep[e.g.,][]{good2019,schwenn2005,subramanian2014,vrsnak2019}. The speed at which CMEs propagate depends on the speed of the surrounding solar wind flow, which has the effect of accelerating slower CMEs and decelerating faster CMEs \citep[e.g.,][]{gopalswamy2000,vrsnak2007}. CME expansion in interplanetary space is believed to usually take place up to ${\sim}10$--$15$~AU, where CMEs reach pressure balance with the solar wind \citep[e.g.,][]{richardson2006,vonsteiger2006}. As CMEs propagate through the outer heliosphere, they may interact with stream interaction regions \citep[SIRs; e.g.,][]{richardson2018} or with other CMEs to produce merged interaction regions \citep[MIRs; e.g.,][]{burlaga1986,burlaga1997}, which are believed to dominate the structure of the heliosphere at large heliocentric distances \citep[e.g.,][]{gazis2006,vonsteiger2006}.

In reality, the sparsity of observations throughout the heliosphere, together with the fact that in-situ measurements typically sample a 1D trajectory through a much larger structure, mean that many aspects of CME evolution are yet to be fully understood \citep[for recent reviews on CME evolution, see][]{luhmann2020,manchester2017}. For example, it is unclear to which extent 1D in-situ measurements are representative of the global CME structure \citep[e.g.,][]{alhaddad2011,owens2017}, mainly because of distortions \citep[e.g.,][]{manchester2004,owens2008,savani2010} and/or the particular sampling distance with respect to the CME nose and central axis \citep[e.g.,][]{cane1997,kilpua2011,marubashi2007}. As they propagate through the solar corona and interplanetary space, CMEs are also known to experience deflections and rotations \citep[e.g.,][]{isavnin2014,kay2015a,vourlidas2011,wang2004b}, which may significantly affect the magnetic configuration that is later measured in situ \citep[e.g.,][]{palmerio2018,yurchyshyn2008}. Furthermore, in addition to the difficulties in understanding CME evolution for single-CME events, cases where CMEs interact with solar wind structures \citep[e.g.,][]{heinemann2019,rouillard2010,winslow2016,winslow2021} or with other CMEs \citep[e.g.,][]{dasso2009,farrugia2004,lugaz2017a,scolini2020} are not infrequent, thus complicating things further.

The properties of interplanetary CMEs \citep[or ICMEs; e.g.,][]{kilpua2017b} have been studied mainly around 1~AU \citep[e.g.,][]{cane2003,jian2006b,jian2018,nieveschinchilla2018a,nieveschinchilla2019,owens2018a,regnault2020,richardson2010}, i.e.\ where continuous in-situ measurements of the solar wind have been available for several decades and up to this day. In particular, the subset of ICME ejecta known as magnetic clouds \citep{burlaga1981} have been analysed extensively around Earth's orbit, both in statistical \citep[e.g.,][]{huttunen2005,janvier2014c,li2014,li2018,lynch2003,nieveschinchilla2005,nieveschinchilla2008,rodriguez2016,wood2017} and in case studies \citep[e.g.,][]{kilpua2009a,lynch2010,mostl2008,mostl2009a,nieveschinchilla2011}. Magnetic clouds are characterised by enhanced magnetic field magnitudes, a large and smooth rotation of the magnetic field over one direction, low proton temperatures, and low plasma beta, and they are of particular interest because their magnetic configuration corresponds to that of a flux rope. Interestingly, the first study that identified and defined magnetic clouds in the solar wind, i.e.\ that of \citet{burlaga1981}, was performed using observations from several spacecraft between 1 and 2~AU, thus highlighting the importance of well-separated multi-spacecraft measurements to understand the intrinsic structure of CMEs in the heliosphere.

Indeed, studying the properties of CMEs at different radial distances as they propagate throughout the heliosphere can provide paramount information and insight into CME evolution. Heliocentric orbiters such as Helios 1/2 and Ulysses have enabled long-term observation and analysis of ICMEs away from 1~AU and/or the ecliptic plane \citep[e.g.,][]{bothmer1994,bothmer1998,du2010,jian2008c,lepri2004,liu2005,richardson2014b,riley2000}. Data from the more recently launched Parker Solar Probe, BepiColombo, and Solar Orbiter have resulted in several studies of CMEs measured between the Sun and 1~AU \citep[e.g.,][]{davies2021,korreck2020,lario2020,nieveschinchilla2020,zhao2020,weiss2021}. Considerable results were also achieved using data from spacecraft, such as Pioneer 10/11 and Voyager 1/2, that were launched on escape trajectories out of the solar system and that have provided valuable information on the behaviour of ICMEs at progressively larger heliocentric distances \citep[e.g.,][]{burlaga2001,wang2001,wang2004a}. Finally, another important contribution in understanding CME properties away from 1~AU has come from planetary missions, which usually spend some part of their orbit outside of the planet's bow shock, thus exposing their instruments to the solar wind \citep[e.g.,][]{collinson2015,good2016,janvier2019,jian2008b,lee2017,palmerio2021,winslow2015}.

In-situ measurements of the same CMEs by multiple spacecraft at different radial distances throughout the heliosphere enable studies of their interplanetary evolution \citep[e.g.,][]{burlaga1981,davies2020,delucas2011,good2018,good2019,lugaz2020,salman2020,vrsnak2019}. Additionally, solar and heliospheric remote-sensing observations can be combined with multi-spacecraft in-situ data to obtain a complete picture of the whole solar--heliospheric system when characterising CME evolution \citep[e.g.,][]{asvestari2021,kilpua2019b,mostl2015,nieveschinchilla2012,palmerio2021,prise2015,richardson2002,rodriguez2008,rouillard2009}. In particular, the use of heliospheric imagery has been proven useful to connect CMEs at the Sun to their in-situ counterparts \citep[e.g.,][]{deforest2013,mostl2009b,mostl2017,palmerio2019,rouillard2011,srivastava2018}. ICME structures can be identified in situ using a number of signatures based on magnetic field, plasma, compositional, and energetic particle data \citep[e.g.,][]{zurbuchen2006}. Furthermore, since spacecraft may lack magnetic field and/or plasma instruments (especially those not dedicated to studying the solar wind), reduction in galactic cosmic rays (GCR) measurements known as Forbush decreases \citep{forbush1937,hess1937} can be used as a proxy for ICME passage \citep[e.g.,][]{cane2000,dumbovic2020,freiherrvonforstner2018,freiherrvonforstner2020,papaioannou2020,richardson2011,winslow2018,witasse2017}.

Despite the amount of spacecraft scattered throughout the heliosphere and decades of research efforts, the number of well-observed multi-spacecraft events covering extended radial distances into the outer heliosphere is still exiguous, even when considering both studies of the solar wind and interplanetary magnetic field \citep[e.g.,][]{burlaga2001,prise2015,witasse2017} as well as multi-planet observing campaigns designed to track the auroral response of giant planetary magnetospheres to CME-driven shocks between 5 and 30~AU \citep[e.g.,][]{branduardiraymont2013,dunn2021,lamy2012,lamy2017b,lamy2020,prange2004}. More so, even fewer studies have focussed on the magnetic structure of CMEs in the outer heliosphere. The Bastille event studied by \citet{burlaga2001} at Voyager 2 at 63~AU is considered the most distant magnetic cloud ever observed, but its handedness was reported to be opposite to that of the corresponding ejecta at 1~AU, thus highlighting the difficulties in associating measurements that are separated by large radial distances. 

\begin{figure}[ht]
\centering
\includegraphics[width=.88\linewidth]{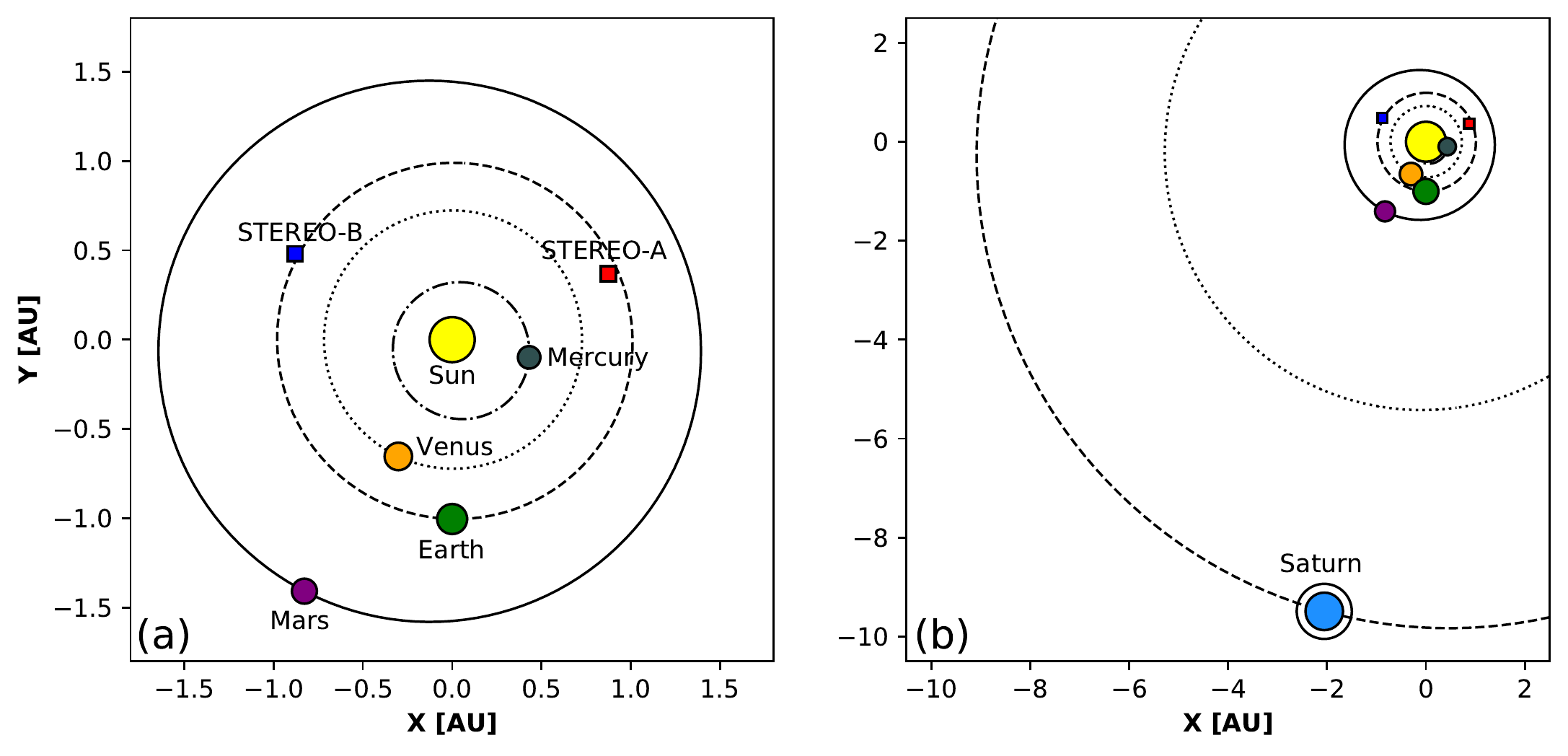}
\caption{Position of various planets and spacecraft until the orbits of (a) Mars and (b) Saturn, on 28~April~2012. The orbits of all planets from Mercury to Saturn are also indicated.}
\label{fig:map}
\end{figure}

This work represents the first in-depth study of the magnetic structure of a (merged) flux rope from the Sun to ${\sim}10$~AU. We analyse here in detail the eruption and evolution of two CMEs that left the Sun on 28~April~2012 and were observed to interact in the inner heliosphere. The favourable position of the spacecraft involved in this study allows us to follow the event in remote-sensing imagery as it propagates away from the Sun and then to study its in-situ signatures and magnetic structure at Venus, Earth, and Saturn. Figure~\ref{fig:map} shows the position of various planets and spacecraft throughout the heliosphere on the day of the eruptions under analysis. This article is organised as follows. In Section~\ref{sec:data}, we enumerate the various instruments that are involved in this study. In Section~\ref{sec:remote}, we describe the CMEs under analysis from a remote-sensing observational perspective. In Section~\ref{sec:models}, we estimate the propagation of the CMEs and their impact at different locations using the remote-sensing observations described in Section~\ref{sec:remote} as input. In Section~\ref{sec:insitu}, we present the in-situ signatures and analyse the magnetic structure of the interacting CMEs at different heliocentric distances. In Section~\ref{sec:discussion}, we discuss different aspects of the event, especially in terms of CME propagation and magnetic structure. Finally, in Section~\ref{sec:conclusions}, we summarise our results and present our conclusions.

\section{Spacecraft and Ground-based Data} \label{sec:data}

We list here the fleet of instruments that are involved in this study, starting from the Sun and moving outwards to Saturn. We combine remote-sensing and in-situ data in order to follow and characterise the event at various locations throughout the heliosphere.

Observations of the solar disc are made from three vantage points, namely Earth and the twin Solar Terrestrial Relations Observatory \citep[STEREO;][]{kaiser2008} spacecraft. The STEREO mission comprises two identical spacecraft that orbit the Sun close to 1~AU, one ahead of Earth in its orbit (STEREO-A) and the other one trailing behind (STEREO-B, which has been out of contact since late 2014). At the time of the events under analysis, STEREO-A and -B were located ${\sim}113^{\circ}$ west and ${\sim}118^{\circ}$ east of the Sun--Earth line, respectively. Extreme ultra-violet (EUV) images are provided by the Atmospheric Imaging Assembly \citep[AIA;][]{lemen2012} onboard the Solar Dynamics Observatory \citep[SDO;][]{pesnell2012} orbiting Earth and the Sun Earth Connection Coronal and Heliospheric Investigation \citep[SECCHI;][]{howard2008a} Extreme UltraViolet Imager (EUVI) onboard STEREO. Line-of-sight magnetograms are available from Earth's viewpoint only, and we use data from the Helioseismic and Magnetic Imager \citep[HMI;][]{scherrer2012} onboard SDO.

Coronagraph observations are made from the same three viewpoints. White-light images from Earth are provided by the Large Angle and Spectrometric Coronagraph \citep[LASCO;][]{brueckner1995} C2 (2.2--6 $R_{\odot}$) and C3 (3.5--30 $R_{\odot}$) cameras onboard the Solar and Heliospheric Observatory \citep[SOHO:][]{domingo1995}. Imagery from STEREO-A and -B is supplied by the SECCHI/COR1 (1.5--4 $R_{\odot}$) and COR2 (2.5 -- 15 $R_{\odot}$) coronagraphs. 

White-light observations of the heliosphere are provided by the Heliospheric Imagers \citep[HI;][]{eyles2009} onboard the twin STEREO spacecraft. Each HI instrument consists of two cameras, HI1 (4--24$^{\circ}$) and HI2 (18--88$^{\circ}$), that image interplanetary space in the vicinity of the Sun--Earth line (the degrees indicate the elongation in helioprojective radial coordinates).

In-situ measurements from Venus are provided by the Venus Express \citep[VEX;][]{svedhem2007} spacecraft. The instruments that we avail ourselves of are the Magnetometer \citep[MAG;][]{zhang2006} and the Analyser of Space Plasmas and Energetic Atoms \citep[ASPERA-4;][]{barabash2007}. From the ASPERA-4 package, we use data taken by the Ion Mass Analyser (IMA) and the Electron Spectrometer (ELS) sensors. VEX terminated operations in early 2015 and was characterised by a 24-hour highly elliptical orbit, most of which was in the solar wind and the remaining couple of hours were spent inside the bow shock of Venus. MAG was operational at all times, whilst ASPERA-4 was turned on for several hours close to periapsis and apoapsis.

In-situ measurements from Earth are provided by the Wind \citep{ogilvie1997} spacecraft, which orbits the Sun from Earth's Lagrange L1 point. Magnetic field and plasma (including solar wind electron distributions) data are taken by the Magnetic Fields Investigation \citep[MFI;][]{lepping1995} and Solar Wind Experiment \citep[SWE;][]{ogilvie1995} instruments. We also use count rate data from the Neutron Monitor Database (NMDB), and specifically from the Thule (THUL), Oulu (OULU), and South Pole (SOPO) observatories on ground, as GCR proxies.

Finally, in-situ measurements from Saturn are provided by the Cassini \citep{matson2002} spacecraft. We use data from the Cassini Magnetic Field Investigation \citep[MAG;][]{dougherty2004}, the Radio and Plasma Wave Science \citep[RPWS;][]{gurnett2004} experiment, and the Magnetosphere Imaging Instrument \citep[MIMI;][]{krimigis2004}. MIMI consists of three different sensors, and we use data from the Low Energy Magnetospheric Measurement System (LEMMS) sensor. The Cassini spacecraft was dismissed in 2017 through a controlled entry into Saturn.


\section{Remote-sensing Observations} \label{sec:remote}

In this section, we provide an overview of the solar events that originated on 28~April~2012 and follow them from the solar disc (Section~\ref{subsec:disc}) through white-light imagery of the corona (Section~\ref{subsec:coronagraph}) and heliosphere (Section~\ref{subsec:hi}).

\subsection{Solar Disc Observations} \label{subsec:disc}

\begin{figure}[ht]
\centering
\includegraphics[width=.99\linewidth]{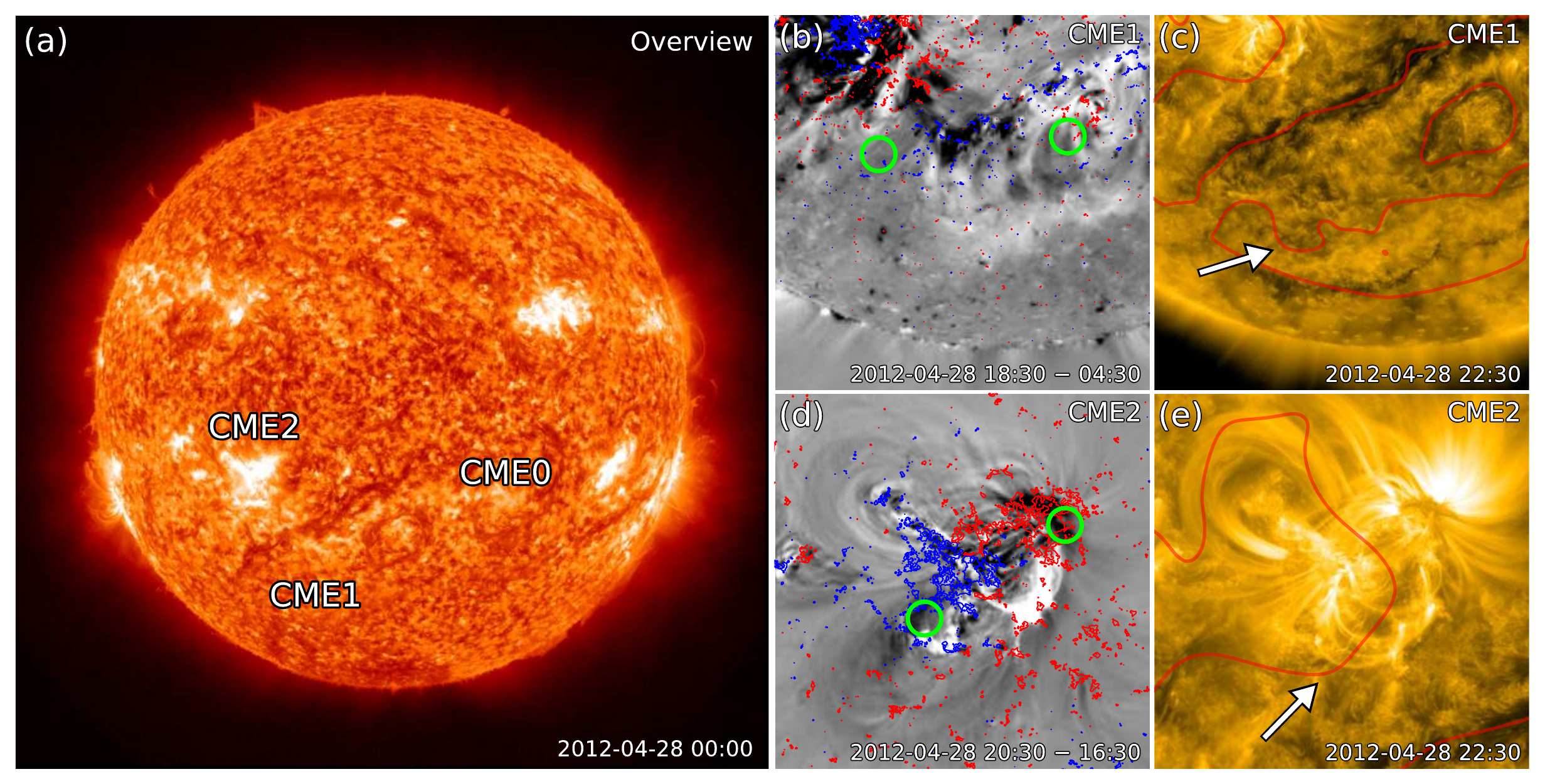}
\caption{Eruptive events on 28~April~2012 as seen in solar disc imagery from SDO. (a) Overview of the solar disc in the 304~{\AA} channel prior to the eruptions, with the approximate source regions of the three CMEs marked in chronological order as `CME0', `CME1', and `CME2', respectively. (b) Base-difference image in the 211~{\AA} channel overlaid with magnetogram contours (red: positive polarity, blue: negative polarity). The footpoints of CME1 are indicated with green circles. (c) Image in the 171~{\AA} channel showing the post-eruption arcade signatures of CME1. The global neutral lines (calculated from smoothed magnetogram data taken at the same time) are overlaid in red, and the main polarity inversion line involved in the eruption is indicated with a white arrow. (d--e) Same as panels (b--c), but for CME2.}
\label{fig:eruption}
\end{figure}

The sequence of events analysed in this article commenced on 28~April~2012, which was a day characterised by several eruptions. Here, we focus on the three major solar events that originated on that day from the Earth-facing disc. Figure~\ref{fig:eruption}a shows the approximate source regions of these CMEs as seen by SDO (i.e., from Earth's perspective). For a complete set of observations of the front-sided eruptions of 28~April~2012 from three viewpoints (STEREO-A, Earth, and STEREO-B), see Movie~S1.

The first major eruption of the day, marked as `CME0' in Figure~\ref{fig:eruption}a, had its onset at ${\sim}$10:00~UT and originated from a quiet-Sun, J-shaped filament located in the southwestern quadrant of the solar disc (approximately at S25W40). Given the large scale of the resulting eruption and the proximity of the filament to the source of the next CME (CME1), it is possible that CME0 triggered the chain of subsequent events \citep[e.g.,][]{schrijver2011,torok2011}. However, this CME was seen to deflect significantly towards the southwest already in the lower corona (see Movie~S1 and Section~\ref{subsec:coronagraph}), making it highly unlikely to interact with the following CMEs or to encounter any observer close to the ecliptic plane. Hence, CME0 will be disregarded for the rest of this study.

The following eruption, marked as `CME1' in Figure~\ref{fig:eruption}a, had its onset at ${\sim}$13:00~UT and originated from a quiet-Sun, U-shaped filament located in the southeastern quadrant of the solar disc (approximately at S45E15). Although this CME erupted from higher latitudes than CME0, it deflected towards the equator (see Movie~S1 and Section~\ref{subsec:coronagraph}), making it more likely to exhibit a significant component along the ecliptic plane and to interact with the following CME2. Hence, we proceed to evaluate the magnetic structure of the flux rope associated with CME1, also known as the ``intrinsic flux rope type''. This can be achieved via a combination of multi-wavelength, remote-sensing observations that yield information on the chirality (or handedness), tilt, and axial field direction of a flux rope during eruption \citep[see][and references therein for a summary of the available proxies]{palmerio2017}. The eruption of CME1 was not characterised by a clear pair of coronal dimmings \citep[usually tracers of a flux rope's footpoints; e.g.,][]{thompson2000,zhukov2004}, but rather by several patches of diffuse dimming regions. Hence, we estimate the CME footpoints to be simply located at either end of the erupting filament's spine, noting however that these are approximate locations because of possible projection effects. The resulting footpoints are shown in Figure~\ref{fig:eruption}b, according to which CME1 was rooted in a positive (negative) polarity to the west (east). Furthermore, we observe the ``roll effect'' (i.e., the sense of bending and twisting) of the filament material off limb in STEREO imagery (see Movie~S1), which can be used as a chirality proxy \citep[e.g.,][]{martin2003,panasenco2008}. We determine that the filament rolled in a right-handed sense, which is also confirmed by the presence of a left-bearing barb \citep[signature of a right-handed CME; e.g.,][]{martin1998a} that we could identify in H$\alpha$ imagery of the solar disc from the Big Bear Solar Observatory (not shown). For the the flux rope tilt, we consider the inclination of the corresponding polarity inversion line and post-eruption arcade \citep[e.g.,][]{marubashi2015}, shown in Figure~\ref{fig:eruption}c. Both are inclined ${\sim}15^{\circ}$ counterclockwise with respect to the solar equator, implying thus a low-inclination flux rope. A right-handed, low-inclination flux rope with an eastward axial field yields a north--east--south (NES) type, following the scheme of \citet{bothmer1998} and \citet{mulligan1998}.

Finally, the last eruption, marked as `CME2' in Figure~\ref{fig:eruption}a, had its onset at ${\sim}$18:00~UT and originated from active region (AR)~11469 (approximately at S15E20), in the vicinity of the eastern footpoint of the CME1 filament. Given the close location of the source region with respect to the centre of the solar disc and noting no major deflections during early evolution (see Movie~S1 and Section~\ref{subsec:coronagraph}), we can expect CME2 to be likely Earth-directed. Hence, we again estimate the intrinsic flux rope type associated with this eruption. Low-coronal signatures of the CME included J-shaped ribbons (shown in Figure~\ref{fig:eruption}d), proxies of right-handed chirality \citep[e.g.,][]{demoulin1996}. Together with coronal dimming pairs, flare ribbons are also tracers of where a flux rope is rooted \cite[e.g.,][]{aulanier2012,janvier2014b}, resulting in the footpoints indicated in Figure~\ref{fig:eruption}d and showing the northern (southern) leg anchored to the positive (negative) magnetic polarity. The polarity inversion line and post-eruption arcade are both inclined ${\sim}55^{\circ}$ counterclockwise with respect to the solar equator, indicating a high-to-intermediate inclination flux rope. A right-handed, high-inclination flux rope with a southward axial field yields an east--south--west (ESW) type, whilst its low-inclination counterpart would be a NES type.

\subsection{Coronagraph Observations} \label{subsec:coronagraph}

\begin{figure}[ht]
\centering
\includegraphics[width=.99\linewidth]{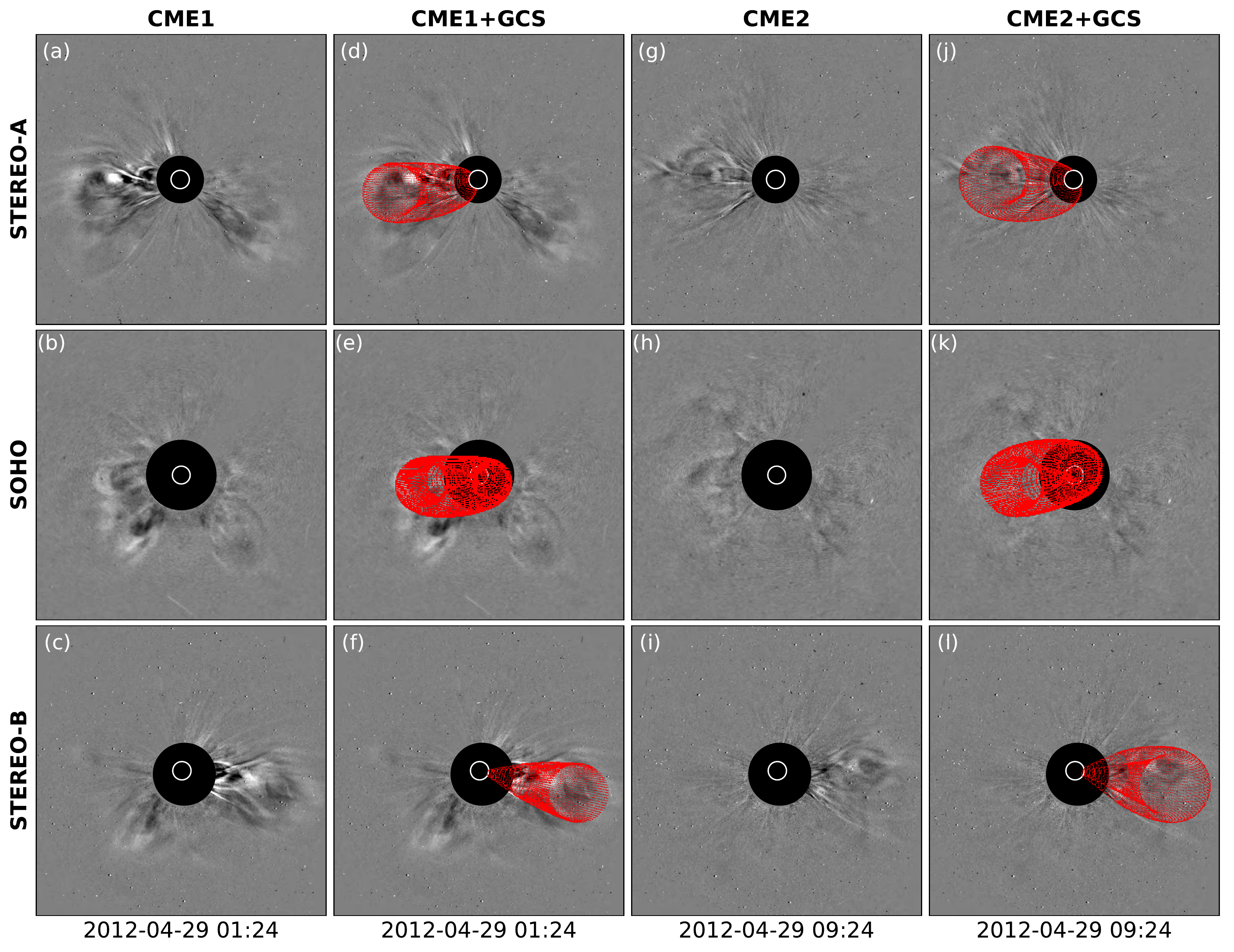}
\caption{Coronagraph observations and GCS reconstructions for CME1 and CME2. (a--c) CME1 shown in difference images from the COR2-A, C3, and COR2-B telescopes, respectively. Background images are taken 1~hour prior to each of the main images. (d--f) Same images as (a--c), with the GCS wireframe overlaid in red. (g--i) Same as (a--c), but for CME2. (j--l) Same as (d--f), but for CME2. All images are taken within 6~minutes of the times reported under each column.}
\label{fig:corona}
\end{figure}

After erupting, the 28~April~2012 CMEs appeared in coronagraph imagery from three locations (STEREO-A, SOHO, and STEREO-B). We provide here an overview of white-light observations of the events throughout the solar corona, with a particular focus on the derivation of geometric and kinematic parameters for the eruptions that we deemed in Section~\ref{subsec:disc} to be possibly Earth-directed (i.e., CME1 and CME2, shown in Figure~\ref{fig:corona}). For a complete set of the coronagraph observations from the three available viewpoints, see Movie~S2.

As mentioned in Section~\ref{subsec:disc}, CME0 (the first to erupt and visible in COR2-A imagery starting at 12:24~UT on 28~April~2012) was seen to propagate mainly below the ecliptic plane (STEREO views) and towards the west (SOHO view), strongly suggesting that the eruption was directed away from the Sun--Earth line and from the following CMEs. This event was particularly faint in STEREO-B imagery, which is not surprising since the CME was propagating fully away with respect to the plane of the sky.

CME1 emerged in coronagraph imagery a few hours after CME0 (around 18:24~UT in COR2-A data) and was seen to propagate at significantly lower latitudes than the previous eruption (see also Section~\ref{subsec:disc}). In addition, the main ``bulk'' of CME1 was preceded by a smaller feature at its southern edge, resulting in a double-lobe structure visible from all viewpoints shown in Movie~S2. We note that asymmetric and complex white-light morphologies of CMEs that originated from filament eruptions have been reported in previous studies \citep[e.g.,][]{palmerio2021,yang2012,zhu2014}. In the case under study, it is likely that the double-lobe structure seen in STEREO imagery resulted from an uneven eruption and disconnection of the corresponding filament legs \citep[e.g.,][]{liu2009,tripathi2006,vourlidas2011}. Furthermore, we note that images of CME1 from the SOHO viewpoint (see e.g.\ Figure~\ref{fig:corona}b) are particularly complex to interpret because of an almost-simultaneous back-sided CME (clearly visible in Figure~\ref{fig:corona}a,c) overlapping due to projection effects. Overall, CME1 did not appear to evolve drastically throughout the solar corona, suggesting that its orientation was maintained similar to that prior to eruption (cf.\ the low-inclination configuration that was inferred from the analysis of solar disc imagery in Section~\ref{subsec:disc}).

Finally, CME2 appeared in coronagraph imagery a few hours after CME1 (around 22:24~UT in COR2-A data) and was seen to propagate mainly along the ecliptic plane. It was the faintest of the three eruptions in white-light data from all available viewpoints (see Movie~S2), possibly because of its passage through an already disturbed corona due to the preceding CME1. The morphology of the eruption in coronagraph images from the two STEREO viewpoints (see also Figure~\ref{fig:corona}g--i) is reminiscent of a low-inclination flux rope seen edge on \citep[see, e.g.,][]{howard2017,krall2006,thernisien2006}, suggesting that the CME rotated slightly upon eruption (cf.\ the high-to-intermediate-inclination configuration that was inferred from the analysis of solar disc imagery in Section~\ref{subsec:disc}).

In order to obtain quantitative estimates of the geometric and kinematic parameters of CME1 and CME2 through the solar corona, we reconstruct both eruptions using the Graduated Cylindrical Shell \citep[GCS;][]{thernisien2009,thernisien2011} model. In the GCS model, a parameterised shell reminiscent of a croissant with its legs attached to the Sun is manually fitted to nearly-simultaneous white-light images until its morphology best matches the observed features. GCS model results are shown in Figure~\ref{fig:corona}d--f for CME1 and Figure~\ref{fig:corona}j--l for CME2. According to our reconstruction results, CME1 had a propagation direction of ($\theta$,$\phi$) = ($-12^{\circ}$,$-18^{\circ}$) and a tilt ($\gamma$) of $5^{\circ}$ with respect to the solar equatorial plane, whilst the corresponding values for CME2 are ($\theta$,$\phi$) = ($-8^{\circ}$,$-24^{\circ}$) and $\gamma = 15^{\circ}$ \citep[here, $\theta$ and $\phi$ are expressed in Stonyhurst coordinates; e.g.,][and a positive $\gamma$ is assumed for counterclockwise rotations]{thompson2006}. Furthermore, we determine the speeds ($v$) of both CMEs through the outer corona by performing GCS reconstructions at two separate times separated by 1~hour, resulting in $v = 405.8$~km$\cdot$s$^{-1}$ for CME1 and $v = 309.2$~km$\cdot$s$^{-1}$ for CME2. Overall, both CMEs propagated slightly towards the southeast as seen from Earth, had low speeds, and were characterised by a low tilt of their axes with respect to the solar equatorial plane. In terms of their resulting magnetic configuration through the corona, it can be assumed that CME1 maintained its NES flux rope type determined in Section~\ref{subsec:disc}, whilst CME2 rotated from its higher-inclination, ESW type inferred in Section~\ref{subsec:disc}. The shortest ``path'' from $\gamma = 55^{\circ}$ to $\gamma = 5^{\circ}$ corresponds to a clockwise motion of the CME axis, which is the expected sense of rotation for a right-handed flux rope \citep[e.g.,][]{green2007,lynch2009}. Hence, these results indicate that both CME1 and CME2 featured a NES flux rope configuration when they left the fields of view of the coronagraphs employed in this study.

\subsection{HI Observations} \label{subsec:hi}

\begin{figure}[ht]
\centering
\includegraphics[width=.99\linewidth]{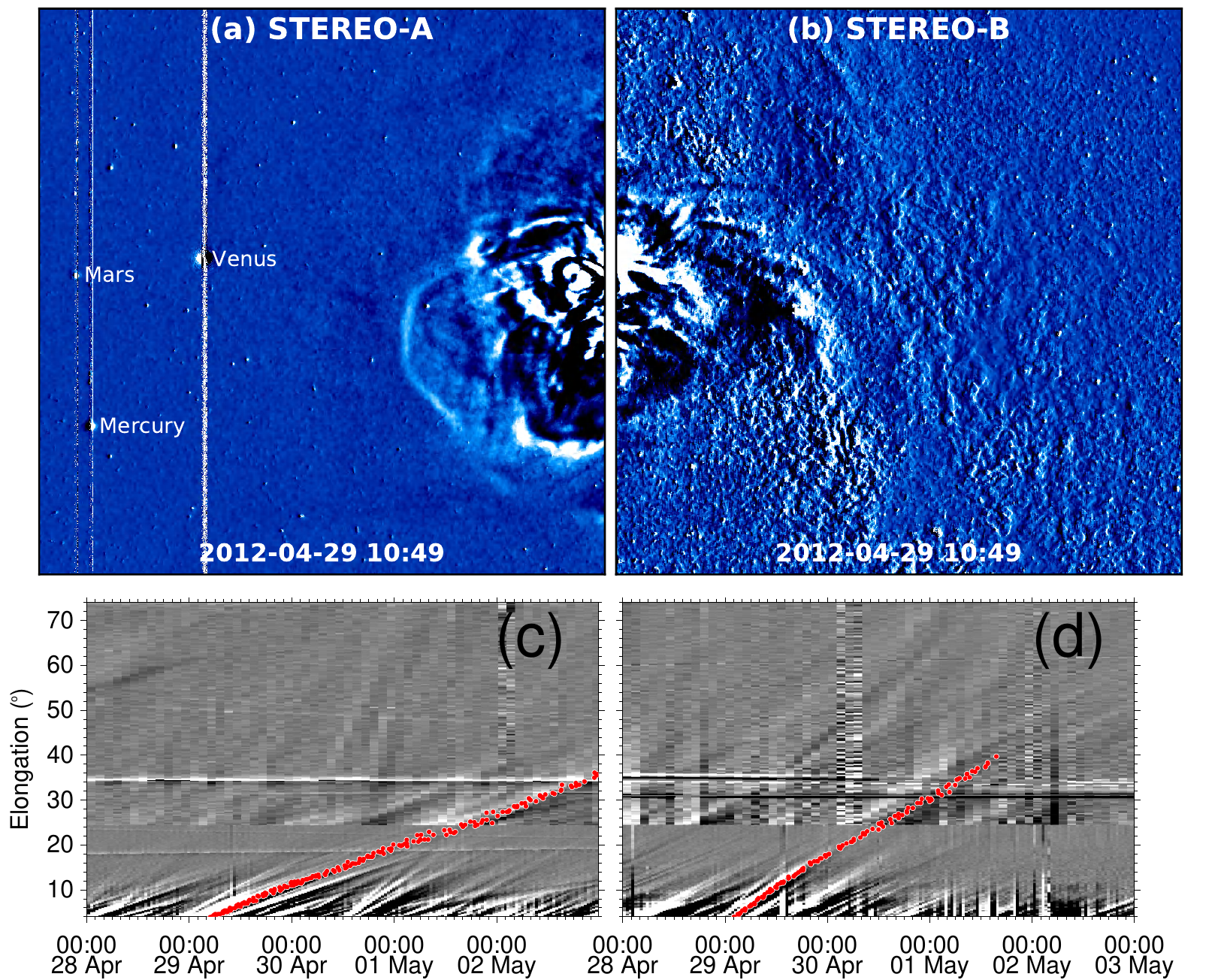}
\caption{CME1 and CME2 seen as a single structure in images from the STEREO/SECCHI/HI cameras. (a--b) The CME seen in running-difference images taken with the (a) HI1-A and (b) HI1-B cameras. The locations of Mercury, Venus, and Mars are marked in panel (a). (c--d) Time--elongation maps from (c) STEREO-A and (d) STEREO-B. The CME is tracked in red. The maps are constructed along position angles of $90^{\circ}$ for STEREO-A and $260^{\circ}$ for STEREO-B.}
\label{fig:hi}
\end{figure}

After leaving the field of view of the COR2 coronagraphs, CME1 and CME2 appeared in images from the HI cameras onboard both STEREO spacecraft. For a complete set of observations from both HI1 cameras, see Movie~S3. An overview of HI observations is shown in Figure~\ref{fig:hi}. CME1 first emerged in HI1-A data at 20:49~UT on 28~April~2012 and in HI1-B data at 22:49~UT on the same day. The complex morphology of CME1, mentioned already in Section~\ref{subsec:coronagraph}, is evident in heliospheric imagery as well. Furthermore, it is difficult to clearly distinguish the front of CME2 amongst the material of the preceding CME1, making the two eruptions appear as a merged, complex structure as they propagate away from the Sun. Figure~\ref{fig:hi}a--b shows a snapshot of the complex structure observed in HI1 imagery, whilst Figure~\ref{fig:hi}c--d shows time--elongation maps constructed using data from both HI1 and HI2 cameras, together with the corresponding tracks of the merged CME.

We note that CME1 and CME2 are also indicated as a single CME in the HELiospheric Cataloguing, Analysis and Techniques Service (HELCATS) catalogues, based on observations made with the STEREO/SECCHI/HI cameras. This event is included in the HICAT catalogue \citep{harrison2018}, which was compiled through visual inspection of HI1 images, and in the HIGeoCAT catalogue \citep{barnes2019}, which was compiled using time--elongation maps and by applying single-spacecraft fitting techniques to derive CME kinematic properties. In both catalogues, CMEs are identified using single-spacecraft data, hence the STEREO-A and STEREO-B observations are presented separately. Amongst the fitting techniques reported in HIGeoCAT, we consider here the results obtained with the Self-Similar Expansion fitting technique \citep[SSEF;][]{davies2012,mostl2013} with a fixed half-width of $30^{\circ}$ applied to time--elongation single-spacecraft data. In the SSEF model, CMEs are assumed to have a circular front and to propagate radially with constant speed and half-width. SSEF results based on STEREO-A observations report a propagation direction of ($\theta$,$\phi$) = ($-16^{\circ}$,$-9^{\circ}$) and a speed of $405$~km$\cdot$s$^{-1}$; whilst SSEF results based on STEREO-B observations report a propagation direction of ($\theta$,$\phi$) = ($-16^{\circ}$,$1^{\circ}$) and a speed of $791$~km$\cdot$s$^{-1}$. Whilst the values for propagation latitude and longitude are consistent with each other and also with the GCS results shown in Section~\ref{subsec:coronagraph}, the SSEF-B speed is approximately twice as large as the SSEF-A one, which is on the other hand basically identical to the speed for CME1 derived using GCS reconstructions.


\section{CME Propagation Modelling} \label{sec:models}

In this section, we estimate the impact locations and arrival times of CME1 and CME2 across the heliosphere. In order to achieve this, we employ two different propagation models that are based on different physical assumptions and observational inputs. The first model (presented in Section~\ref{subsec:ssse}) is based uniquely on HI observations (see also Section~\ref{subsec:hi}) and treats CME1 and CME2 as a single, merged structure. In the second model (presented in Section~\ref{subsec:enlil}) CME1 and CME2 are inserted separately and their input parameters are derived from coronagraph observations (see also Section~\ref{subsec:coronagraph}).

\subsection{SSSE Model} \label{subsec:ssse}

\begin{figure}[ht]
\centering
\includegraphics[width=.99\linewidth]{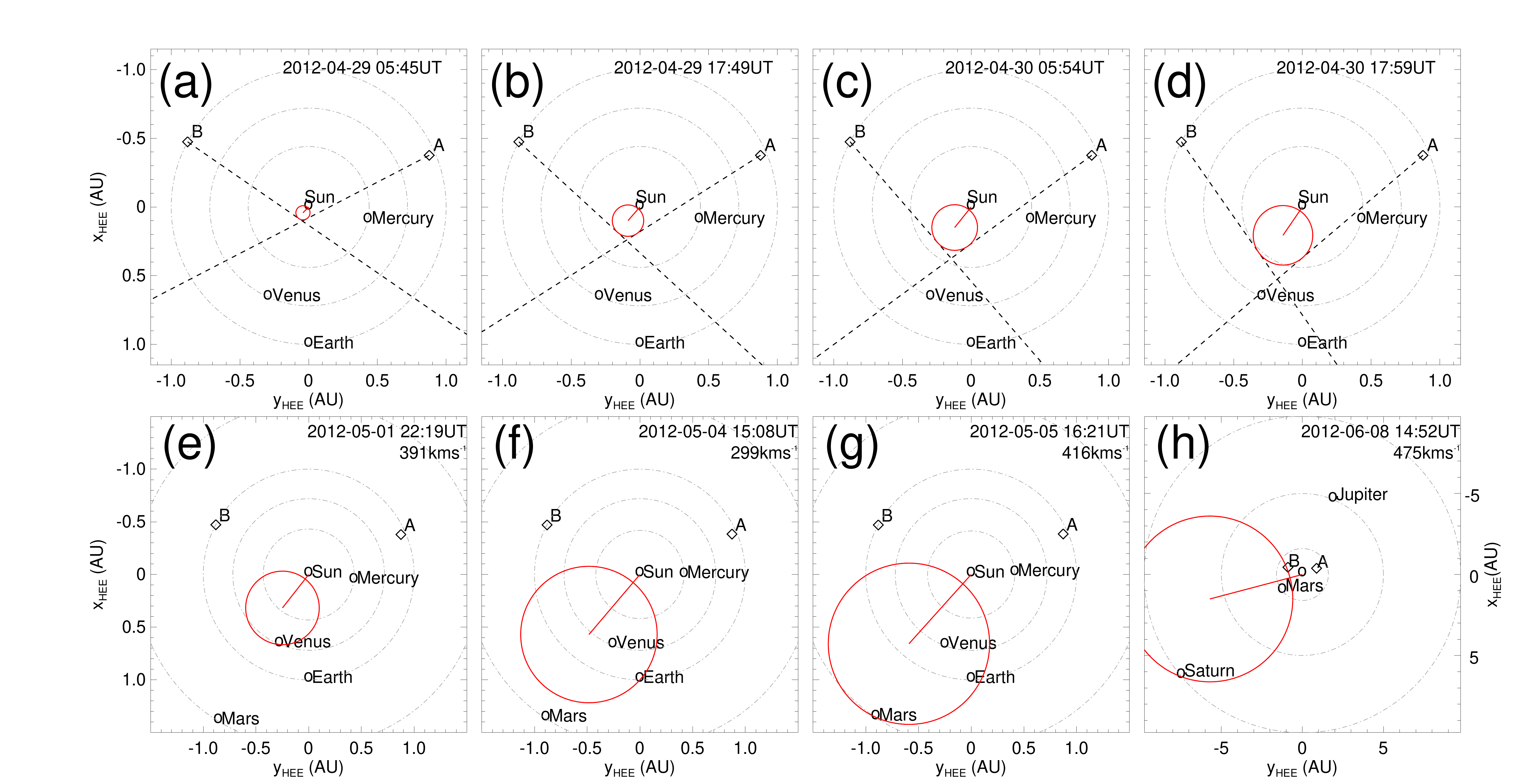}
\caption{Schematic representation of the SSSE method applied to the merged CME1 and CME2 observed in HI data. (a–d) Position of the CME front (red circle) within the ecliptic plane triangulated from the observed leading edge in the HI1 cameras (dashed lines), shown at $\sim$12-hour intervals. (e–h) CME position extrapolated from the last observations to predict arrival times at four planets: (e) Venus, (f) Earth, (g) Mars, and (h) Saturn. Plots are shown in the Heliocentric Earth Ecliptic (HEE) coordinate system.}
\label{fig:ssse}
\end{figure}

The first CME propagation model that we employ in this study is based on HI observations and fittings such as the SSEF results presented in Section~\ref{subsec:hi}. The SSEF results reported in the HELCATS catalogues, however, consider observations made with each spacecraft separately. Since the merged CME1 and CME2 were observed from both STEREO probes, we use here the two-spacecraft version of the SSE model, i.e.\ the Stereoscopic Self-Similar Expansion \citep[SSSE;][]{davies2013} model. With the aid of time--elongation data (see Figure~\ref{fig:hi}) from both spacecraft, we initially triangulate the CME front as it propagates away from the Sun assuming a circular cross-section and fixed half-width \citep[set in this case as $\omega/2 = 60^{\circ}$, since the model was found by][to perform better with angular extents $\gg 30^{\circ}$]{barnes2020}. Then, in order to propagate the CME beyond its last observation time in both HI cameras, we fit a second-order polynomial to the CME apex as a function of time, assuming a constant propagation direction. As a result, the merged CME is predicted to impact Venus (2012-05-01T22:19), Earth (2012-05-04T15:08), Mars (2012-05-05T16:21), and Saturn (2012-06-08T14:52). Figure~\ref{fig:ssse} shows the position of the tracked CME front (top row) and the impact location at the four planets with respect to the apex (bottom row).

\subsection{Enlil Simulation} \label{subsec:enlil}

\begin{figure}[!ht]
\centering
\includegraphics[width=.99\linewidth]{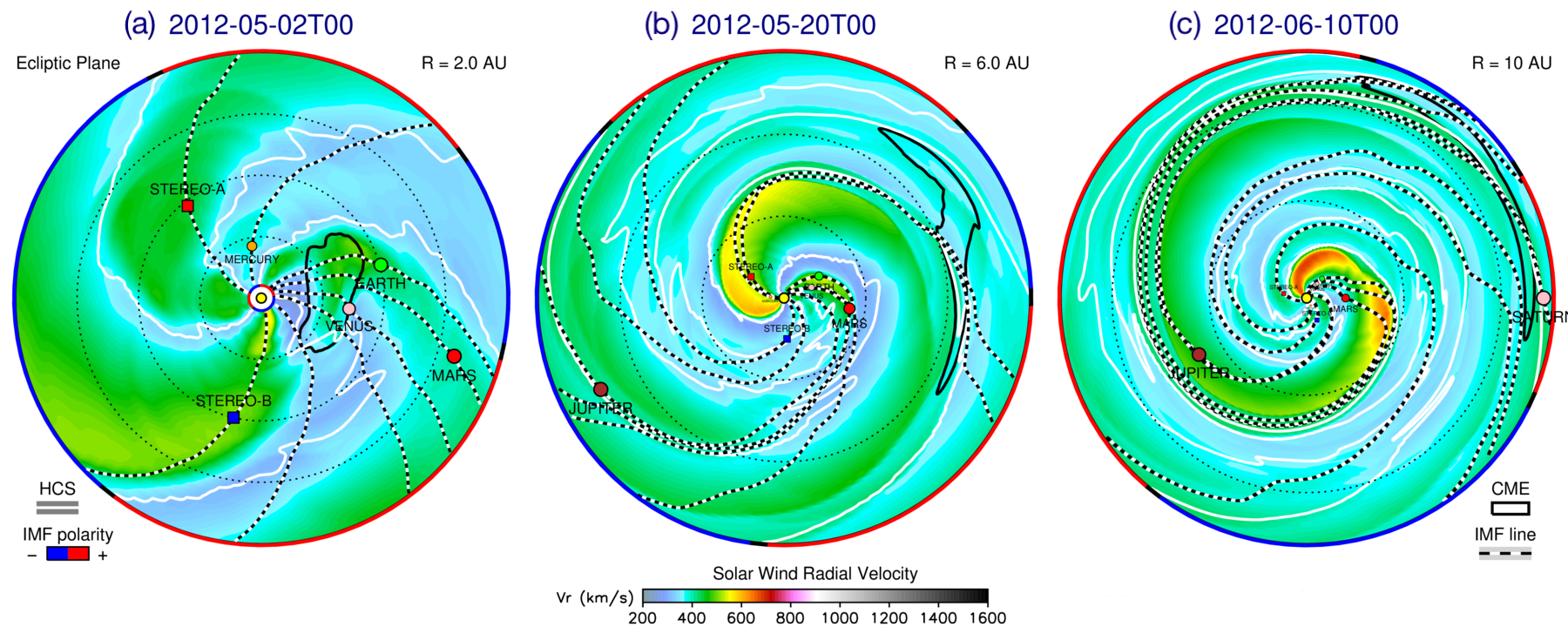}
\caption{Screenshots from the WSA--Enlil+Cone simulation. The parameter shown in the plots is the solar wind radial speed in the ecliptic plane on (a) 2~May~2012, (b) 20~May~2012, and (c) 10~June~2012. The merged CME ejecta is represented with a black countour.}
\label{fig:enlil}
\end{figure}

The second CME propagation model that we use in this work is the 3D heliospheric magnetohydrodynamic (MHD) Enlil \citep{odstrcil2003,odstrcil2004} model. Enlil uses the Wang--Sheeley--Arge \citep[WSA;][]{arge2004} coronal model to generate a background solar wind from its inner boundary (placed at $21.5$\,R$_{\odot}$ or 0.1~AU) onwards. Here, we set the outer boundary of the simulation domain at 10~AU. CMEs are launched through the heliospheric domain at the inner boundary as spherical hydrodynamic structures, i.e.\ lacking an internal magnetic field. The input parameters for CME1 and CME2 are entirely derived from the GCS reconstructions reported in Section~\ref{subsec:coronagraph}. The two CMEs that we inject have an elliptical cross-section and their angular extent is obtained by ``cutting'' a slice out of the GCS shell \citep[see Figure~\ref{fig:corona} and][]{thernisien2011}. The speeds and injection times are derived from the last observations in coronagraph data and by propagating the CMEs up to $21.5$\,R$_{\odot}$ under the assumption of constant speed. As a result, CME1 is launched on 29~April~2012 at 04:32~UT, at ($\theta$,$\phi$) = ($-12^{\circ}$,$-18^{\circ}$) and with axis tilt $\gamma = 5^{\circ}$, speed $v=405.8$~km$\cdot$s$^{-1}$, and half-angular width ($R_\mathrm{max}$,$R_\mathrm{min}$) = ($41.6^{\circ}$,$16.9^{\circ}$); CME2 is launched on 29~April~2012 at 12:42~UT, at ($\theta$,$\phi$) = ($-8^{\circ}$,$-24^{\circ}$) and with $\gamma = 15^{\circ}$, $v=309.2$~km$\cdot$s$^{-1}$, and ($R_\mathrm{max}$,$R_\mathrm{min}$) = ($44.8^{\circ}$,$20.5^{\circ}$). Three screenshots from the simulation, corresponding to when the merged CME was at approximately 1, 5, and 10~AU, are shown in Figure~\ref{fig:enlil}. We note that, although CME2 is slightly slower than CME1, the two eruptions merge early on and appear to propagate away from the Sun as a single structure, possibly due to solar wind preconditioning caused by the prior CME (i.e., CME1), which is a phenomenon seen both in observations \citep[e.g.,][]{liu2014,liu2019,temmer2015} and in simulations \citep[e.g.,][]{desai2020,scolini2020}. The resulting CME-driven shock and/or sheath is predicted to impact Venus (2012-05-01T19:35), Earth (2012-05-02T06:15), Mars (2012-05-05T23:33), and Saturn (2012-06-10T14:10). The merged ejecta is predicted to impact Venus (2012-05-02T00:42), Earth (2012-05-02T13:53), Mars (2012-05-06T06:31), and Saturn (2012-06-11T14:26).

\section{In-situ Measurements} \label{sec:insitu}

In this section, we present and analyse in-situ data from Venus at 0.7~AU (Section~\ref{subsec:venus}), Earth at 1.0~AU (Section~\ref{subsec:earth}), and finally Saturn at 9.7~AU (Section~\ref{subsec:saturn}). We remark that, although both propagation models shown in Section~\ref{sec:models} estimated an impact at Mars, the main focus of this work is the study of the magnetic structure of the CMEs from eruption through heliospheric propagation. Since there was no spacecraft equipped with a solar wind-sampling magnetometer in orbit around Mars at the time of the events presented here, this location is not included in our investigation.

\subsection{Measurements at Venus} \label{subsec:venus}

\begin{figure}[ht]
\centering
\includegraphics[width=.99\linewidth]{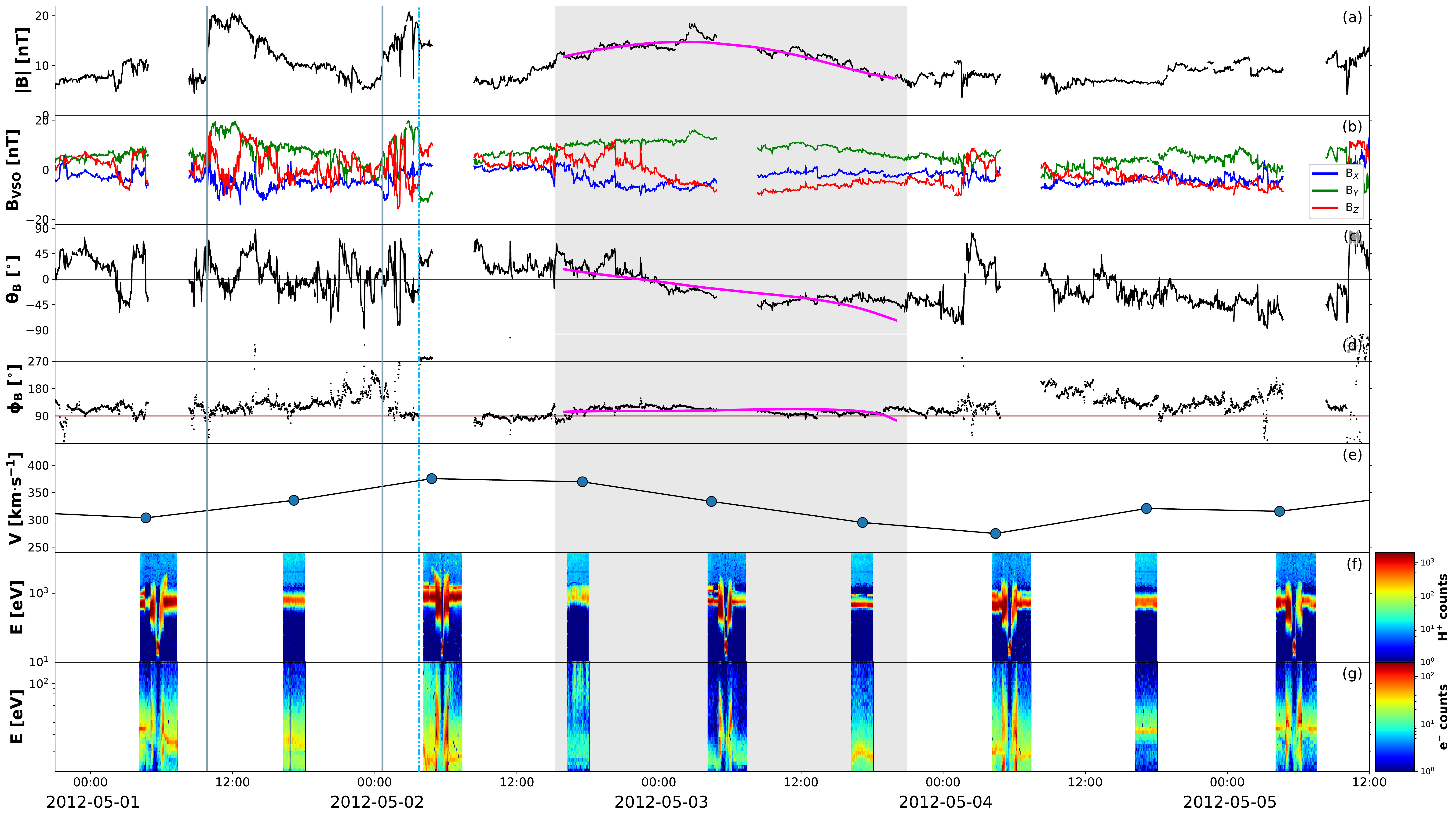}
\caption{Measurements at Venus around the expected arrival time of the merged CME1 and CME2. The parameters shown are: (a) magnetic field magnitude, (b) magnetic field components in Venus Solar Orbital (VSO) Cartesian coordinates, (c) $\theta$ and (d) $\phi$ angles of the magnetic field in VSO angular coordinates, (e) solar wind speed, and (f) proton and (g) electron energy distribution. The solid grey lines indicate two possible interplanetary shocks. The dash-dotted blue line represents a possible start of the magnetic field rotation. The shaded green area marks the period characterised by flux rope signatures and enhanced magnetic field, corresponding to a magnetic ejecta. The magenta curves over magnetic field data within the ejecta show fitting results from applying the Elliptic--Cylindrical flux rope model of \citet{nieveschinchilla2018b}.}
\label{fig:venus}
\end{figure}

The first impact location predicted by both models presented in Section~\ref{sec:models} is Venus. On 28~April~2012, Venus was located ${\sim}25^{\circ}$ east of the Sun--Earth line, at 0.72~AU (see Figure~\ref{fig:map}). In-situ measurements at Venus taken around the expected arrival time of CME1 and CME2 are shown in Figure~\ref{fig:venus}, revealing the passage of a clear, albeit weak, interplanetary disturbance. In particular, two sudden increases in the magnetic field magnitude (at 2012-05-01T09:50 and 2012-05-02T00:39, marked by solid lines in Figure~\ref{fig:venus}) may correspond to two interplanetary shocks, but it is not possible to establish this with certainty because of the lack of high-cadence plasma data. Nevertheless, the solar wind speed displays an increase after each magnetic field jump, suggesting that the two structures may indeed coincide with shocks. These are followed by a period of enhanced magnetic field featuring a rotation in the $\theta_{B}$ component and a steady $\phi_{B}$ (shaded area in Figure~\ref{fig:venus}), characteristic of a flux rope configuration. The exact boundaries (i.e., leading and trailing edge) of this magnetic ejecta are not straightforward to identify, in particular because the rotation seem to extend beyond the enhancement in the magnetic field magnitude (see also the dash-dotted line in Figure~\ref{fig:venus}, which appears to mark the beginning of a smoothly rotating region but is followed by a data gap). Hence, the identified ICME ejecta region (from 2012-05-02T15:14 to 2012-05-03T20:56) is based largely on the magnetic field magnitude (assumed to be greater than its surrounding material). Furthermore, it is unclear whether this structure corresponds to a single ejecta or to the merged CME1 and CME2, especially because of the lack of high-cadence plasma measurements and a data gap in the middle of the flux rope. It has been shown, in fact, that the interaction of two CMEs may result in an ejecta that resembles a coherent, isolated magnetic flux rope \citep[e.g.,][]{kilpua2019b,lugaz2014}.

Visual inspection of the magnetic field components within the ejecta reveals a (weak) rotation from north to south and a constant eastward direction, corresponding to a right-handed, NES flux rope type. This was also the magnetic configuration of both CME1 and CME2 in the outer corona as inferred from the analysis of solar disc and coronagraph imagery (see Sections~\ref{subsec:disc}--\ref{subsec:coronagraph}). We also fit the structure using the Elliptic--Cylindrical (EC) analytical model of \citet{nieveschinchilla2018b}, which is able to describe a magnetic flux rope topology with a distorted cross-section, which may result from interactions with e.g.\ the solar wind. The results are shown over the magnetic field measurements in in Figure~\ref{fig:venus} and the full set of parameters obtained are reported in Table~S1. Apart from vector magnetic field data, the model requires as input the CME average speed, for which we choose a value $v = 333$~km$\cdot$s$^{-1}$ based on the three data points available. According to the EC model, the flux rope is right-handed and has axis orientation ($\theta$,$\phi$) = ($-10^{\circ}$,$128^{\circ}$), fully consistent with a NES-type structure. Furthermore, its cross-section is rather distorted, with a distortion parameter $\delta = 0.47$ ($\delta$ is defined to be 1 for a circular cross-section and 0 for maximum distortion).

\subsection{Measurements at Earth} \label{subsec:earth}

\begin{figure}[ht]
\centering
\includegraphics[width=.99\linewidth]{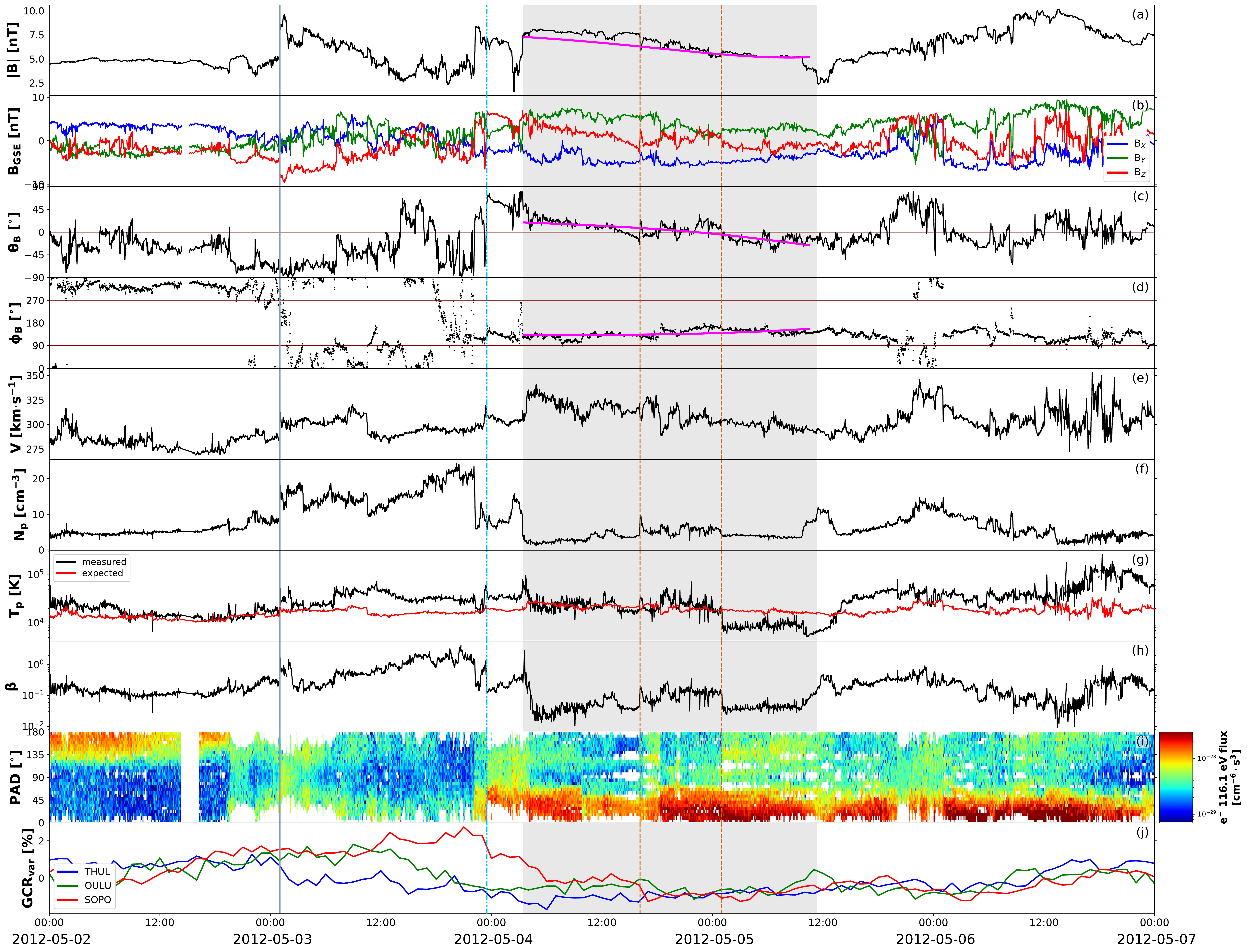}
\caption{Measurements at Earth around the expected arrival time of the merged CME1 and CME2. The parameters shown are: (a) magnetic field magnitude, (b) magnetic field components in Geocentric Solar Ecliptic (GSE) Cartesian coordinates, (c) $\theta$ and (d) $\phi$ angles of the magnetic field in GSE angular coordinates, (e) solar wind speed, (f) proton density, (g) proton temperature \citep[with the expected temperature defined by][shown in red]{richardson1995}, (h) plasma beta, (i) pitch angle distribution, and (j) galactic cosmic rays percentage variation (measured by three different neutron monitors on ground). The solid grey line marks the arrival of the interplanetary shock. The dash-dotted blue line represents the beginning of the magnetic field rotation. The shaded grey area marks the magnetic cloud, with an interaction region delimited by dashed orange lines. The magenta curves over magnetic field data within the ejecta show fitting results from applying the Elliptic--Cylindrical flux rope model of \citet{nieveschinchilla2018b}.}
\label{fig:earth}
\end{figure}

The next impact location predicted by the models presented in Section~\ref{sec:models} is Earth, situated at 1.01~AU on 28~April~2012 (see Figure~\ref{fig:map}). In-situ measurements at Earth taken around the expected arrival time of CME1 and CME2 are shown in Figure~\ref{fig:earth}. The sequence of events starts with an interplanetary shock (at 2012-05-03T01:01, marked by a solid vertical line in Figure~\ref{fig:earth}), followed by a decreasing magnetic field profile similarly to the first structure encountered at Venus (cf. Figure~\ref{fig:venus}). We note that the shock is remarkably slow ($v\sim 300$~km$\cdot$s$^{-1}$), which has however been shown to be possible for slow CMEs characterised by significant expansion that travel through a slow upstream solar wind with a low magnetosonic speed \citep{lugaz2017b}. We do not find signatures of a second shock at Earth. A second sharp increase in the magnetic field magnitude (similar to the one detected at Venus) is not associated with a solar wind speed jump and, furthermore, density and temperature feature a decrease. This suggests that the second discontinuity at Venus was also not a shock or that the second shock dissipated between 0.7 and 1.0~AU. The structure (between 2012-05-04T03:25 and 2012-05-05T11:23) following the second sharp magnetic field increase displays clear signatures of a magnetic cloud (shaded region in Figure~\ref{fig:earth}), albeit with a low magnetic field strength and several complex characteristics. In particular, the centre of the identified flux rope (between the dashed orange lines in Figure~\ref{fig:earth}) features a region characterised by less smooth magnetic field, an irregular speed profile, and enhanced density and plasma beta. Furthermore, the proton temperature is relatively high throughout the first portion of the ejecta and finally drops below expected levels only after the high-beta region (the same trend is observed for alpha particles, not shown here). Unfortunately, heavy ion composition or charge state data from the Advanced Composition Explorer (ACE) spacecraft are not available because of a data gap. Nevertheless, these characteristics are consistent with the merging CMEs or multiple magnetic clouds scenario described by, e.g., \citet{lugaz2014} and \citet{wang2003}, in which the central portion represents the interaction region between the two original ejecta (in this case, CME1 and CME2). Finally, we note that the interplanetary disturbance is associated with a weak (${\sim}2$\% variation) Forbush decrease, registered at all neutron monitors considered here but with different onset times, ranging from the arrival of the interplanetary shock to the beginning of the magnetic field rotation (dash-dotted line in Figure~\ref{fig:earth}).

The magnetic cloud boundaries identified here coincide with the ones defined in the NASA--Wind ICME list \citep{nieveschinchilla2018a}. Visual inspection of the magnetic field components within the ejecta reveal a very similar configuration to the one encountered at Venus (see Section~\ref{subsec:venus}), i.e.\ characterised by a north--south rotation of the helical field and an eastward axial field, forming a NES-type flux rope. This picture is consistent with CME1 and CME2 (both of NES type in the outer corona, see Section~\ref{subsec:coronagraph}) reconnecting in interplanetary space before reaching Venus and coalescing into a single NES flux rope. Again, we fit the structure using the EC model and assuming a CME average speed of $v = 310$~km$\cdot$s$^{-1}$ directly from in-situ data. The results are shown over the magnetic field measurements in Figure~\ref{fig:earth} and the full set of parameters obtained are reported in Table~S1. The resulting flux rope is right-handed, has axis orientation ($\theta$,$\phi$) = ($2^{\circ}$,$145^{\circ}$), and its distortion parameter is $\delta = 0.46$. We note that these results are compatible with those at Venus, i.e.\ consistent with a NES flux rope that is rather distorted.

\subsection{Measurements at Saturn} \label{subsec:saturn}

\begin{figure}[ht]
\centering
\includegraphics[width=.99\linewidth]{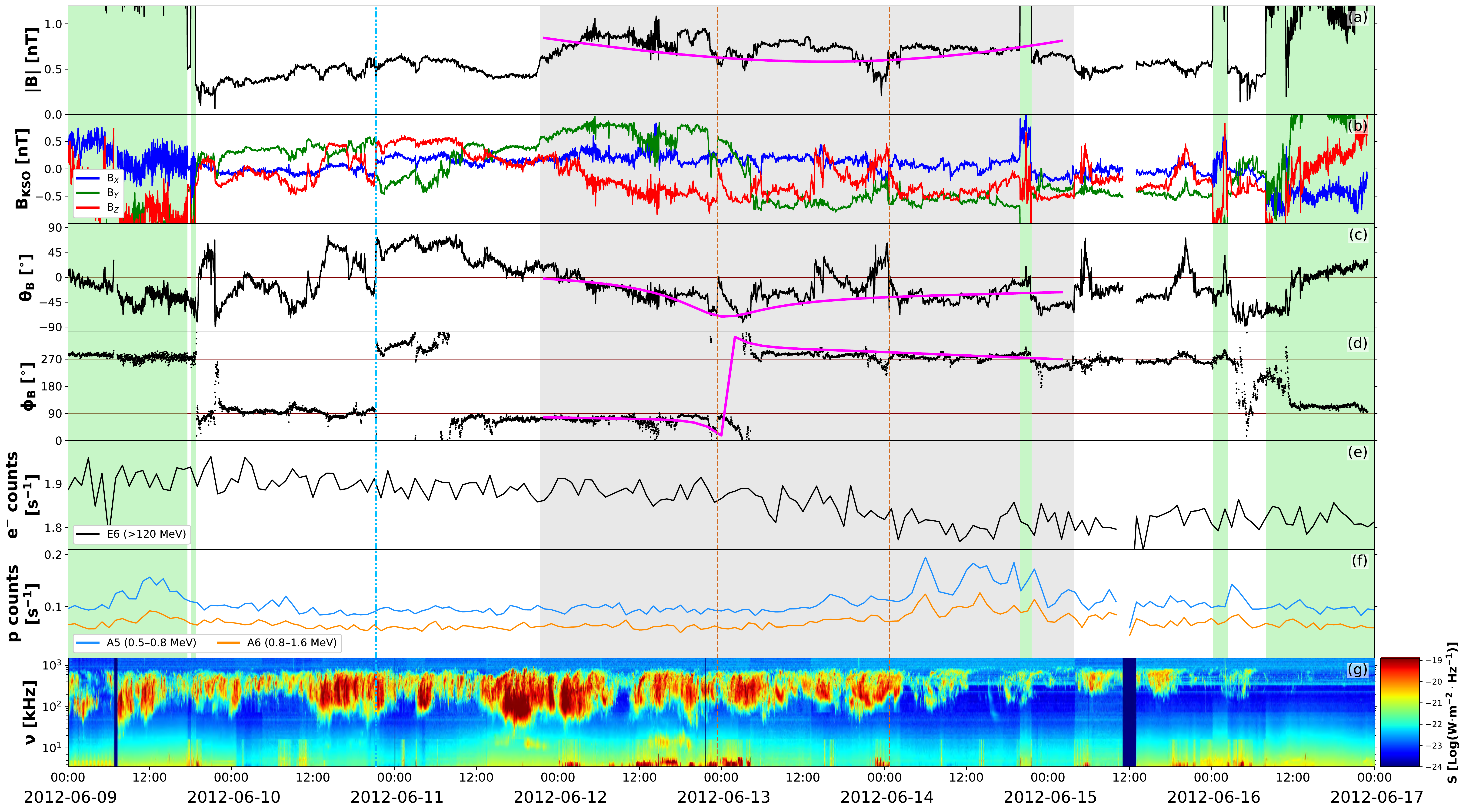}
\caption{Measurements at Saturn around the expected arrival time of the merged CME1 and CME2. The parameters shown are: (a) magnetic field magnitude, (b) magnetic field components in Kronocentric Solar Orbital (KSO) Cartesian coordinates, (c) $\theta$ and (d) $\phi$ angles of the magnetic field in KSO angular coordinates, (e) electron (including penetrating GCR protons) and (f) proton count rates (with the corresponding LEMMS channels and proton energy ranges indicated), and (g) dynamic spectrum of SKR spectral flux density normalised to 1~AU. Green-shaded regions correspond to periods in which Cassini was inside the Kronian magnetosheath. The beginning of the period featuring magnetic field rotation is marked with the dash-dotted blue line, and the period of enhanced magnetic field magnitude is shaded in grey, with a possible interaction region indicated by the dashed orange lines. The magenta curves over magnetic field data within the ejecta show fitting results from the Elliptic--Cylindrical flux rope model of \citet{nieveschinchilla2018b}.}
\label{fig:saturn}
\end{figure}

Finally, the last location where an arrival of CME1 and CME2 is predicted is Saturn, which was positioned at 9.73~AU from the Sun and ${\sim}12^{\circ}$ east of Earth on 28~April~2012 (see Figure~\ref{fig:map}). Even though CMEs usually take a relatively long time to reach ${\sim}10$~AU \citep[about a month; e.g.,][]{prange2004}, Saturn's 29-year orbit results in the planet moving about $1^{\circ}$ in longitude per month and, thus, can be considered essentially fixed in space throughout the analysed period. In-situ measurements at Saturn taken around the expected arrival time of CME1 and CME2 are shown in Figure~\ref{fig:saturn}. Providentially, Cassini exited the Kronian magnetosheath between 9--16~June~2012 and was fully immersed in the solar wind during most of this time interval. On the other hand, Cassini's plasma spectrometer was permanently turned off on 2~June~2012 (i.e., just a week before the start of the data shown in Figure~\ref{fig:saturn}) due to short circuits in the instrument, hence there are no measurements such as solar wind speed and proton density available. Nevertheless, we complement magnetic field data with particle measurements and radio observations of Saturn's Kilometric Radiation \citep[SKR; e.g.,][]{kaiser1984,lamy2008,warwick1981}. The SKR is Saturn's primary radio emission and is generated by auroral electrons that are accelerated along field lines rooted around Saturn's auroral oval \citep[e.g.,][]{lamy2009,lamy2017a}. The emission is highly dependent on solar wind conditions \citep[e.g.,][]{bradley2020,clarke2009,desch1982,kurth2005,lamy2018}, and previous studies have reported SKR enhancements concurrently with the passage of ICMEs \citep[e.g.,][]{crary2005,palmaerts2018} and SIRs \citep[e.g.,][]{badman2008,kurth2016}.

First of all, the magnetic field measurements shown in Figure~\ref{fig:saturn} reveal the passage of a (weak) disturbance, since the (ambient) interplanetary magnetic field around Saturn's orbit has usual magnitudes below 0.5~nT \citep[e.g.,][]{jackman2008,echer2019}. In this case, the magnetic field magnitude reaches values just under 1~nT, which is rather low \citep[cf. the ${\sim}$2~nT measurements reported by][for an ICME in 2014 November]{witasse2017}, but is on the other hand consistent with the ``weakness'' of the transient measured at Earth (${\sim}$7.5~nT, see Figure~\ref{fig:earth}). We do not find signatures of an interplanetary shock, but we could identify a period of smoothly rotating magnetic field (starting at 2012-06-10T21:24, marked by the dash-dotted blue line in Figure~\ref{fig:saturn}) that includes the interval of enhanced magnetic field magnitude (between 2012-06-11T21:24 and 2012-06-15T03:50, shaded grey region in Figure~\ref{fig:saturn}). We note that the central portion of this interval (bounded between the dashed orange lines in Figure~\ref{fig:saturn}) is characterised by a more irregular $\theta_{B}$ component that is reminiscent of the one encountered at Earth (see Figure~\ref{fig:earth}), i.e.\ indicating a possible interaction region. In order to confirm that an interplanetary disturbance has indeed impacted Saturn, we also examine energetic particle observations. In this regard, \citet{roussos2018,roussos2020} showed that Cassini observations of solar energetic particle (SEP) and GCR transients can be used to identify disturbed solar wind conditions around Saturn, especially due to CMEs and SIRs. This is because SEPs and GCRs are able to penetrate Saturn's magnetosphere, hence they can be monitored at all times \citep[e.g.,][]{roussos2008,roussos2011}. During our period of interest, we find signatures of a Forbush decrease in the GCR data that are indirectly monitored with the LEMMS electron channel E6 \citep[due to penetrating GCR protons;][]{roussos2019}. This event was included by \citet{roussos2018} in their list of SEP and GCR transients at Saturn between 2004 and 2016. The decrease commenced around 10~June~2012 and reached its minimum around 14~June, concurrently with an increase in the LEMMS proton channels A5/A6 \citep[the recovery phase of the Forbush decrease follows the plotted interval and is shown in its full extent by][]{roussos2018}. Furthermore, SKR data feature increased emission between 2012-06-10T13:08 and 2012-06-14T02:16, i.e.\ over an interval that is in agreement with the Forbush decrease. The temporal extent of the SKR intensification is consistent with enhancements triggered by solar wind pressure fronts, as opposed to transient ones triggered by planetary rotation \citep[lasting a few hours; e.g.,][]{reed2018}. Finally, we note a peak in proton counts occurring on 9~June~2012 shortly before the interval of enhanced SKR emission. It is unclear whether this structure has a magnetospheric or solar origin, especially since it was observed when Cassini was inside the Kronian magnetosheath. Nevertheless, the features detected during 10--14~June~2012 (in magnetic field, particles, and SKR) occur approximately at the same time, and strongly suggest that a solar transient, and in particular an ICME, impacted Saturn during the observed period.

The magnetic configuration of the entire structure characterised by smoother magnetic field components (i.e., from the dash-dotted blue line to the end of the shaded grey area in Figure~\ref{fig:saturn}) exhibits a larger rotation than that expected for an axial-symmetric flux rope (the $\phi_{B}$ component rotates as west--east--west, resulting in a $360^{\circ}$ rotation in the $B_{Y}$--$B_{Z}$ hodogram). Similar events displaying a rotation greater than $180^{\circ}$ in the magnetic field were analysed at 1~AU by \citet{nieveschinchilla2019}, who suggested that such events can be interpreted as flux ropes with significant curvature and/or distortion or as more complex topologies, such as a spheromak or a double flux rope. Since it is not unusual for interplanetary transients to interact and possibly merge by 1~AU, one may expect even more interaction and complexity in structures detected in the outer heliosphere \citep[e.g.,][]{hanlon2004,prise2015}. Hence, it is not possible to establish with certainty whether the observed configuration stems from a single, intrinsically complex structure, from the interaction of different structures, or if the region preceding the period of enhanced magnetic field magnitude corresponds to a ``smoothed'' sheath. Nevertheless, considering only the shaded grey region in Figure~\ref{fig:saturn}, visual inspection of the magnetic field yields an ESW flux rope type, still right-handed but corresponding to a ${\sim}90^{\circ}$ counterclockwise rotation of the configuration found at Venus and Earth (NES). Again, we fit the structure using the EC model. In this case, due to the lack of plasma data at Saturn, we assume an average CME speed of $v = 380$~km$\cdot$s$^{-1}$ by considering the transit times of the flux rope leading and trailing edges from Earth to Saturn. Fitting results are shown over the magnetic field measurements in in Figure~\ref{fig:saturn} and the full set of parameters obtained are reported in Table~S1. The resulting structure is right-handed (as expected), but its axis has orientation ($\theta$,$\phi$) = ($-9^{\circ}$,$3^{\circ}$), i.e.\ consistent with a low-inclination flux rope. Although the reconstructed magnetic field components fit well to the data, the observed east--west rotation is attributed in the model to a crossing more parallel to the central axis of a low-inclination flux rope, rather than a crossing perpendicular to the axis of a high-inclination one. Furthermore, the resulting distortion parameter is $\delta = 0.21$, indicating that the structure is highly distorted.


\section{Discussion} \label{sec:discussion}

In this section, we synthesise the multi-spacecraft observations, modelling results, and interpretations presented in Sections~\ref{sec:remote}, \ref{sec:models}, and \ref{sec:insitu} and discuss them in the context of two main aspects: the propagation of the interacting CME1 and CME2 to 10~AU (Section~\ref{subsec:propagation}) and the evolution of their magnetic structure from the Sun to Saturn (Section~\ref{subsec:mag}).

\subsection{CME Propagation} \label{subsec:propagation}

CME1 and CME2 erupted ${\sim}$5~hours apart from the southeastern quadrant of the Earth-facing disc, and their source regions were separated by ${\sim}30^{\circ}$ in latitude (see Section~\ref{subsec:disc}). Through the solar corona (see Section~\ref{subsec:coronagraph}), CME1 could be seen to deflect significantly towards the solar equator (its apex position changed from approximately S45E15 to S12E18), whilst CME2 propagated radially (its apex direction could be considered basically unchanged, from S15E20 at the Sun to S12E24 in the corona). This resulted in the two CMEs travelling away from the Sun in close succession and on a very similar trajectory. Despite CME2 appearing slightly slower (by ${\sim}100$~km$\cdot$s$^{-1}$) than CME1 at an altitude of ${\sim}15$\,$R_{\odot}$, the two eruptions could not be clearly distinguished in HI imagery (see Section~\ref{subsec:hi}), possibly indicating that they interacted somewhere in the HI1 field of view. This outcome occurs also in the Enlil simulation (Section~\ref{subsec:enlil}), where the CMEs appear as a merged structure throughout the heliospheric domain even if they were inserted separately at the inner boundary of $21.5$\,$R_{\odot}$. We suggested that this scenario can be attributed to solar wind preconditioning \citep[e.g.,][]{temmer2017}, which allowed CME2 to travel through a rarefied background experiencing little to no drag and thus to run into CME1. As a consequence, we considered CME1 and CME2 as a single, interacting structure when estimating their propagation throughout the heliosphere.

In Section~\ref{sec:models}, we used two models to evaluate the arrival times of the interacting CME1 and CME2 at different locations and to aid interpretation of the in-situ measurements shown in Section~\ref{sec:insitu}. The two propagation models that we employed are substantially different in their physics, assumptions, and observational input: the SSSE model (Section~\ref{subsec:ssse}) consists of a 2D circular cross-section reconstructed using HI data and is then propagated outwards assuming constant acceleration, whilst Enlil (Section~\ref{subsec:enlil}) is a full 3D MHD model of the heliosphere in which we inserted CME1 and CME2 as hydrodynamic pulses and with input parameters based on coronagraph imagery. Nevertheless, both techniques estimated impacts at the same locations (Venus, Earth, Mars, and Saturn) and within reasonable temporal windows compared to the actual in-situ observations. In particular, the predicted arrival time at Saturn was remarkably accurate (with same-day precision) in the case of Enlil, and about 3~days off for SSSE (corresponding to a 7\% error for a ${\sim}$43-day propagation). We note that, regardless of these encouraging results, both models have made a number of simplifying assumptions. For example, the treatment of CMEs as hydrodynamic pulses within Enlil is likely to affect the predicted arrival times due to the absence of internal magnetic forces that can contribute to the acceleration profile. However, as a CME travels away from the Sun these forces are expected to be less influential, especially at radial distances of a few AU and through the outer heliosphere. The SSSE model, on the other hand, treats CMEs as self-similarly propagating spherical fronts under the assumption of constant half-width and acceleration, from which it would be unrealistic to expect precise estimates at 10~AU. Nevertheless, these results suggest that even simplifying models can be used to successfully approximate a CME arrival time window up to the outer heliosphere.

\begin{figure}[t!]
\centering
\includegraphics[width=.99\linewidth]{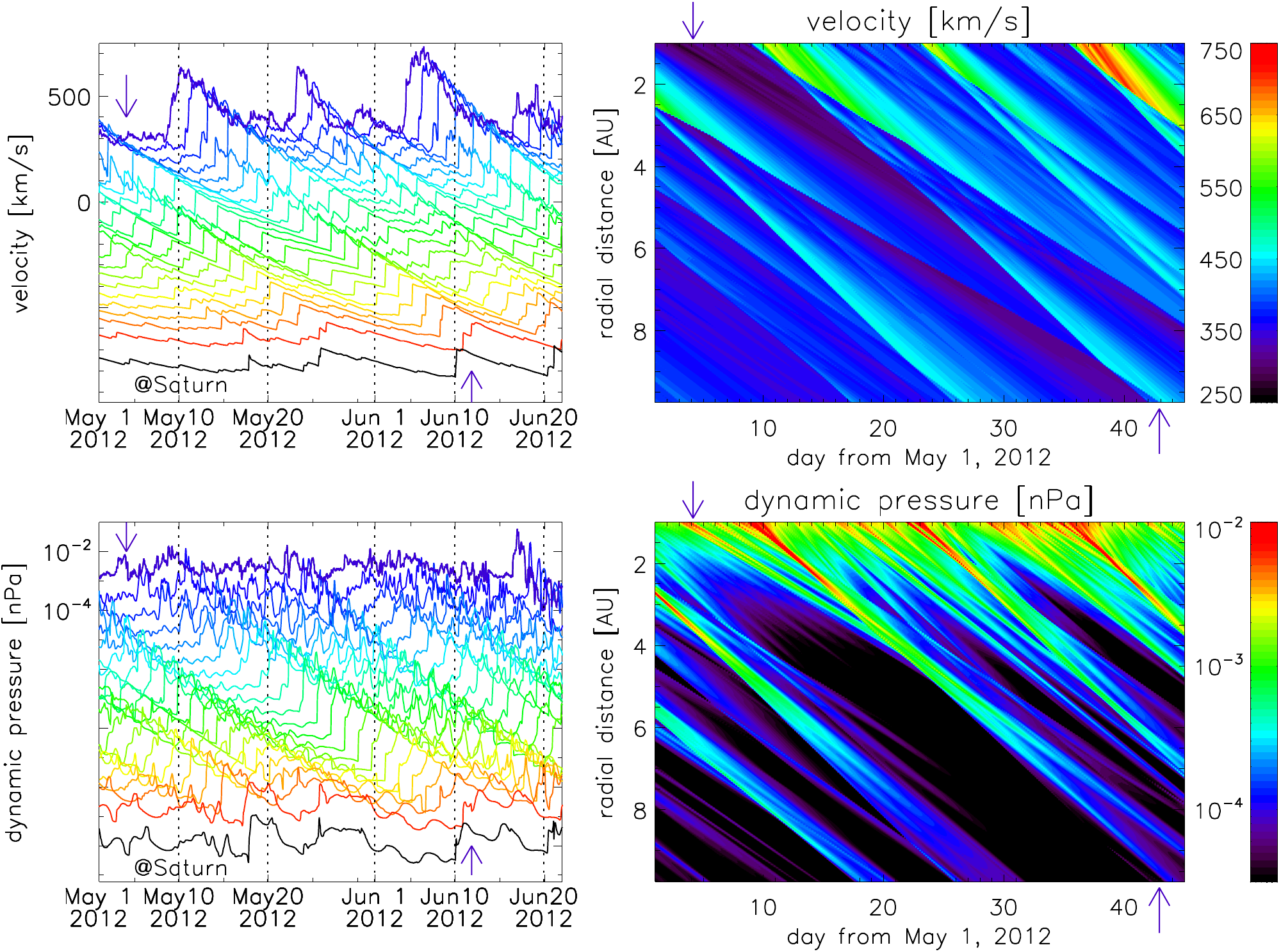}
\caption{Solar wind propagated from 1 to 9.75~AU using the \citet{tao2005} model. The top panels show solar wind speed, whilst the bottom panels show dynamic pressure. Results are shown in both line (left) and contour (right) formats. In the line plots, the values for speed and dynamic pressure beyond 1~AU are progressively shifted towards lower values for visibility (values at different heliocentric distances can be read in the contour plots). In the line plots, the blue lines show measurements at Earth, the red lines show solar wind propagated to 9.75~AU along the Sun--Earth line, and the black lines at the bottom show model results shifted to Saturn's position. In all panels, the downward-pointing arrows mark the ICME ejecta arrival time at Earth (see Figure~\ref{fig:earth}), whilst the upward-pointing arrows mark the ICME ejecta arrival time at Saturn (see Figure~\ref{fig:saturn}). }
\label{fig:taomodel}
\end{figure}

As a further indication of the solar wind propagation that occurred between 1 and 10~AU, we employ the 1D numerical MHD model of \citet{tao2005}, which uses in-situ measurements from Earth or the STEREO spacecraft to estimate the interplanetary conditions further out in the heliosphere, such as the orbits of Jupiter \citep[e.g.,][]{dunn2020}, Saturn \citep[e.g.,][]{provan2015}, and Uranus \citep[e.g.,][]{lamy2017b}. In this case, we use data at Earth (closest to Saturn in longitude, see Figure~\ref{fig:map}) and propagate the measurements up to ${\sim}$10~AU along the Sun--Earth line. The results for speed and dynamic pressure are shown in Figure~\ref{fig:taomodel}. The ICME at Earth was followed by a high-speed stream (HSS), which is estimated to have arrived at Saturn around 2012 June~10. The solar wind preceding this HSS (which includes the ICME measurements shown in Figure~\ref{fig:earth}) is expected to be caught up by the fast stream around 25~May~2012 at a heliocentric distance of ${\sim}5$~AU, possibly resulting in compression and acceleration of the ICME from behind (this is consistent with the Enlil simulation, see Figure~\ref{fig:enlil}b). The arrival time of these features at 10~AU matches quite well with the Enlil results (see Section~\ref{subsec:enlil}) and the in-situ measurements at Saturn (see Section~\ref{subsec:saturn}), further confirming the likelihood of our connection. 

We also explored the level of interaction and possible merging between CME1 and CME2 through interplanetary space. We emphasise that, whilst the Enlil simulation estimated the two eruptions to merge well before reaching Mercury's orbit, it is not possible to fully model the nature and outcome of their interaction due to the lack of an internal magnetic field in the purely hydrodynamic representation of CMEs. Hence, our interpretation is entirely based on the in-situ observations presented in Section~\ref{sec:insitu}. At Venus, we did not find clear indications of interaction, possibly because of the lack of high-cadence plasma data and a data gap exactly at the centre of the flux rope interval that we identified (see Figure~\ref{fig:venus}). At Earth, a seemingly single magnetic cloud structure displayed less smooth magnetic field, variable speed, and enhanced plasma parameters at its centre, which could be considered signatures of the interaction between CME1 and CME2 (see Figure~\ref{fig:earth}). At Saturn, the region that we selected as the magnetic ejecta featured similar magnetic field characteristics at its centre as the ones encountered at Earth (see Figure~\ref{fig:saturn}). However, we could not determine with confidence whether the observed structure corresponded entirely to the ejecta detected at Venus and Earth, or whether additional material was added to it via further interaction with nearby structures in the solar wind. At all locations, fitting the entire ejecta interval as a single flux rope yielded the best results, suggesting that, if interaction was present, it resulted in the two eruptions slowly merging and travelling as a single magnetic structure, rather than in more extreme outcomes such as the second CME overcoming and compressing the first. This is plausible, considering that CME1 and CME2 were ejected close in time and with comparable speeds. An alternative interpretation is that CME1 skimmed the inner planets and its signatures were observed following the first interplanetary shock, and the entire ejecta intervals at Venus and Earth belonged on the other hand to CME2.

\subsection{CME Magnetic Structure} \label{subsec:mag}

Both CME1 and CME2 erupted from the Sun as right-handed flux ropes, but with slightly different orientations according to our analysis of solar disc imagery (Section~\ref{subsec:disc}): CME1 featured a low-inclination NES type, whilst CME2 displayed a high-to-intermediate inclination between an ESW and a NES type. In the solar corona, CME1 was observed to largely maintain its orientation, whilst CME2 appeared to have slightly rotated in a clockwise direction, from which we inferred that both eruptions consisted of NES-type flux ropes based on coronagraph imagery and reconstructions (Section~\ref{subsec:coronagraph}). The two CMEs were hard to distinguish as separate structures in HI data (Section~\ref{subsec:hi}), suggesting that they interacted in some capacity in the inner heliosphere. At Venus (Section~\ref{subsec:venus}), we found a single flux rope signature of NES type (based on both visual inspection and flux rope fitting), which did not display evident indications of interaction (possibly because of lacking high-cadence plasma data and magnetic field measurements at the centre of the structure). Nevertheless, the whole interplanetary distubance (including shock and sheath material) was rather long in duration (${\sim}$2.5~days), suggesting that it may be linked to more than one parent eruption. At Earth (Section~\ref{subsec:earth}), we again found a long-duration (${\sim}$2.5~days) disturbance culminating in a seemingly single magnetic cloud of NES type (based on both visual inspection and flux rope fitting). However, we observed characteristic interaction signatures at the centre of the flux rope, including enhanced plasma density, temperature, and beta, as well as an irregular speed profile and a less smooth magnetic field in the $\theta_{B}$ direction. Finally, at Saturn (Section~\ref{subsec:saturn}), we found an extended (${\sim}$4~days in duration) disturbance mostly characterised by smoothly rotating field. We associated a period of enhanced magnetic field magnitude with a magnetic ejecta, noting that it featured at its centre an irregular magnetic field profile (especially in $\theta_{B}$) reminiscent of the one encountered at Earth and attributed to interaction signatures. Its corresponding flux rope type was determined to be ESW based on visual inspection, whilst flux rope fitting yielded a low-inclination structure with its axis pointing back towards the Sun. The chirality was found to be consistently right-handed at all in-situ locations.

The overall picture extrapolated from the set of observations described above is that CME1 and CME2 started to interact in the inner heliosphere (roughly in the HI1 field of view and before the orbit of Venus). The distinct NES ejecta measured at Venus and Earth suggests that reconnection between the trailing edge of CME1 and the leading edge of CME2 resulted in a single flux rope that maintained the original orientation of its constituent parts. The resulting magnetic structure of interacting CMEs has been analysed via simulations by, e.g., \citet{lugaz2013} and \citet{schmidt2004}. Two flux ropes with the same twist and orientation (in this case, NES) are favourably configured for magnetic reconnection during interaction, at least at the interface between the two, which is where antiparallel fields meet. These conclusions were also found by \citet{kilpua2019b}, who analysed two interacting CMEs in June~2012 that left the solar corona as NES types and were observed at Venus at the beginning of their interaction and at Earth as a single NES flux rope. The structure observed at Saturn, if corresponding in its entirely to the merged flux ropes detected in the inner heliosphere, would be consistent either with a modest (${<}90^{\circ}$) counterclockwise rotation in the north--south direction (based on the ESW configuration retrieved from visual inspection) or with a ${\sim}130^{\circ}$ rotation along the equatorial plane (based on flux rope fitting) between 1 and 10~AU, possibly as a result of the pressure exerted by a following HSS (see also the discussion in Section~\ref{subsec:propagation}). 

\begin{figure}[ht]
\centering
\includegraphics[width=.99\linewidth]{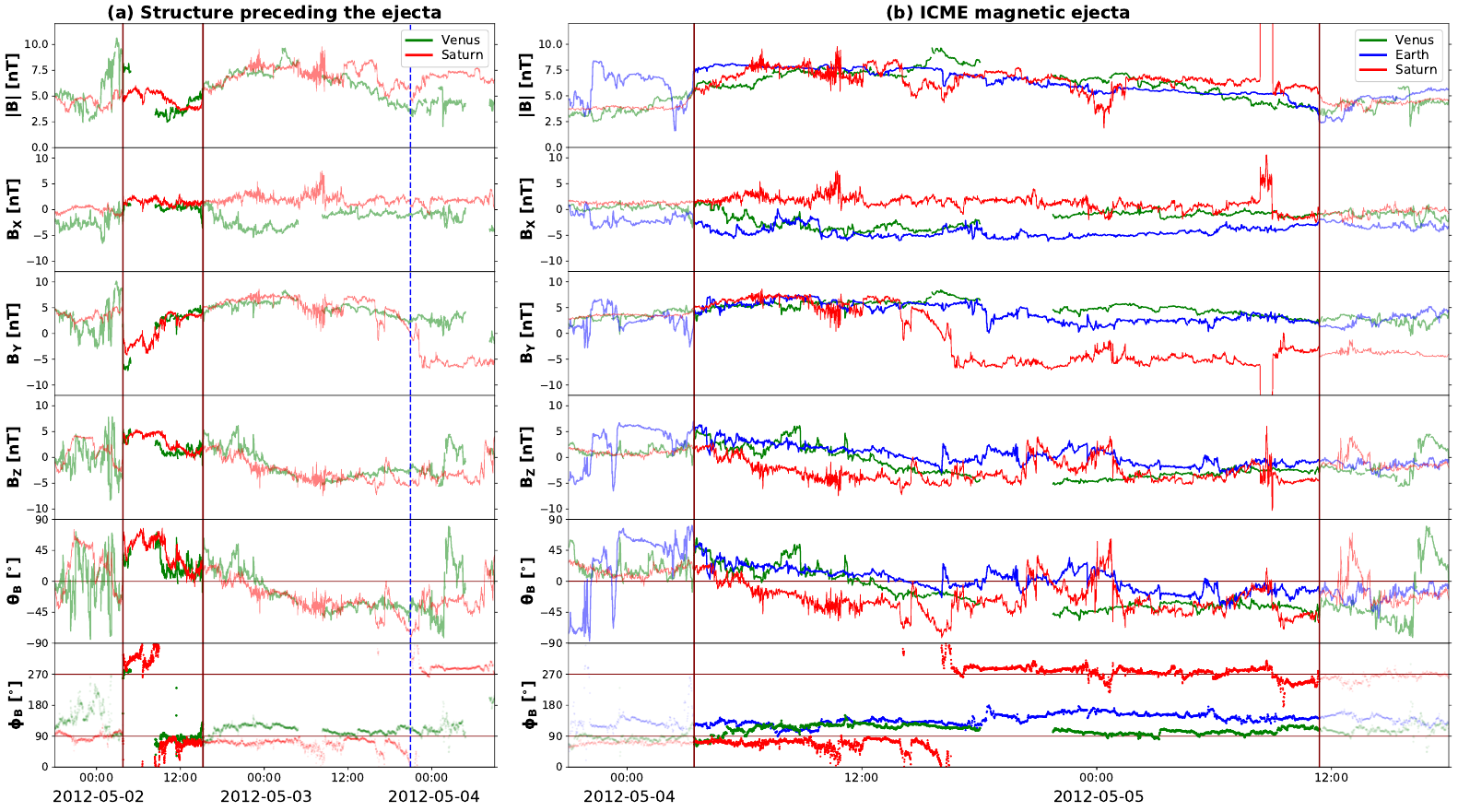}
\caption{Magnetic field mapping between (a) Venus and Saturn for the small structure preceding the ejecta, and (b) Venus, Earth, and Saturn for the full ICME ejecta, using the method in \citet{good2018}. In (a) measurements at Saturn have been temporally shifted and scaled to match those at Venus. The dashed line marks the identified ejecta trailing edge at Venus. In (b), measurements at Venus and Saturn have been temporally shifted and their magnitudes have been scaled to match data taken near Earth.}
\label{fig:magmapping}
\end{figure}

Finally, in order to investigate how the observed structure at Saturn relates to the ones at Venus and Earth, we employ the magnetic field mapping technique of \citet{good2018}, used to determine the arrival time and magnetic field of different plasma parcels in an ejecta between radially separated spacecraft. The technique takes into account expansion as a CME travels through interplanetary space, and assumes that each ejecta features a monotonically increasing/decreasing speed profile determined by its leading and trailing edge speeds (calculated on the basis of their arrival times from one location to the next). Magnetic field mapping results for two different structures are shown in Figure~\ref{fig:magmapping}. The full identified magnetic ejecta at Venus, Earth, and Saturn (i.e., the shaded grey areas in Figures~\ref{fig:venus}, \ref{fig:earth}, and \ref{fig:saturn}) is mapped in panel (b), whilst in panel (a) the smaller magnetic structure preceding the ejecta (i.e., from the dash-dotted blue lines to the ejecta leading edges in Figures~\ref{fig:venus} and \ref{fig:saturn}) is mapped between Venus and Saturn. In both panels, the measurements bounded by solid vertical lines are mapped according to the \citet{good2018} technique, whilst the preceding and following data are propagated at the resulting leading and trailing edge speeds, respectively. In Figure~\ref{fig:magmapping}a, it is evident that, despite a data gap at Venus, the magnetic field mapping shows good agreement between the two spacecraft in all the magnetic field components. We also note that, by propagating the following measurements at Saturn at the resulting trailing edge speed (387~km$\cdot$s$^{-1}$), good agreement between the two data sets continues to be displayed roughly until the ejecta trailing edge at Venus and the start of the identified interaction region at Saturn (i.e., close to the first dashed orange line in Figure~\ref{fig:saturn}). This would suggest that the original ejecta resulting from the interaction of CME1 and CME2 may be linked to approximately the first third of the flux rope interval identified at Saturn, and that the full ejecta may on the other hand correspond to a MIR, which as mentioned in the Introduction is a dominating structure in the outer heliosphere. Contrarily, the magnetic field mapping in Figure~\ref{fig:magmapping}b shows good agreement in all the magnetic field components between Venus and Earth, but a substantial difference in the $\phi_{B}$ component at Saturn in the second half of the ejecta (which rotates from east to west). This is however not surprising, since the structure at Saturn was found to display a different orientation from that at Venus and Earth. Furthermore, the central, more turbulent portion of magnetic field measurements found at Earth and Saturn appears to match relatively well. According to these results, it is not possible to establish with certainty the relationship between measurements at Saturn and those at Venus and Earth (i.e., whether the original flux rope corresponds to the full ejecta interval at Saturn or to a portion of it). We emphasise that the magnetic field mapping technique of \citet{good2018} assumes monotonically increasing/decreasing speed, which is not a realistic assumption as far from the Sun as 10~AU, where it is likely for a CME to have accelerated/decelerated multiple times via continued interactions with the ambient solar wind (evident also in the high distortion parameter found at Saturn). Nevertheless, this method appears to work well at least for a qualitative comparison of structures observed in the inner and outer heliosphere.

\section{Conclusions} \label{sec:conclusions}

In this work, we have analysed the eruption and evolution of two CMEs (CME1 and CME2) that left the Sun on 28~April~2012 just a few hours apart. After observing their early evolution through the solar corona, we found indications of interaction in HI imagery, and we eventually identified signatures of a single ejecta at the in-situ locations that we considered (Venus, Earth, and Saturn), suggesting that the two eruptions had merged. This study represents the first detailed analysis of the magnetic structure of CME flux ropes from the Sun to 10~AU, taking advantage of remote-sensing observations of the Sun, its corona, and interplanetary space, as well as in-situ measurements at three planets. Whilst the chirality of CME1 and CME2 at the Sun and of the flux rope ejecta detected in situ was found to be consistently right-handed, the flux rope type presented some changes. At the Sun, CME1 erupted as a NES flux rope and CME2 displayed a high-to-intermediate inclination (between ESW and NES). At Venus and Earth, the single observed ejecta was of type NES, whilst at Saturn we found a structure featuring an ESW rotation that is consistent with a rotation on either the meridional or the equatorial plane. We could not determine with certainty whether the ejecta that we identified at Saturn consisted entirely of material from CME1 and CME2, or whether additional material was gathered on the way to 10~AU, thus forming a MIR.

The CME propagation models that we used to estimate the impact locations of CME1 and CME2 (which were treated as a single, merged structure) across the heliosphere provided useful insights necessary to interpret the in-situ observations, and were reasonably well-timed compared to the actual measurements. However, in-depth understanding of how CME magnetic fields evolve throughout interplanetary space and interact with other CMEs or the ambient solar wind still remains an arduous task. In the case of the events under study, an ``intermediate'' observer between Earth and Saturn, perhaps around 5~AU, may have helped shed more light on the evolution of solar transients beyond 1~AU, especially as it may have been possible to catch the ICME ejecta at the beginning of its interaction with the following HSS. Nevertheless, this study remarks the importance of multi-point studies of CMEs and the advantages to be gained by analysing data from both heliospheric and planetary missions, which are necessary steps to undertake in order to deepen our current understanding of the structure and evolution of solar transients in the outer heliosphere.


\section*{Data Availability Statement}

The HELCATS catalogues are available at \url{https://www.helcats-fp7.eu}. Images and additional information on the 28~April~2012 CME(s) are available at \url{https://www.helcats-fp7.eu/catalogues/event_page.html?id=HCME_A__20120428_01} (STEREO-A viewpoint) and \url{https://www.helcats-fp7.eu/catalogues/event_page.html?id=HCME_B__20120428_02} (STEREO-B viewpoint).
Enlil simulation results have been provided by the Community Coordinated Modeling Center (CCMC) at NASA Goddard Space Flight Center through their public Runs on Request system (\url{http://ccmc.gsfc.nasa.gov}). The full Enlil simulation results are available at \url{https://ccmc.gsfc.nasa.gov/database_SH/Erika_Palmerio_031021_SH_3.php} (run id: \texttt{Erika\_Palmerio\_031021\_SH\_3}).
The NASA--Wind ICME list can be found at \url{https://wind.nasa.gov/ICMEindex.php}.
Solar disc and coronagraph data from SDO, SOHO, and STEREO are openly available at the Virtual Solar Observatory (VSO; \url{https://sdac.virtualsolar.org/}). These data were processed and analysed trough SunPy \citep{sunpy2015,sunpy2020}, IDL SolarSoft \citep{freeland1998}, and the ESA JHelioviewer software \citep{muller2017}. 
Level-2 processed STEREO/HI data were obtained from the UK Solar System Data Centre (UKSSDC; \url{https://www.ukssdc.ac.uk/solar/stereo/data.html}).
VEX data are openly available at ESA's Planetary Science Archive \citep[PSA;][]{besse2018}, accessible at \url{https://archives.esac.esa.int/psa}. These data were processed and analysed with the aid of the \texttt{irfpy} library (\url{https://irfpy.irf.se/irfpy/index.html}).
Wind data are publicly available at NASA's Coordinated Data Analysis Web (CDAWeb) database (\url{https://cdaweb.sci.gsfc.nasa.gov/index.html/}).
NMDB data are publicly available at \url{http://www.nmdb.eu}.
Cassini data are openly available at the Planetary Plasma Interactions (PPI) Node of NASA's Planetary Data System (PDS), accessible at \url{https://pds-ppi.igpp.ucla.edu}.
All in-situ spacecraft data can also be found on the Automated Multi-Dataset Analysis \citep[AMDA;][]{genot2021} tool at the address \url{http://amda.irap.omp.eu}.
Solar wind data propagated with the \citet{tao2005} model can be browsed at the AMDA website and at the HelioPropa tool, accessible at \url{http://heliopropa.irap.omp.eu}. 


\acknowledgments
E.~Palmerio is supported by the NASA Living With a Star (LWS) Jack Eddy Postdoctoral Fellowship Program, administered by UCAR's Cooperative Programs for the Advancement of Earth System Science (CPAESS) under award no. NNX16AK22G.
E.~Kilpua acknowledges the SolMAG project (ERC-COG 724391) funded by the European Research Council (ERC) in the framework of the Horizon 2020 Research and Innovation Programme, Academy of Finland project SMASH 310445, and the Finnish Centre of Excellence in Research of Sustainable Space (Academy of Finland grant no. 312390). 
A.~Zhukov thanks the Belgian Federal Science Policy Office (BELSPO) for the provision of financial support in the framework of the PRODEX Programme of the European Space Agency (ESA) under contract no. 4000117262.
L.~Jian acknowledges support from the NASA LWS and Heliophysics Supporting Research (HSR) programs.
G.~Provan and B.~S{\'a}nchez-Cano acknowledge support through UK‐STFC Grant ST/S000429/1.
L.~Lamy acknowledges support from CNES and CNRS/INSU programs of planetology and heliophysics.
C.~{M\"o}stl thanks the Austrian Science Fund (FWF): P31521-N27.
A.~Masters was supported by a Royal Society University Research Fellowship.
We acknowledge support from the European Union FP7-SPACE-2013-1 programme for the HELCATS project (grant no. 606692).
The HI instruments on STEREO were developed by a consortium that comprised the Rutherford Appleton Laboratory (UK), the University of Birmingham (UK), Centre Spatial de Li{\`e}ge (CSL, Belgium) and the Naval Research Laboratory (NRL, USA). The STEREO/SECCHI project, of which HI is a part, is an international consortium led by NRL. We recognise the support of the UK Space Agency for funding STEREO/HI operations in the UK.
The WSA model was developed by C. N. Arge (currently at NASA/GSFC), and the Enlil model was developed by D. Odstrcil (currently at GMU). We thank the model developers, R. Colaninno, and the CCMC staff.
We acknowledge the NMDB, founded under the European Union's FP7 programme (contract no. 213007), for providing neutron monitor data.
Finally, we thank the instrument teams of all the spacecraft involved in this study.

\bibliography{bibliography.bib} 

\begin{thebibliography}{227}
\providecommand{\natexlab}[1]{#1}
\expandafter\ifx\csname urlstyle\endcsname\relax
  \providecommand{\doi}[1]{doi:\discretionary{}{}{}#1}\else
  \providecommand{\doi}{doi:\discretionary{}{}{}\begingroup
  \urlstyle{rm}\Url}\fi

\bibitem[{\textit{{Al-Haddad} et~al.}(2011)\textit{{Al-Haddad}, {Roussev},
  {M{\"o}stl}, {Jacobs}, {Lugaz}, {Poedts}, and {Farrugia}}}]{alhaddad2011}
{Al-Haddad}, N., {Roussev}, I.~I., {M{\"o}stl}, C., {Jacobs}, C., {Lugaz}, N.,
  {Poedts}, S., and {Farrugia}, C.~J. (2011), {On the Internal Structure of the
  Magnetic Field in Magnetic Clouds and Interplanetary Coronal Mass Ejections:
  Writhe versus Twist}, \textit{\apjl}, \textit{738}(2), L18,
  \doi{10.1088/2041-8205/738/2/L18}.

\bibitem[{\textit{{Arge} et~al.}(2004)\textit{{Arge}, {Luhmann}, {Odstrcil},
  {Schrijver}, and {Li}}}]{arge2004}
{Arge}, C.~N., {Luhmann}, J.~G., {Odstrcil}, D., {Schrijver}, C.~J., and {Li},
  Y. (2004), {Stream structure and coronal sources of the solar wind during the
  May 12th, 1997 CME}, \textit{\jastp}, \textit{66}, 1295--1309,
  \doi{10.1016/j.jastp.2004.03.018}.

\bibitem[{\textit{{Asvestari} et~al.}(2021)\textit{{Asvestari}, {Pomoell},
  {Kilpua}, {Good}, {Chatzistergos}, {Temmer}, {Palmerio}, {Poedts}, and
  {Magdalenic}}}]{asvestari2021}
{Asvestari}, E., {Pomoell}, J., {Kilpua}, E., {Good}, S., {Chatzistergos}, T.,
  {Temmer}, M., {Palmerio}, E., {Poedts}, S., and {Magdalenic}, J. (2021),
  {Modelling a multi-spacecraft coronal mass ejection encounter with EUHFORIA},
  \textit{\aap}, \textit{652}, A27, \doi{10.1051/0004-6361/202140315}.

\bibitem[{\textit{{Aulanier} et~al.}(2012)\textit{{Aulanier}, {Janvier}, and
  {Schmieder}}}]{aulanier2012}
{Aulanier}, G., {Janvier}, M., and {Schmieder}, B. (2012), {The standard flare
  model in three dimensions. I. Strong-to-weak shear transition in post-flare
  loops}, \textit{\aap}, \textit{543}, A110, \doi{10.1051/0004-6361/201219311}.

\bibitem[{\textit{{Badman} et~al.}(2008)\textit{{Badman}, {Cowley}, {Lamy},
  {Cecconi}, and {Zarka}}}]{badman2008}
{Badman}, S.~V., {Cowley}, S.~W.~H., {Lamy}, L., {Cecconi}, B., and {Zarka}, P.
  (2008), {Relationship between solar wind corotating interaction regions and
  the phasing and intensity of Saturn kilometric radiation bursts},
  \textit{\angeo}, \textit{26}(12), 3641--3651,
  \doi{10.5194/angeo-26-3641-2008}.

\bibitem[{\textit{{Barabash} et~al.}(2007)\textit{{Barabash}, {Sauvaud},
  {Gunell}, {Andersson}, {Grigoriev}, {Brinkfeldt}, {Holmstr{\"o}m}, {Lundin},
  {Yamauchi}, {Asamura}, {Baumjohann}, {Zhang}, {Coates}, {Linder}, {Kataria},
  {Curtis}, {Hsieh}, {Sandel}, {Fedorov}, {Mazelle}, {Thocaven}, {Grande},
  {Koskinen}, {Kallio}, {S{\"a}les}, {Riihela}, {Kozyra}, {Krupp}, {Woch},
  {Luhmann}, {McKenna-Lawlor}, {Orsini}, {Cerulli-Irelli}, {Mura}, {Milillo},
  {Maggi}, {Roelof}, {Brandt}, {Russell}, {Szego}, {Winningham}, {Frahm},
  {Scherrer}, {Sharber}, {Wurz}, and {Bochsler}}}]{barabash2007}
{Barabash}, S., {Sauvaud}, J.-A., {Gunell}, H., {Andersson}, H., {Grigoriev},
  A., {Brinkfeldt}, K., {Holmstr{\"o}m}, M., {Lundin}, R., {Yamauchi}, M.,
  {Asamura}, K., {Baumjohann}, W., {Zhang}, T.~L., {Coates}, A.~J., {Linder},
  D.~R., {Kataria}, D.~O., {Curtis}, C.~C., {Hsieh}, K.~C., {Sandel}, B.~R.,
  {Fedorov}, A., {Mazelle}, C., {Thocaven}, J.-J., {Grande}, M., {Koskinen},
  H.~E.~J., {Kallio}, E., {S{\"a}les}, T., {Riihela}, P., {Kozyra}, J.,
  {Krupp}, N., {Woch}, J., {Luhmann}, J., {McKenna-Lawlor}, S., {Orsini}, S.,
  {Cerulli-Irelli}, R., {Mura}, M., {Milillo}, M., {Maggi}, M., {Roelof}, E.,
  {Brandt}, P., {Russell}, C.~T., {Szego}, K., {Winningham}, J.~D., {Frahm},
  R.~A., {Scherrer}, J., {Sharber}, J.~R., {Wurz}, P., and {Bochsler}, P.
  (2007), {The Analyser of Space Plasmas and Energetic Atoms (ASPERA-4) for the
  Venus Express mission}, \textit{\planss}, \textit{55}, 1772--1792,
  \doi{10.1016/j.pss.2007.01.014}.

\bibitem[{\textit{{Barnes} et~al.}(2019)\textit{{Barnes}, {Davies}, {Harrison},
  {Byrne}, {Perry}, {Bothmer}, {Eastwood}, {Gallagher}, {Kilpua}, and
  {M{\"o}stl}}}]{barnes2019}
{Barnes}, D., {Davies}, J.~A., {Harrison}, R.~A., {Byrne}, J.~P., {Perry},
  C.~H., {Bothmer}, V., {Eastwood}, J.~P., {Gallagher}, P.~T., {Kilpua},
  E.~K.~J., and {M{\"o}stl}, C. (2019), {CMEs in the Heliosphere: II. A
  Statistical Analysis of the Kinematic Properties Derived from
  Single-Spacecraft Geometrical Modelling Techniques Applied to CMEs Detected
  in the Heliosphere from 2007 to 2017 by STEREO/HI-1}, \textit{\solphys},
  \textit{294}(5), 57, \doi{10.1007/s11207-019-1444-4}.

\bibitem[{\textit{{Barnes} et~al.}(2020)\textit{{Barnes}, {Davies}, {Harrison},
  {Byrne}, {Perry}, {Bothmer}, {Eastwood}, {Gallagher}, {Kilpua}, {M{\"o}stl},
  {Rodriguez}, {Rouillard}, and {Odstr{\v{c}}il}}}]{barnes2020}
{Barnes}, D., {Davies}, J.~A., {Harrison}, R.~A., {Byrne}, J.~P., {Perry},
  C.~H., {Bothmer}, V., {Eastwood}, J.~P., {Gallagher}, P.~T., {Kilpua},
  E.~K.~J., {M{\"o}stl}, C., {Rodriguez}, L., {Rouillard}, A.~P., and
  {Odstr{\v{c}}il}, D. (2020), {CMEs in the Heliosphere: III. A Statistical
  Analysis of the Kinematic Properties Derived from Stereoscopic Geometrical
  Modelling Techniques Applied to CMEs Detected in the Heliosphere from 2008 to
  2014 by STEREO/HI-1}, \textit{\solphys}, \textit{295}(11), 150,
  \doi{10.1007/s11207-020-01717-w}.

\bibitem[{\textit{{Besse} et~al.}(2018)\textit{{Besse}, {Vallat}, {Barthelemy},
  {Coia}, {Costa}, {De Marchi}, {Fraga}, {Grotheer}, {Heather}, {Lim},
  {Martinez}, {Arviset}, {Barbarisi}, {Docasal}, {Macfarlane}, {Rios}, {Saiz},
  and {Vallejo}}}]{besse2018}
{Besse}, S., {Vallat}, C., {Barthelemy}, M., {Coia}, D., {Costa}, M., {De
  Marchi}, G., {Fraga}, D., {Grotheer}, E., {Heather}, D., {Lim}, T.,
  {Martinez}, S., {Arviset}, C., {Barbarisi}, I., {Docasal}, R., {Macfarlane},
  A., {Rios}, C., {Saiz}, J., and {Vallejo}, F. (2018), {ESA's Planetary
  Science Archive: Preserve and present reliable scientific data sets},
  \textit{\planss}, \textit{150}, 131--140, \doi{10.1016/j.pss.2017.07.013}.

\bibitem[{\textit{{Bothmer} and {Schwenn}}(1994)}]{bothmer1994}
{Bothmer}, V., and {Schwenn}, R. (1994), {Eruptive prominences as sources of
  magnetic clouds in the solar wind}, \textit{\ssr}, \textit{70}, 215--220,
  \doi{10.1007/BF00777872}.

\bibitem[{\textit{{Bothmer} and {Schwenn}}(1998)}]{bothmer1998}
{Bothmer}, V., and {Schwenn}, R. (1998), {The structure and origin of magnetic
  clouds in the solar wind}, \textit{\angeo}, \textit{16}, 1--24,
  \doi{10.1007/s00585-997-0001-x}.

\bibitem[{\textit{{Bradley} et~al.}(2020)\textit{{Bradley}, {Cowley}, {Bunce},
  {Melin}, {Provan}, {Nichols}, {Dougherty}, {Roussos}, {Krupp}, {Tao}, {Lamy},
  {Pryor}, and {Hunt}}}]{bradley2020}
{Bradley}, T.~J., {Cowley}, S.~W.~H., {Bunce}, E.~J., {Melin}, H., {Provan},
  G., {Nichols}, J.~D., {Dougherty}, M.~K., {Roussos}, E., {Krupp}, N., {Tao},
  C., {Lamy}, L., {Pryor}, W.~R., and {Hunt}, G.~J. (2020), {Saturn's Nightside
  Dynamics During Cassini's F Ring and Proximal Orbits: Response to Solar Wind
  and Planetary Period Oscillation Modulations}, \textit{\jgra},
  \textit{125}(9), e27907, \doi{10.1029/2020JA027907}.

\bibitem[{\textit{{Branduardi-Raymont}
  et~al.}(2013)\textit{{Branduardi-Raymont}, {Ford}, {Hansen}, {Lamy},
  {Masters}, {Cecconi}, {Coates}, {Dougherty}, {Gladstone}, and
  {Zarka}}}]{branduardiraymont2013}
{Branduardi-Raymont}, G., {Ford}, P.~G., {Hansen}, K.~C., {Lamy}, L.,
  {Masters}, A., {Cecconi}, B., {Coates}, A.~J., {Dougherty}, M.~K.,
  {Gladstone}, G.~R., and {Zarka}, P. (2013), {Search for Saturn's X-ray
  aurorae at the arrival of a solar wind shock}, \textit{\jgra},
  \textit{118}(5), 2145--2156, \doi{10.1002/jgra.50112}.

\bibitem[{\textit{{Brueckner} et~al.}(1995)\textit{{Brueckner}, {Howard},
  {Koomen}, {Korendyke}, {Michels}, {Moses}, {Socker}, {Dere}, {Lamy},
  {Llebaria}, {Bout}, {Schwenn}, {Simnett}, {Bedford}, and
  {Eyles}}}]{brueckner1995}
{Brueckner}, G.~E., {Howard}, R.~A., {Koomen}, M.~J., {Korendyke}, C.~M.,
  {Michels}, D.~J., {Moses}, J.~D., {Socker}, D.~G., {Dere}, K.~P., {Lamy},
  P.~L., {Llebaria}, A., {Bout}, M.~V., {Schwenn}, R., {Simnett}, G.~M.,
  {Bedford}, D.~K., and {Eyles}, C.~J. (1995), {The Large Angle Spectroscopic
  Coronagraph (LASCO)}, \textit{\solphys}, \textit{162}, 357--402,
  \doi{10.1007/BF00733434}.

\bibitem[{\textit{{Burlaga} et~al.}(1981)\textit{{Burlaga}, {Sittler},
  {Mariani}, and {Schwenn}}}]{burlaga1981}
{Burlaga}, L., {Sittler}, E., {Mariani}, F., and {Schwenn}, R. (1981),
  {Magnetic loop behind an interplanetary shock - Voyager, Helios, and IMP 8
  observations}, \textit{\jgr}, \textit{86}, 6673--6684,
  \doi{10.1029/JA086iA08p06673}.

\bibitem[{\textit{{Burlaga} et~al.}(1986)\textit{{Burlaga}, {McDonald}, and
  {Schwenn}}}]{burlaga1986}
{Burlaga}, L.~F., {McDonald}, F.~B., and {Schwenn}, R. (1986), {Formation of a
  compound stream between 0.85 AU and 6.2 AU and its effects on solar energetic
  particles and galactic cosmic rays}, \textit{\jgr}, \textit{91}(A12),
  13,331--13,340, \doi{10.1029/JA091iA12p13331}.

\bibitem[{\textit{{Burlaga} et~al.}(1997)\textit{{Burlaga}, {Ness}, and
  {Belcher}}}]{burlaga1997}
{Burlaga}, L.~F., {Ness}, N.~F., and {Belcher}, J.~W. (1997), {Radial evolution
  of corotating merged interaction regions and flows between
  \raisebox{-0.5ex}\textasciitilde14AU and
  \raisebox{-0.5ex}\textasciitilde43AU}, \textit{\jgr}, \textit{102}(A3),
  4661--4672, \doi{10.1029/96JA03629}.

\bibitem[{\textit{{Burlaga} et~al.}(2001)\textit{{Burlaga}, {Ness},
  {Richardson}, and {Lepping}}}]{burlaga2001}
{Burlaga}, L.~F., {Ness}, N.~F., {Richardson}, J.~D., and {Lepping}, R.~P.
  (2001), {The Bastille day Shock and Merged Interaction Region at 63 au:
  Voyager 2 Observations}, \textit{\solphys}, \textit{204}, 399--411,
  \doi{10.1023/A:1014269926730}.

\bibitem[{\textit{{Cane}}(2000)}]{cane2000}
{Cane}, H.~V. (2000), {Coronal Mass Ejections and Forbush Decreases},
  \textit{\ssr}, \textit{93}, 55--77, \doi{10.1023/A:1026532125747}.

\bibitem[{\textit{{Cane} and {Richardson}}(2003)}]{cane2003}
{Cane}, H.~V., and {Richardson}, I.~G. (2003), {Interplanetary coronal mass
  ejections in the near-Earth solar wind during 1996-2002}, \textit{\jgr},
  \textit{108}(A4), 1156, \doi{10.1029/2002JA009817}.

\bibitem[{\textit{{Cane} et~al.}(1997)\textit{{Cane}, {Richardson}, and
  {Wibberenz}}}]{cane1997}
{Cane}, H.~V., {Richardson}, I.~G., and {Wibberenz}, G. (1997), {Helios 1 and 2
  observations of particle decreases, ejecta, and magnetic clouds},
  \textit{\jgr}, \textit{102}, 7075--7086, \doi{10.1029/97JA00149}.

\bibitem[{\textit{{Chen}}(2011)}]{chen2011}
{Chen}, P.~F. (2011), {Coronal Mass Ejections: Models and Their Observational
  Basis}, \textit{\lrsp}, \textit{8}, 1, \doi{10.12942/lrsp-2011-1}.

\bibitem[{\textit{{Clarke} et~al.}(2009)\textit{{Clarke}, {Nichols},
  {G{\'e}rard}, {Grodent}, {Hansen}, {Kurth}, {Gladstone}, {Duval},
  {Wannawichian}, {Bunce}, {Cowley}, {Crary}, {Dougherty}, {Lamy}, {Mitchell},
  {Pryor}, {Retherford}, {Stallard}, {Zieger}, {Zarka}, and
  {Cecconi}}}]{clarke2009}
{Clarke}, J.~T., {Nichols}, J., {G{\'e}rard}, J.~C., {Grodent}, D., {Hansen},
  K.~C., {Kurth}, W., {Gladstone}, G.~R., {Duval}, J., {Wannawichian}, S.,
  {Bunce}, E., {Cowley}, S.~W.~H., {Crary}, F., {Dougherty}, M., {Lamy}, L.,
  {Mitchell}, D., {Pryor}, W., {Retherford}, K., {Stallard}, T., {Zieger}, B.,
  {Zarka}, P., and {Cecconi}, B. (2009), {Response of Jupiter's and Saturn's
  auroral activity to the solar wind}, \textit{\jgr}, \textit{114}(A5), A05210,
  \doi{10.1029/2008JA013694}.

\bibitem[{\textit{{Collinson} et~al.}(2015)\textit{{Collinson}, {Grebowsky},
  {Sibeck}, {Jian}, {Boardsen}, {Espley}, {Hartle}, {Zhang}, {Barabash}, and
  {Futaana}}}]{collinson2015}
{Collinson}, G.~A., {Grebowsky}, J., {Sibeck}, D.~G., {Jian}, L.~K.,
  {Boardsen}, S., {Espley}, J., {Hartle}, D., {Zhang}, T.~L., {Barabash}, S.,
  and {Futaana}, Y. (2015), {The impact of a slow interplanetary coronal mass
  ejection on Venus}, \textit{\jgra}, \textit{120}(5), 3489--3502,
  \doi{10.1002/2014JA020616}.

\bibitem[{\textit{{Crary} et~al.}(2005)\textit{{Crary}, {Clarke}, {Dougherty},
  {Hanlon}, {Hansen}, {Steinberg}, {Barraclough}, {Coates}, {G{\'e}rard},
  {Grodent}, {Kurth}, {Mitchell}, {Rymer}, and {Young}}}]{crary2005}
{Crary}, F.~J., {Clarke}, J.~T., {Dougherty}, M.~K., {Hanlon}, P.~G., {Hansen},
  K.~C., {Steinberg}, J.~T., {Barraclough}, B.~L., {Coates}, A.~J.,
  {G{\'e}rard}, J.~C., {Grodent}, D., {Kurth}, W.~S., {Mitchell}, D.~G.,
  {Rymer}, A.~M., and {Young}, D.~T. (2005), {Solar wind dynamic pressure and
  electric field as the main factors controlling Saturn's aurorae},
  \textit{\nat}, \textit{433}(7027), 720--722, \doi{10.1038/nature03333}.

\bibitem[{\textit{{Dasso} et~al.}(2009)\textit{{Dasso}, {Mandrini},
  {Schmieder}, {Cremades}, {Cid}, {Cerrato}, {Saiz}, {D{\'e}moulin}, {Zhukov},
  {Rodriguez}, {Aran}, {Menvielle}, and {Poedts}}}]{dasso2009}
{Dasso}, S., {Mandrini}, C.~H., {Schmieder}, B., {Cremades}, H., {Cid}, C.,
  {Cerrato}, Y., {Saiz}, E., {D{\'e}moulin}, P., {Zhukov}, A.~N., {Rodriguez},
  L., {Aran}, A., {Menvielle}, M., and {Poedts}, S. (2009), {Linking two
  consecutive nonmerging magnetic clouds with their solar sources},
  \textit{Journal of Geophysical Research (Space Physics)}, \textit{114}(A2),
  A02109, \doi{10.1029/2008JA013102}.

\bibitem[{\textit{{Davies} et~al.}(2020)\textit{{Davies}, {Forsyth}, {Good},
  and {Kilpua}}}]{davies2020}
{Davies}, E.~E., {Forsyth}, R.~J., {Good}, S.~W., and {Kilpua}, E. K.~J.
  (2020), {On the Radial and Longitudinal Variation of a Magnetic Cloud: ACE,
  Wind, ARTEMIS and Juno Observations}, \textit{\solphys}, \textit{295}(11),
  157, \doi{10.1007/s11207-020-01714-z}.

\bibitem[{\textit{{Davies} et~al.}(2021)\textit{{Davies}, {M{\"o}stl}, {Owens},
  {Weiss}, {Amerstorfer}, {Hinterreiter}, {Bauer}, {Bailey}, {Reiss},
  {Forsyth}, {Horbury}, {O'Brien}, {Evans}, {Angelini}, {Heyner}, {Richter},
  {Auster}, {Magnes}, {Baumjohann}, {Fischer}, {Barnes}, {Davies}, and
  {Harrison}}}]{davies2021}
{Davies}, E.~E., {M{\"o}stl}, C., {Owens}, M.~J., {Weiss}, A.~J.,
  {Amerstorfer}, T., {Hinterreiter}, J., {Bauer}, M., {Bailey}, R.~L., {Reiss},
  M.~A., {Forsyth}, R.~J., {Horbury}, T.~S., {O'Brien}, H., {Evans}, V.,
  {Angelini}, V., {Heyner}, D., {Richter}, I., {Auster}, H.-U., {Magnes}, W.,
  {Baumjohann}, W., {Fischer}, D., {Barnes}, D., {Davies}, J.~A., and
  {Harrison}, R.~A. (2021), {In-Situ Multi-Spacecraft and Remote Imaging
  Observations of the First CME Detected by Solar Orbiter and BepiColombo},
  \textit{\aap}, \textit{in press}, \doi{10.1051/0004-6361/202040113}.

\bibitem[{\textit{{Davies} et~al.}(2012)\textit{{Davies}, {Harrison}, {Perry},
  {M{\"o}stl}, {Lugaz}, {Rollett}, {Davis}, {Crothers}, {Temmer}, {Eyles}, and
  {Savani}}}]{davies2012}
{Davies}, J.~A., {Harrison}, R.~A., {Perry}, C.~H., {M{\"o}stl}, C., {Lugaz},
  N., {Rollett}, T., {Davis}, C.~J., {Crothers}, S.~R., {Temmer}, M., {Eyles},
  C.~J., and {Savani}, N.~P. (2012), {A Self-similar Expansion Model for Use in
  Solar Wind Transient Propagation Studies}, \textit{\apj}, \textit{750}, 23,
  \doi{10.1088/0004-637X/750/1/23}.

\bibitem[{\textit{{Davies} et~al.}(2013)\textit{{Davies}, {Perry}, {Trines},
  {Harrison}, {Lugaz}, {M{\"o}stl}, {Liu}, and {Steed}}}]{davies2013}
{Davies}, J.~A., {Perry}, C.~H., {Trines}, R.~M.~G.~M., {Harrison}, R.~A.,
  {Lugaz}, N., {M{\"o}stl}, C., {Liu}, Y.~D., and {Steed}, K. (2013),
  {Establishing a Stereoscopic Technique for Determining the Kinematic
  Properties of Solar Wind Transients based on a Generalized Self-similarly
  Expanding Circular Geometry}, \textit{\apj}, \textit{777}, 167,
  \doi{10.1088/0004-637X/777/2/167}.

\bibitem[{\textit{{de Lucas} et~al.}(2011)\textit{{de Lucas}, {Dal Lago},
  {Schwenn}, and {Cl{\'u}a de Gonzalez}}}]{delucas2011}
{de Lucas}, A., {Dal Lago}, A., {Schwenn}, R., and {Cl{\'u}a de Gonzalez},
  A.~L. (2011), {Multi-spacecraft observed magnetic clouds as seen by Helios
  mission}, \textit{\jastp}, \textit{73}(11-12), 1361--1371,
  \doi{10.1016/j.jastp.2011.02.007}.

\bibitem[{\textit{{DeForest} et~al.}(2013)\textit{{DeForest}, {Howard}, and
  {McComas}}}]{deforest2013}
{DeForest}, C.~E., {Howard}, T.~A., and {McComas}, D.~J. (2013), {Tracking
  Coronal Features from the Low Corona to Earth: A Quantitative Analysis of the
  2008 December 12 Coronal Mass Ejection}, \textit{\apj}, \textit{769}, 43,
  \doi{10.1088/0004-637X/769/1/43}.

\bibitem[{\textit{{D{\'e}moulin} and {Dasso}}(2009)}]{demoulin2009a}
{D{\'e}moulin}, P., and {Dasso}, S. (2009), {Causes and consequences of
  magnetic cloud expansion}, \textit{\aap}, \textit{498}(2), 551--566,
  \doi{10.1051/0004-6361/200810971}.

\bibitem[{\textit{{D{\'e}moulin} et~al.}(1996)\textit{{D{\'e}moulin}, {Priest},
  and {Lonie}}}]{demoulin1996}
{D{\'e}moulin}, P., {Priest}, E.~R., and {Lonie}, D.~P. (1996),
  {Three-dimensional magnetic reconnection without null points 2. Application
  to twisted flux tubes}, \textit{\jgr}, \textit{101}, 7631--7646,
  \doi{10.1029/95JA03558}.

\bibitem[{\textit{{Desai} et~al.}(2020)\textit{{Desai}, {Zhang}, {Davies},
  {Stawarz}, {Mico-Gomez}, and {Iv{\'a}{\~n}ez-Ballesteros}}}]{desai2020}
{Desai}, R.~T., {Zhang}, H., {Davies}, E.~E., {Stawarz}, J.~E., {Mico-Gomez},
  J., and {Iv{\'a}{\~n}ez-Ballesteros}, P. (2020), {Three-Dimensional
  Simulations of Solar Wind Preconditioning and the 23 July 2012 Interplanetary
  Coronal Mass Ejection}, \textit{\solphys}, \textit{295}(9), 130,
  \doi{10.1007/s11207-020-01700-5}.

\bibitem[{\textit{{Desch}}(1982)}]{desch1982}
{Desch}, M.~D. (1982), {Evidence for solar wind control of solar radio
  emission}, \textit{\jgr}, \textit{87}(A6), 4549--4554,
  \doi{10.1029/JA087iA06p04549}.

\bibitem[{\textit{{Domingo} et~al.}(1995)\textit{{Domingo}, {Fleck}, and
  {Poland}}}]{domingo1995}
{Domingo}, V., {Fleck}, B., and {Poland}, A.~I. (1995), {The SOHO Mission: an
  Overview}, \textit{\solphys}, \textit{162}, 1--37, \doi{10.1007/BF00733425}.

\bibitem[{\textit{{Dougherty} et~al.}(2004)\textit{{Dougherty}, {Kellock},
  {Southwood}, {Balogh}, {Smith}, {Tsurutani}, {Gerlach}, {Glassmeier},
  {Gleim}, {Russell}, {Erdos}, {Neubauer}, and {Cowley}}}]{dougherty2004}
{Dougherty}, M.~K., {Kellock}, S., {Southwood}, D.~J., {Balogh}, A., {Smith},
  E.~J., {Tsurutani}, B.~T., {Gerlach}, B., {Glassmeier}, K.-H., {Gleim}, F.,
  {Russell}, C.~T., {Erdos}, G., {Neubauer}, F.~M., and {Cowley}, S.~W.~H.
  (2004), {The Cassini Magnetic Field Investigation}, \textit{\ssr},
  \textit{114}, 331--383, \doi{10.1007/s11214-004-1432-2}.

\bibitem[{\textit{{Du} et~al.}(2010)\textit{{Du}, {Zuo}, and {Zhang}}}]{du2010}
{Du}, D., {Zuo}, P.~B., and {Zhang}, X.~X. (2010), {Interplanetary Coronal Mass
  Ejections Observed by Ulysses Through Its Three Solar Orbits},
  \textit{\solphys}, \textit{262}(1), 171--190,
  \doi{10.1007/s11207-009-9505-8}.

\bibitem[{\textit{{Dumbovi{\'c}} et~al.}(2020)\textit{{Dumbovi{\'c}},
  {Vr{\v{s}}nak}, {Guo}, {Heber}, {Dissauer}, {Carcaboso}, {Temmer}, {Veronig},
  {Podladchikova}, {M{\"o}stl}, {Amerstorfer}, and {Kirin}}}]{dumbovic2020}
{Dumbovi{\'c}}, M., {Vr{\v{s}}nak}, B., {Guo}, J., {Heber}, B., {Dissauer}, K.,
  {Carcaboso}, F., {Temmer}, M., {Veronig}, A., {Podladchikova}, T.,
  {M{\"o}stl}, C., {Amerstorfer}, T., and {Kirin}, A. (2020), {Evolution of
  Coronal Mass Ejections and the Corresponding Forbush Decreases: Modeling vs.
  Multi-Spacecraft Observations}, \textit{\solphys}, \textit{295}(7), 104,
  \doi{10.1007/s11207-020-01671-7}.

\bibitem[{\textit{{Dunn} et~al.}(2020)\textit{{Dunn}, {Gray}, {Wibisono},
  {Lamy}, {Louis}, {Badman}, {Branduardi-Raymont}, {Elsner}, {Gladstone},
  {Ebert}, {Ford}, {Foster}, {Tao}, {Ray}, {Yao}, {Rae}, {Bunce}, {Rodriguez},
  {Jackman}, {Nicolaou}, {Clarke}, {Nichols}, {Elliott}, and
  {Kraft}}}]{dunn2020}
{Dunn}, W.~R., {Gray}, R., {Wibisono}, A.~D., {Lamy}, L., {Louis}, C.,
  {Badman}, S.~V., {Branduardi-Raymont}, G., {Elsner}, R., {Gladstone}, G.~R.,
  {Ebert}, R., {Ford}, P., {Foster}, A., {Tao}, C., {Ray}, L.~C., {Yao}, Z.,
  {Rae}, I.~J., {Bunce}, E.~J., {Rodriguez}, P., {Jackman}, C.~M., {Nicolaou},
  G., {Clarke}, J., {Nichols}, J., {Elliott}, H., and {Kraft}, R. (2020),
  {Comparisons Between Jupiter's X-ray, UV and Radio Emissions and In-Situ
  Solar Wind Measurements During 2007}, \textit{\jgra}, \textit{125}(6),
  e27222, \doi{10.1029/2019JA027222}.

\bibitem[{\textit{{Dunn} et~al.}(2021)\textit{{Dunn}, {Ness}, {Lamy},
  {Tremblay}, {Branduardi-Raymont}, {Snios}, {Kraft}, {Yao}, and
  {Wibisono}}}]{dunn2021}
{Dunn}, W.~R., {Ness}, J.~U., {Lamy}, L., {Tremblay}, G.~R.,
  {Branduardi-Raymont}, G., {Snios}, B., {Kraft}, R.~P., {Yao}, Z., and
  {Wibisono}, A.~D. (2021), {A Low Signal Detection of X Rays From Uranus},
  \textit{\jgra}, \textit{126}(4), e28739, \doi{10.1029/2020JA028739}.

\bibitem[{\textit{{Echer}}(2019)}]{echer2019}
{Echer}, E. (2019), {Solar wind and interplanetary shock parameters near
  Saturn's orbit ({\ensuremath{\sim}}10 AU)}, \textit{\planss}, \textit{165},
  210--220, \doi{10.1016/j.pss.2018.10.006}.

\bibitem[{\textit{{Eyles} et~al.}(2009)\textit{{Eyles}, {Harrison}, {Davis},
  {Waltham}, {Shaughnessy}, {Mapson-Menard}, {Bewsher}, {Crothers}, {Davies},
  {Simnett}, {Howard}, {Moses}, {Newmark}, {Socker}, {Halain}, {Defise},
  {Mazy}, and {Rochus}}}]{eyles2009}
{Eyles}, C.~J., {Harrison}, R.~A., {Davis}, C.~J., {Waltham}, N.~R.,
  {Shaughnessy}, B.~M., {Mapson-Menard}, H.~C.~A., {Bewsher}, D., {Crothers},
  S.~R., {Davies}, J.~A., {Simnett}, G.~M., {Howard}, R.~A., {Moses}, J.~D.,
  {Newmark}, J.~S., {Socker}, D.~G., {Halain}, J.-P., {Defise}, J.-M., {Mazy},
  E., and {Rochus}, P. (2009), {The Heliospheric Imagers Onboard the STEREO
  Mission}, \textit{\solphys}, \textit{254}, 387--445,
  \doi{10.1007/s11207-008-9299-0}.

\bibitem[{\textit{{Farrugia} and {Berdichevsky}}(2004)}]{farrugia2004}
{Farrugia}, C., and {Berdichevsky}, D. (2004), {Evolutionary signatures in
  complex ejecta and their driven shocks}, \textit{\angeo}, \textit{22}(10),
  3679--3698, \doi{10.5194/angeo-22-3679-2004}.

\bibitem[{\textit{{Forbes}}(2000)}]{forbes2000}
{Forbes}, T.~G. (2000), {A review on the genesis of coronal mass ejections},
  \textit{\jgr}, \textit{105}, 23,153--23,166, \doi{10.1029/2000JA000005}.

\bibitem[{\textit{{Forbush}}(1937)}]{forbush1937}
{Forbush}, S.~E. (1937), {On the Effects in Cosmic-Ray Intensity Observed
  During the Recent Magnetic Storm}, \textit{Physical Review}, \textit{51},
  1108--1109, \doi{10.1103/PhysRev.51.1108.3}.

\bibitem[{\textit{{Freeland} and {Handy}}(1998)}]{freeland1998}
{Freeland}, S.~L., and {Handy}, B.~N. (1998), {Data Analysis with the SolarSoft
  System}, \textit{\solphys}, \textit{182}, 497--500,
  \doi{10.1023/A:1005038224881}.

\bibitem[{\textit{{Freiherr von Forstner} et~al.}(2018)\textit{{Freiherr von
  Forstner}, {Guo}, {Wimmer-Schweingruber}, {Hassler}, {Temmer},
  {Dumbovi{\'c}}, {Jian}, {Appel}, {{\v{C}}alogovi{\'c}}, and
  {Ehresmann}}}]{freiherrvonforstner2018}
{Freiherr von Forstner}, J.~L., {Guo}, J., {Wimmer-Schweingruber}, R.~F.,
  {Hassler}, D.~M., {Temmer}, M., {Dumbovi{\'c}}, M., {Jian}, L.~K., {Appel},
  J.~K., {{\v{C}}alogovi{\'c}}, J., and {Ehresmann}, B. (2018), {Using Forbush
  Decreases to Derive the Transit Time of ICMEs Propagating from 1 AU to Mars},
  \textit{\jgra}, \textit{123}(1), 39--56, \doi{10.1002/2017JA024700}.

\bibitem[{\textit{{Freiherr von Forstner} et~al.}(2020)\textit{{Freiherr von
  Forstner}, {Guo}, {Wimmer-Schweingruber}, {Dumbovi{\'c}}, {Janvier},
  {D{\'e}moulin}, {Veronig}, {Temmer}, {Papaioannou}, {Dasso}, {Hassler}, and
  {Zeitlin}}}]{freiherrvonforstner2020}
{Freiherr von Forstner}, J.~L., {Guo}, J., {Wimmer-Schweingruber}, R.~F.,
  {Dumbovi{\'c}}, M., {Janvier}, M., {D{\'e}moulin}, P., {Veronig}, A.,
  {Temmer}, M., {Papaioannou}, A., {Dasso}, S., {Hassler}, D.~M., and
  {Zeitlin}, C.~J. (2020), {Comparing the Properties of ICME-Induced Forbush
  Decreases at Earth and Mars}, \textit{\jgra}, \textit{125}(3), e2019JA027662,
  \doi{10.1029/2019JA027662}.

\bibitem[{\textit{{Gazis} et~al.}(2006)\textit{{Gazis}, {Balogh}, {Dalla},
  {Decker}, {Heber}, {Horbury}, {Kilchenmann}, {Kota}, {Kucharek}, and
  {Kunow}}}]{gazis2006}
{Gazis}, P.~R., {Balogh}, A., {Dalla}, S., {Decker}, R., {Heber}, B.,
  {Horbury}, T., {Kilchenmann}, A., {Kota}, J., {Kucharek}, H., and {Kunow}, H.
  (2006), {ICMEs at High Latitudes and in the Outer Heliosphere. Report of
  Working Group H}, \textit{\ssr}, \textit{123}(1-3), 417--451,
  \doi{10.1007/s11214-006-9023-z}.

\bibitem[{\textit{{G{\'e}not} et~al.}(2021)\textit{{G{\'e}not}, {Budnik},
  {Jacquey}, {Bouchemit}, {Renard}, {Dufourg}, {Andr{\'e}}, {Cecconi},
  {Pitout}, {Lavraud}, {Fedorov}, {Ganfloff}, {Plotnikov}, {Modolo}, {Lormant},
  {Mohand}, {Tao}, {Besson}, {Heulet}, {Boucon}, {Durand}, {Bourrel},
  {Brzustowski}, {Jourdane}, {Hitier}, {Garnier}, {Grison}, {Aunai}, {Jeandet},
  and {Cabrolie}}}]{genot2021}
{G{\'e}not}, V., {Budnik}, E., {Jacquey}, C., {Bouchemit}, M., {Renard}, B.,
  {Dufourg}, N., {Andr{\'e}}, N., {Cecconi}, B., {Pitout}, F., {Lavraud}, B.,
  {Fedorov}, A., {Ganfloff}, M., {Plotnikov}, I., {Modolo}, R., {Lormant}, N.,
  {Mohand}, H. S.~H., {Tao}, C., {Besson}, B., {Heulet}, D., {Boucon}, D.,
  {Durand}, J., {Bourrel}, N., {Brzustowski}, Q., {Jourdane}, N., {Hitier}, R.,
  {Garnier}, P., {Grison}, B., {Aunai}, N., {Jeandet}, A., and {Cabrolie}, F.
  (2021), {Automated Multi-Dataset Analysis (AMDA): An on-line database and
  analysis tool for heliospheric and planetary plasma data}, \textit{\planss},
  \textit{201}, 105214, \doi{10.1016/j.pss.2021.105214}.

\bibitem[{\textit{{Good} and {Forsyth}}(2016)}]{good2016}
{Good}, S.~W., and {Forsyth}, R.~J. (2016), {Interplanetary Coronal Mass
  Ejections Observed by MESSENGER and Venus Express}, \textit{\solphys},
  \textit{291}, 239--263, \doi{10.1007/s11207-015-0828-3}.

\bibitem[{\textit{{Good} et~al.}(2018)\textit{{Good}, {Forsyth}, {Eastwood},
  and {M{\"o}stl}}}]{good2018}
{Good}, S.~W., {Forsyth}, R.~J., {Eastwood}, J.~P., and {M{\"o}stl}, C. (2018),
  {Correlation of ICME Magnetic Fields at Radially Aligned Spacecraft},
  \textit{\solphys}, \textit{293}, 52, \doi{10.1007/s11207-018-1264-y}.

\bibitem[{\textit{{Good} et~al.}(2019)\textit{{Good}, {Kilpua}, {LaMoury},
  {Forsyth}, {Eastwood}, and {M{\"o}stl}}}]{good2019}
{Good}, S.~W., {Kilpua}, E.~K.~J., {LaMoury}, A.~T., {Forsyth}, R.~J.,
  {Eastwood}, J.~P., and {M{\"o}stl}, C. (2019), {Self-Similarity of ICME Flux
  Ropes: Observations by Radially Aligned Spacecraft in the Inner Heliosphere},
  \textit{\jgra}, \textit{124}(7), 4960--4982, \doi{10.1029/2019JA026475}.

\bibitem[{\textit{{Gopalswamy} et~al.}(2000)\textit{{Gopalswamy}, {Lara},
  {Lepping}, {Kaiser}, {Berdichevsky}, and {St.~Cyr}}}]{gopalswamy2000}
{Gopalswamy}, N., {Lara}, A., {Lepping}, R.~P., {Kaiser}, M.~L.,
  {Berdichevsky}, D., and {St.~Cyr}, O.~C. (2000), {Interplanetary acceleration
  of coronal mass ejections}, \textit{\grl}, \textit{27}, 145--148,
  \doi{10.1029/1999GL003639}.

\bibitem[{\textit{{Green} et~al.}(2007)\textit{{Green}, {Kliem},
  {T{\"o}r{\"o}k}, {van Driel-Gesztelyi}, and {Attrill}}}]{green2007}
{Green}, L.~M., {Kliem}, B., {T{\"o}r{\"o}k}, T., {van Driel-Gesztelyi}, L.,
  and {Attrill}, G.~D.~R. (2007), {Transient Coronal Sigmoids and Rotating
  Erupting Flux Ropes}, \textit{\solphys}, \textit{246}, 365--391,
  \doi{10.1007/s11207-007-9061-z}.

\bibitem[{\textit{{Green} et~al.}(2018)\textit{{Green}, {T{\"o}r{\"o}k},
  {Vr{\v{s}}nak}, {Manchester}, and {Veronig}}}]{green2018}
{Green}, L.~M., {T{\"o}r{\"o}k}, T., {Vr{\v{s}}nak}, B., {Manchester}, W., and
  {Veronig}, A. (2018), {The Origin, Early Evolution and Predictability of
  Solar Eruptions}, \textit{\ssr}, \textit{214}, 46,
  \doi{10.1007/s11214-017-0462-5}.

\bibitem[{\textit{{Gurnett} et~al.}(2004)\textit{{Gurnett}, {Kurth},
  {Kirchner}, {Hospodarsky}, {Averkamp}, {Zarka}, {Lecacheux}, {Manning},
  {Roux}, {Canu}, {Cornilleau-Wehrlin}, {Galopeau}, {Meyer}, {Bostr{\"o}m},
  {Gustafsson}, {Wahlund}, {{\r{A}}hlen}, {Rucker}, {Ladreiter}, {Macher},
  {Woolliscroft}, {Alleyne}, {Kaiser}, {Desch}, {Farrell}, {Harvey}, {Louarn},
  {Kellogg}, {Goetz}, and {Pedersen}}}]{gurnett2004}
{Gurnett}, D.~A., {Kurth}, W.~S., {Kirchner}, D.~L., {Hospodarsky}, G.~B.,
  {Averkamp}, T.~F., {Zarka}, P., {Lecacheux}, A., {Manning}, R., {Roux}, A.,
  {Canu}, P., {Cornilleau-Wehrlin}, N., {Galopeau}, P., {Meyer}, A.,
  {Bostr{\"o}m}, R., {Gustafsson}, G., {Wahlund}, J.~E., {{\r{A}}hlen}, L.,
  {Rucker}, H.~O., {Ladreiter}, H.~P., {Macher}, W., {Woolliscroft}, L.~J.~C.,
  {Alleyne}, H., {Kaiser}, M.~L., {Desch}, M.~D., {Farrell}, W.~M., {Harvey},
  C.~C., {Louarn}, P., {Kellogg}, P.~J., {Goetz}, K., and {Pedersen}, A.
  (2004), {The Cassini Radio and Plasma Wave Investigation}, \textit{\ssr},
  \textit{114}(1-4), 395--463, \doi{10.1007/s11214-004-1434-0}.

\bibitem[{\textit{{Hanlon} et~al.}(2004)\textit{{Hanlon}, {Dougherty},
  {Forsyth}, {Owens}, {Hansen}, {T{\'o}th}, {Crary}, and {Young}}}]{hanlon2004}
{Hanlon}, P.~G., {Dougherty}, M.~K., {Forsyth}, R.~J., {Owens}, M.~J.,
  {Hansen}, K.~C., {T{\'o}th}, G., {Crary}, F.~J., and {Young}, D.~T. (2004),
  {On the evolution of the solar wind between 1 and 5 AU at the time of the
  Cassini Jupiter flyby: Multispacecraft observations of interplanetary coronal
  mass ejections including the formation of a merged interaction region},
  \textit{\jgr}, \textit{109}(A9), A09S03, \doi{10.1029/2003JA010112}.

\bibitem[{\textit{{Harrison} et~al.}(2018)\textit{{Harrison}, {Davies},
  {Barnes}, {Byrne}, {Perry}, {Bothmer}, {Eastwood}, {Gallagher}, {Kilpua}, and
  {M{\"o}stl}}}]{harrison2018}
{Harrison}, R.~A., {Davies}, J.~A., {Barnes}, D., {Byrne}, J.~P., {Perry},
  C.~H., {Bothmer}, V., {Eastwood}, J.~P., {Gallagher}, P.~T., {Kilpua},
  E.~K.~J., and {M{\"o}stl}, C. (2018), {CMEs in the Heliosphere: I. A
  Statistical Analysis of the Observational Properties of CMEs Detected in the
  Heliosphere from 2007 to 2017 by STEREO/HI-1}, \textit{\solphys},
  \textit{293}(5), 77, \doi{10.1007/s11207-018-1297-2}.

\bibitem[{\textit{{Heinemann} et~al.}(2019)\textit{{Heinemann}, {Temmer},
  {Farrugia}, {Dissauer}, {Kay}, {Wiegelmann}, {Dumbovi{\'c}}, {Veronig},
  {Podladchikova}, {Hofmeister}, {Lugaz}, and {Carcaboso}}}]{heinemann2019}
{Heinemann}, S.~G., {Temmer}, M., {Farrugia}, C.~J., {Dissauer}, K., {Kay}, C.,
  {Wiegelmann}, T., {Dumbovi{\'c}}, M., {Veronig}, A.~M., {Podladchikova}, T.,
  {Hofmeister}, S.~J., {Lugaz}, N., and {Carcaboso}, F. (2019), {CME-HSS
  Interaction and Characteristics Tracked from Sun to Earth},
  \textit{\solphys}, \textit{294}(9), 121, \doi{10.1007/s11207-019-1515-6}.

\bibitem[{\textit{{Hess} and {Demmelmair}}(1937)}]{hess1937}
{Hess}, V.~F., and {Demmelmair}, A. (1937), {World-wide Effect in Cosmic Ray
  Intensity, as Observed during a Recent Magnetic Storm}, \textit{\nat},
  \textit{140}, 316--317, \doi{10.1038/140316a0}.

\bibitem[{\textit{{Howard} et~al.}(2008)\textit{{Howard}, {Moses}, {Vourlidas},
  {Newmark}, {Socker}, {Plunkett}, {Korendyke}, {Cook}, {Hurley}, {Davila},
  {Thompson}, {St Cyr}, {Mentzell}, {Mehalick}, {Lemen}, {Wuelser}, {Duncan},
  {Tarbell}, {Wolfson}, {Moore}, {Harrison}, {Waltham}, {Lang}, {Davis},
  {Eyles}, {Mapson-Menard}, {Simnett}, {Halain}, {Defise}, {Mazy}, {Rochus},
  {Mercier}, {Ravet}, {Delmotte}, {Auchere}, {Delaboudiniere}, {Bothmer},
  {Deutsch}, {Wang}, {Rich}, {Cooper}, {Stephens}, {Maahs}, {Baugh},
  {McMullin}, and {Carter}}}]{howard2008a}
{Howard}, R.~A., {Moses}, J.~D., {Vourlidas}, A., {Newmark}, J.~S., {Socker},
  D.~G., {Plunkett}, S.~P., {Korendyke}, C.~M., {Cook}, J.~W., {Hurley}, A.,
  {Davila}, J.~M., {Thompson}, W.~T., {St Cyr}, O.~C., {Mentzell}, E.,
  {Mehalick}, K., {Lemen}, J.~R., {Wuelser}, J.~P., {Duncan}, D.~W., {Tarbell},
  T.~D., {Wolfson}, C.~J., {Moore}, A., {Harrison}, R.~A., {Waltham}, N.~R.,
  {Lang}, J., {Davis}, C.~J., {Eyles}, C.~J., {Mapson-Menard}, H., {Simnett},
  G.~M., {Halain}, J.~P., {Defise}, J.~M., {Mazy}, E., {Rochus}, P., {Mercier},
  R., {Ravet}, M.~F., {Delmotte}, F., {Auchere}, F., {Delaboudiniere}, J.~P.,
  {Bothmer}, V., {Deutsch}, W., {Wang}, D., {Rich}, N., {Cooper}, S.,
  {Stephens}, V., {Maahs}, G., {Baugh}, R., {McMullin}, D., and {Carter}, T.
  (2008), {Sun Earth Connection Coronal and Heliospheric Investigation
  (SECCHI)}, \textit{\ssr}, \textit{136}, 67--115,
  \doi{10.1007/s11214-008-9341-4}.

\bibitem[{\textit{{Howard} et~al.}(2017)\textit{{Howard}, {DeForest},
  {Schneck}, and {Alden}}}]{howard2017}
{Howard}, T.~A., {DeForest}, C.~E., {Schneck}, U.~G., and {Alden}, C.~R.
  (2017), {Challenging Some Contemporary Views of Coronal Mass Ejections. II.
  The Case for Absent Filaments}, \textit{\apj}, \textit{834}(1), 86,
  \doi{10.3847/1538-4357/834/1/86}.

\bibitem[{\textit{{Huttunen} et~al.}(2005)\textit{{Huttunen}, {Schwenn},
  {Bothmer}, and {Koskinen}}}]{huttunen2005}
{Huttunen}, K.~E.~J., {Schwenn}, R., {Bothmer}, V., and {Koskinen}, H.~E.~J.
  (2005), {Properties and geoeffectiveness of magnetic clouds in the rising,
  maximum and early declining phases of solar cycle 23}, \textit{\angeo},
  \textit{23}, 625--641, \doi{10.5194/angeo-23-625-2005}.

\bibitem[{\textit{{Isavnin} et~al.}(2014)\textit{{Isavnin}, {Vourlidas}, and
  {Kilpua}}}]{isavnin2014}
{Isavnin}, A., {Vourlidas}, A., and {Kilpua}, E.~K.~J. (2014),
  {Three-Dimensional Evolution of Flux-Rope CMEs and Its Relation to the Local
  Orientation of the Heliospheric Current Sheet}, \textit{\solphys},
  \textit{289}, 2141--2156, \doi{10.1007/s11207-013-0468-4}.

\bibitem[{\textit{{Jackman} et~al.}(2008)\textit{{Jackman}, {Forsyth}, and
  {Dougherty}}}]{jackman2008}
{Jackman}, C.~M., {Forsyth}, R.~J., and {Dougherty}, M.~K. (2008), {The overall
  configuration of the interplanetary magnetic field upstream of Saturn as
  revealed by Cassini observations}, \textit{\jgr}, \textit{113}(A8), A08114,
  \doi{10.1029/2008JA013083}.

\bibitem[{\textit{{Janvier} et~al.}(2014{\natexlab{a}})\textit{{Janvier},
  {D{\'e}moulin}, and {Dasso}}}]{janvier2014c}
{Janvier}, M., {D{\'e}moulin}, P., and {Dasso}, S. (2014{\natexlab{a}}), {In
  situ properties of small and large flux ropes in the solar wind},
  \textit{\jgra}, \textit{119}(9), 7088--7107, \doi{10.1002/2014JA020218}.

\bibitem[{\textit{{Janvier} et~al.}(2014{\natexlab{b}})\textit{{Janvier},
  {Aulanier}, {Bommier}, {Schmieder}, {D{\'e}moulin}, and
  {Pariat}}}]{janvier2014b}
{Janvier}, M., {Aulanier}, G., {Bommier}, V., {Schmieder}, B., {D{\'e}moulin},
  P., and {Pariat}, E. (2014{\natexlab{b}}), {Electric Currents in Flare
  Ribbons: Observations and Three-dimensional Standard Model}, \textit{\apj},
  \textit{788}(1), 60, \doi{10.1088/0004-637X/788/1/60}.

\bibitem[{\textit{{Janvier} et~al.}(2019)\textit{{Janvier}, {Winslow}, {Good},
  {Bonhomme}, {D{\'e}moulin}, {Dasso}, {M{\"o}stl}, {Lugaz}, {Amerstorfer},
  {Soubri{\'e}}, and {Boakes}}}]{janvier2019}
{Janvier}, M., {Winslow}, R.~M., {Good}, S., {Bonhomme}, E., {D{\'e}moulin},
  P., {Dasso}, S., {M{\"o}stl}, C., {Lugaz}, N., {Amerstorfer}, T.,
  {Soubri{\'e}}, E., and {Boakes}, P.~D. (2019), {Generic Magnetic Field
  Intensity Profiles of Interplanetary Coronal Mass Ejections at Mercury,
  Venus, and Earth From Superposed Epoch Analyses}, \textit{\jgra},
  \textit{124}(2), 812--836, \doi{10.1029/2018JA025949}.

\bibitem[{\textit{{Jian} et~al.}(2006)\textit{{Jian}, {Russell}, {Luhmann}, and
  {Skoug}}}]{jian2006b}
{Jian}, L., {Russell}, C.~T., {Luhmann}, J.~G., and {Skoug}, R.~M. (2006),
  {Properties of Interplanetary Coronal Mass Ejections at One AU During 1995 --
  2004}, \textit{\solphys}, \textit{239}, 393--436,
  \doi{10.1007/s11207-006-0133-2}.

\bibitem[{\textit{{Jian} et~al.}(2008{\natexlab{a}})\textit{{Jian}, {Russell},
  {Luhmann}, {Skoug}, and {Steinberg}}}]{jian2008c}
{Jian}, L.~K., {Russell}, C.~T., {Luhmann}, J.~G., {Skoug}, R.~M., and
  {Steinberg}, J.~T. (2008{\natexlab{a}}), {Stream Interactions and
  Interplanetary Coronal Mass Ejections at 5.3 AU near the Solar Ecliptic
  Plane}, \textit{\solphys}, \textit{250}(2), 375--402,
  \doi{10.1007/s11207-008-9204-x}.

\bibitem[{\textit{{Jian} et~al.}(2008{\natexlab{b}})\textit{{Jian}, {Russell},
  {Luhmann}, {Skoug}, and {Steinberg}}}]{jian2008b}
{Jian}, L.~K., {Russell}, C.~T., {Luhmann}, J.~G., {Skoug}, R.~M., and
  {Steinberg}, J.~T. (2008{\natexlab{b}}), {Stream Interactions and
  Interplanetary Coronal Mass Ejections at 0.72 AU}, \textit{\solphys},
  \textit{249}(1), 85--101, \doi{10.1007/s11207-008-9161-4}.

\bibitem[{\textit{{Jian} et~al.}(2018)\textit{{Jian}, {Russell}, {Luhmann}, and
  {Galvin}}}]{jian2018}
{Jian}, L.~K., {Russell}, C.~T., {Luhmann}, J.~G., and {Galvin}, A.~B. (2018),
  {STEREO Observations of Interplanetary Coronal Mass Ejections in 2007-2016},
  \textit{\apj}, \textit{855}, 114, \doi{10.3847/1538-4357/aab189}.

\bibitem[{\textit{{Kaiser} et~al.}(1984)\textit{{Kaiser}, {Desch}, {Kurth},
  {Lecacheux}, {Genova}, {Pedersen}, and {Evans}}}]{kaiser1984}
{Kaiser}, M.~L., {Desch}, M.~D., {Kurth}, W.~S., {Lecacheux}, A., {Genova}, F.,
  {Pedersen}, B.~M., and {Evans}, D.~R. (1984), {Saturn as a radio source}, in
  \textit{Saturn}, edited by T.~{Gehrels} and M.~S. {Matthews}, pp. 378--415,
  University of Arizona Press.

\bibitem[{\textit{{Kaiser} et~al.}(2008)\textit{{Kaiser}, {Kucera}, {Davila},
  {St.~Cyr}, {Guhathakurta}, and {Christian}}}]{kaiser2008}
{Kaiser}, M.~L., {Kucera}, T.~A., {Davila}, J.~M., {St.~Cyr}, O.~C.,
  {Guhathakurta}, M., and {Christian}, E. (2008), {The STEREO Mission: An
  Introduction}, \textit{\ssr}, \textit{136}, 5--16,
  \doi{10.1007/s11214-007-9277-0}.

\bibitem[{\textit{{Kay} et~al.}(2015)\textit{{Kay}, {Opher}, and
  {Evans}}}]{kay2015a}
{Kay}, C., {Opher}, M., and {Evans}, R.~M. (2015), {Global Trends of CME
  Deflections Based on CME and Solar Parameters}, \textit{\apj}, \textit{805},
  168, \doi{10.1088/0004-637X/805/2/168}.

\bibitem[{\textit{{Kilpua} et~al.}(2017)\textit{{Kilpua}, {Koskinen}, and
  {Pulkkinen}}}]{kilpua2017b}
{Kilpua}, E., {Koskinen}, H.~E.~J., and {Pulkkinen}, T.~I. (2017), {Coronal
  mass ejections and their sheath regions in interplanetary space},
  \textit{\lrsp}, \textit{14}, 5, \doi{10.1007/s41116-017-0009-6}.

\bibitem[{\textit{{Kilpua} et~al.}(2009)\textit{{Kilpua}, {Liewer}, {Farrugia},
  {Luhmann}, {M{\"o}stl}, {Li}, {Liu}, {Lynch}, {Russell}, {Vourlidas},
  {Acuna}, {Galvin}, {Larson}, and {Sauvaud}}}]{kilpua2009a}
{Kilpua}, E.~K.~J., {Liewer}, P.~C., {Farrugia}, C., {Luhmann}, J.~G.,
  {M{\"o}stl}, C., {Li}, Y., {Liu}, Y., {Lynch}, B.~J., {Russell}, C.~T.,
  {Vourlidas}, A., {Acuna}, M.~H., {Galvin}, A.~B., {Larson}, D., and
  {Sauvaud}, J.~A. (2009), {Multispacecraft Observations of Magnetic Clouds and
  Their Solar Origins between 19 and 23 May 2007}, \textit{\solphys},
  \textit{254}, 325--344, \doi{10.1007/s11207-008-9300-y}.

\bibitem[{\textit{{Kilpua} et~al.}(2011)\textit{{Kilpua}, {Jian}, {Li},
  {Luhmann}, and {Russell}}}]{kilpua2011}
{Kilpua}, E.~K.~J., {Jian}, L.~K., {Li}, Y., {Luhmann}, J.~G., and {Russell},
  C.~T. (2011), {Multipoint ICME encounters: Pre-STEREO and STEREO
  observations}, \textit{\jastp}, \textit{73}, 1228--1241,
  \doi{10.1016/j.jastp.2010.10.012}.

\bibitem[{\textit{{Kilpua} et~al.}(2019)\textit{{Kilpua}, {Good}, {Palmerio},
  {Asvestari}, {Lumme}, {Ala-Lahti}, {Kalliokoski}, {Morosan}, {Pomoell},
  {Price}, {Magdaleni{\'c}}, {Poedts}, and {Futaana}}}]{kilpua2019b}
{Kilpua}, E. K.~J., {Good}, S.~W., {Palmerio}, E., {Asvestari}, E., {Lumme},
  E., {Ala-Lahti}, M., {Kalliokoski}, M. M.~H., {Morosan}, D.~E., {Pomoell},
  J., {Price}, D.~J., {Magdaleni{\'c}}, J., {Poedts}, S., and {Futaana}, Y.
  (2019), {Multipoint Observations of the June 2012 Interacting Interplanetary
  Flux Ropes}, \textit{\frass}, \textit{6}, 50, \doi{10.3389/fspas.2019.00050}.

\bibitem[{\textit{{Klimchuk}}(2001)}]{klimchuk2001}
{Klimchuk}, J.~A. (2001), {Theory of Coronal Mass Ejections},
  \textit{Washington DC American Geophysical Union Geophysical Monograph
  Series}, \textit{125}, 143, \doi{10.1029/GM125p0143}.

\bibitem[{\textit{{Korreck} et~al.}(2020)\textit{{Korreck}, {Szabo}, {Nieves
  Chinchilla}, {Lavraud}, {Luhmann}, {Niembro}, {Higginson}, {Alzate},
  {Wallace}, {Paulson}, {Rouillard}, {Kouloumvakos}, {Poirier}, {Kasper},
  {Case}, {Stevens}, {Bale}, {Pulupa}, {Whittlesey}, {Livi}, {Goetz}, {Larson},
  {Malaspina}, {Morgan}, {Narock}, {Schwadron}, {Bonnell}, {Harvey}, and
  {Wygant}}}]{korreck2020}
{Korreck}, K.~E., {Szabo}, A., {Nieves Chinchilla}, T., {Lavraud}, B.,
  {Luhmann}, J., {Niembro}, T., {Higginson}, A., {Alzate}, N., {Wallace}, S.,
  {Paulson}, K., {Rouillard}, A., {Kouloumvakos}, A., {Poirier}, N., {Kasper},
  J.~C., {Case}, A.~W., {Stevens}, M.~L., {Bale}, S.~D., {Pulupa}, M.,
  {Whittlesey}, P., {Livi}, R., {Goetz}, K., {Larson}, D., {Malaspina}, D.~M.,
  {Morgan}, H., {Narock}, A.~A., {Schwadron}, N.~A., {Bonnell}, J., {Harvey},
  P., and {Wygant}, J. (2020), {Source and Propagation of a Streamer Blowout
  Coronal Mass Ejection Observed by the Parker Solar Probe}, \textit{\apjs},
  \textit{246}(2), 69, \doi{10.3847/1538-4365/ab6ff9}.

\bibitem[{\textit{{Krall} and {St. Cyr}}(2006)}]{krall2006}
{Krall}, J., and {St. Cyr}, O.~C. (2006), {Flux-Rope Coronal Mass Ejection
  Geometry and Its Relation to Observed Morphology}, \textit{\apj},
  \textit{652}(2), 1740--1746, \doi{10.1086/508337}.

\bibitem[{\textit{{Krimigis} et~al.}(2004)\textit{{Krimigis}, {Mitchell},
  {Hamilton}, {Livi}, {Dandouras}, {Jaskulek}, {Armstrong}, {Boldt}, {Cheng},
  {Gloeckler}, {Hayes}, {Hsieh}, {Ip}, {Keath}, {Kirsch}, {Krupp},
  {Lanzerotti}, {Lundgren}, {Mauk}, {McEntire}, {Roelof}, {Schlemm}, {Tossman},
  {Wilken}, and {Williams}}}]{krimigis2004}
{Krimigis}, S.~M., {Mitchell}, D.~G., {Hamilton}, D.~C., {Livi}, S.,
  {Dandouras}, J., {Jaskulek}, S., {Armstrong}, T.~P., {Boldt}, J.~D., {Cheng},
  A.~F., {Gloeckler}, G., {Hayes}, J.~R., {Hsieh}, K.~C., {Ip}, W.-H., {Keath},
  E.~P., {Kirsch}, E., {Krupp}, N., {Lanzerotti}, L.~J., {Lundgren}, R.,
  {Mauk}, B.~H., {McEntire}, R.~W., {Roelof}, E.~C., {Schlemm}, C.~E.,
  {Tossman}, B.~E., {Wilken}, B., and {Williams}, D.~J. (2004), {Magnetosphere
  Imaging Instrument (MIMI) on the Cassini Mission to Saturn/Titan},
  \textit{\ssr}, \textit{114}, 233--329, \doi{10.1007/s11214-004-1410-8}.

\bibitem[{\textit{{Kurth} et~al.}(2005)\textit{{Kurth}, {Gurnett}, {Clarke},
  {Zarka}, {Desch}, {Kaiser}, {Cecconi}, {Lecacheux}, {Farrell}, {Galopeau},
  {G{\'e}rard}, {Grodent}, {Prang{\'e}}, {Dougherty}, and {Crary}}}]{kurth2005}
{Kurth}, W.~S., {Gurnett}, D.~A., {Clarke}, J.~T., {Zarka}, P., {Desch}, M.~D.,
  {Kaiser}, M.~L., {Cecconi}, B., {Lecacheux}, A., {Farrell}, W.~M.,
  {Galopeau}, P., {G{\'e}rard}, J.~C., {Grodent}, D., {Prang{\'e}}, R.,
  {Dougherty}, M.~K., and {Crary}, F.~J. (2005), {An Earth-like correspondence
  between Saturn's auroral features and radio emission}, \textit{\nat},
  \textit{433}(7027), 722--725, \doi{10.1038/nature03334}.

\bibitem[{\textit{{Kurth} et~al.}(2016)\textit{{Kurth}, {Hospodarsky},
  {Gurnett}, {Lamy}, {Dougherty}, {Nichols}, {Bunce}, {Pryor}, {Baines},
  {Stallard}, {Melin}, and {Crary}}}]{kurth2016}
{Kurth}, W.~S., {Hospodarsky}, G.~B., {Gurnett}, D.~A., {Lamy}, L.,
  {Dougherty}, M.~K., {Nichols}, J., {Bunce}, E.~J., {Pryor}, W., {Baines}, K.,
  {Stallard}, T., {Melin}, H., and {Crary}, F.~J. (2016), {Saturn kilometric
  radiation intensities during the Saturn auroral campaign of 2013},
  \textit{\icarus}, \textit{263}, 2--9, \doi{10.1016/j.icarus.2015.01.003}.

\bibitem[{\textit{{Lamy}}(2017)}]{lamy2017a}
{Lamy}, L. (2017), {The Saturnian kilometric radiation before the Cassini Grand
  Finale}, in \textit{Planetary Radio Emissions VIII}, edited by G.~{Fischer},
  G.~{Mann}, M.~{Panchenko}, and P.~{Zarka}, pp. 171--190,
  \doi{10.1553/PRE8s171}.

\bibitem[{\textit{{Lamy}}(2020)}]{lamy2020}
{Lamy}, L. (2020), {Auroral emissions from Uranus and Neptune},
  \textit{\rspta}, \textit{378}(2187), 20190481, \doi{10.1098/rsta.2019.0481}.

\bibitem[{\textit{{Lamy} et~al.}(2008)\textit{{Lamy}, {Zarka}, {Cecconi},
  {Prang{\'e}}, {Kurth}, and {Gurnett}}}]{lamy2008}
{Lamy}, L., {Zarka}, P., {Cecconi}, B., {Prang{\'e}}, R., {Kurth}, W.~S., and
  {Gurnett}, D.~A. (2008), {Saturn kilometric radiation: Average and
  statistical properties}, \textit{\jgr}, \textit{113}(A7), A07201,
  \doi{10.1029/2007JA012900}.

\bibitem[{\textit{{Lamy} et~al.}(2009)\textit{{Lamy}, {Cecconi}, {Prang{\'e}},
  {Zarka}, {Nichols}, and {Clarke}}}]{lamy2009}
{Lamy}, L., {Cecconi}, B., {Prang{\'e}}, R., {Zarka}, P., {Nichols}, J.~D., and
  {Clarke}, J.~T. (2009), {An auroral oval at the footprint of Saturn's
  kilometric radio sources, colocated with the UV aurorae}, \textit{\jgr},
  \textit{114}(A10), A10212, \doi{10.1029/2009JA014401}.

\bibitem[{\textit{{Lamy} et~al.}(2012)\textit{{Lamy}, {Prang{\'e}}, {Hansen},
  {Clarke}, {Zarka}, {Cecconi}, {Aboudarham}, {Andr{\'e}},
  {Branduardi-Raymont}, {Gladstone}, {Barth{\'e}l{\'e}my}, {Achilleos}, {Guio},
  {Dougherty}, {Melin}, {Cowley}, {Stallard}, {Nichols}, and
  {Ballester}}}]{lamy2012}
{Lamy}, L., {Prang{\'e}}, R., {Hansen}, K.~C., {Clarke}, J.~T., {Zarka}, P.,
  {Cecconi}, B., {Aboudarham}, J., {Andr{\'e}}, N., {Branduardi-Raymont}, G.,
  {Gladstone}, R., {Barth{\'e}l{\'e}my}, M., {Achilleos}, N., {Guio}, P.,
  {Dougherty}, M.~K., {Melin}, H., {Cowley}, S.~W.~H., {Stallard}, T.~S.,
  {Nichols}, J.~D., and {Ballester}, G. (2012), {Earth-based detection of
  Uranus' aurorae}, \textit{\grl}, \textit{39}(7), L07105,
  \doi{10.1029/2012GL051312}.

\bibitem[{\textit{{Lamy} et~al.}(2017)\textit{{Lamy}, {Prang{\'e}}, {Hansen},
  {Tao}, {Cowley}, {Stallard}, {Melin}, {Achilleos}, {Guio}, {Badman}, {Kim},
  and {Pogorelov}}}]{lamy2017b}
{Lamy}, L., {Prang{\'e}}, R., {Hansen}, K.~C., {Tao}, C., {Cowley}, S.~W.~H.,
  {Stallard}, T.~S., {Melin}, H., {Achilleos}, N., {Guio}, P., {Badman}, S.~V.,
  {Kim}, T., and {Pogorelov}, N. (2017), {The aurorae of Uranus past equinox},
  \textit{\jgra}, \textit{122}(4), 3997--4008, \doi{10.1002/2017JA023918}.

\bibitem[{\textit{{Lamy} et~al.}(2018)\textit{{Lamy}, {Prang{\'e}}, {Tao},
  {Kim}, {Badman}, {Zarka}, {Cecconi}, {Kurth}, {Pryor}, {Bunce}, and
  {Radioti}}}]{lamy2018}
{Lamy}, L., {Prang{\'e}}, R., {Tao}, C., {Kim}, T., {Badman}, S.~V., {Zarka},
  P., {Cecconi}, B., {Kurth}, W.~S., {Pryor}, W., {Bunce}, E., and {Radioti},
  A. (2018), {Saturn's Northern Aurorae at Solstice From HST Observations
  Coordinated With Cassini's Grand Finale}, \textit{\grl}, \textit{45}(18),
  9353--9362, \doi{10.1029/2018GL078211}.

\bibitem[{\textit{{Lario} et~al.}(2020)\textit{{Lario}, {Balmaceda}, {Alzate},
  {Mays}, {Richardson}, {Allen}, {Florido-Llinas}, {Nieves-Chinchilla},
  {Koval}, {Lugaz}, {Jian}, {Arge}, {Macneice}, {Odstrcil}, {Morgan}, {Szabo},
  {Desai}, {Whittlesey}, {Stevens}, {Ho}, and {Luhmann}}}]{lario2020}
{Lario}, D., {Balmaceda}, L., {Alzate}, N., {Mays}, M.~L., {Richardson}, I.~G.,
  {Allen}, R.~C., {Florido-Llinas}, M., {Nieves-Chinchilla}, T., {Koval}, A.,
  {Lugaz}, N., {Jian}, L.~K., {Arge}, C.~N., {Macneice}, P.~J., {Odstrcil}, D.,
  {Morgan}, H., {Szabo}, A., {Desai}, M.~I., {Whittlesey}, P.~L., {Stevens},
  M.~L., {Ho}, G.~C., and {Luhmann}, J.~G. (2020), {The Streamer Blowout Origin
  of a Flux Rope and Energetic Particle Event Observed by Parker Solar Probe at
  0.5 au}, \textit{\apj}, \textit{897}(2), 134, \doi{10.3847/1538-4357/ab9942}.

\bibitem[{\textit{{Lee} et~al.}(2017)\textit{{Lee}, {Hara}, {Halekas},
  {Thiemann}, {Chamberlin}, {Eparvier}, {Lillis}, {Larson}, {Dunn}, and
  {Espley}}}]{lee2017}
{Lee}, C.~O., {Hara}, T., {Halekas}, J.~S., {Thiemann}, E., {Chamberlin}, P.,
  {Eparvier}, F., {Lillis}, R.~J., {Larson}, D.~E., {Dunn}, P.~A., and
  {Espley}, J.~R. (2017), {MAVEN observations of the solar cycle 24 space
  weather conditions at Mars}, \textit{\jgra}, \textit{122}(3), 2768--2794,
  \doi{10.1002/2016JA023495}.

\bibitem[{\textit{{Lemen} et~al.}(2012)\textit{{Lemen}, {Title}, {Akin},
  {Boerner}, {Chou}, {Drake}, {Duncan}, {Edwards}, {Friedlaender}, {Heyman},
  {Hurlburt}, {Katz}, {Kushner}, {Levay}, {Lindgren}, {Mathur}, {McFeaters},
  {Mitchell}, {Rehse}, {Schrijver}, {Springer}, {Stern}, {Tarbell}, {Wuelser},
  {Wolfson}, {Yanari}, {Bookbinder}, {Cheimets}, {Caldwell}, {Deluca}, {Gates},
  {Golub}, {Park}, {Podgorski}, {Bush}, {Scherrer}, {Gummin}, {Smith}, {Auker},
  {Jerram}, {Pool}, {Soufli}, {Windt}, {Beardsley}, {Clapp}, {Lang}, and
  {Waltham}}}]{lemen2012}
{Lemen}, J.~R., {Title}, A.~M., {Akin}, D.~J., {Boerner}, P.~F., {Chou}, C.,
  {Drake}, J.~F., {Duncan}, D.~W., {Edwards}, C.~G., {Friedlaender}, F.~M.,
  {Heyman}, G.~F., {Hurlburt}, N.~E., {Katz}, N.~L., {Kushner}, G.~D., {Levay},
  M., {Lindgren}, R.~W., {Mathur}, D.~P., {McFeaters}, E.~L., {Mitchell}, S.,
  {Rehse}, R.~A., {Schrijver}, C.~J., {Springer}, L.~A., {Stern}, R.~A.,
  {Tarbell}, T.~D., {Wuelser}, J.-P., {Wolfson}, C.~J., {Yanari}, C.,
  {Bookbinder}, J.~A., {Cheimets}, P.~N., {Caldwell}, D., {Deluca}, E.~E.,
  {Gates}, R., {Golub}, L., {Park}, S., {Podgorski}, W.~A., {Bush}, R.~I.,
  {Scherrer}, P.~H., {Gummin}, M.~A., {Smith}, P., {Auker}, G., {Jerram}, P.,
  {Pool}, P., {Soufli}, R., {Windt}, D.~L., {Beardsley}, S., {Clapp}, M.,
  {Lang}, J., and {Waltham}, N. (2012), {The Atmospheric Imaging Assembly (AIA)
  on the Solar Dynamics Observatory (SDO)}, \textit{\solphys}, \textit{275},
  17--40, \doi{10.1007/s11207-011-9776-8}.

\bibitem[{\textit{{Lepping} et~al.}(1995)\textit{{Lepping}, {Ac{\~u}na},
  {Burlaga}, {Farrell}, {Slavin}, {Schatten}, {Mariani}, {Ness}, {Neubauer},
  {Whang}, {Byrnes}, {Kennon}, {Panetta}, {Scheifele}, and
  {Worley}}}]{lepping1995}
{Lepping}, R.~P., {Ac{\~u}na}, M.~H., {Burlaga}, L.~F., {Farrell}, W.~M.,
  {Slavin}, J.~A., {Schatten}, K.~H., {Mariani}, F., {Ness}, N.~F., {Neubauer},
  F.~M., {Whang}, Y.~C., {Byrnes}, J.~B., {Kennon}, R.~S., {Panetta}, P.~V.,
  {Scheifele}, J., and {Worley}, E.~M. (1995), {The Wind Magnetic Field
  Investigation}, \textit{\ssr}, \textit{71}, 207--229,
  \doi{10.1007/BF00751330}.

\bibitem[{\textit{{Lepri} and {Zurbuchen}}(2004)}]{lepri2004}
{Lepri}, S.~T., and {Zurbuchen}, T.~H. (2004), {Iron charge state distributions
  as an indicator of hot ICMEs: Possible sources and temporal and spatial
  variations during solar maximum}, \textit{\jgr}, \textit{109}(A1), A01112,
  \doi{10.1029/2003JA009954}.

\bibitem[{\textit{{Li} et~al.}(2014)\textit{{Li}, {Luhmann}, {Lynch}, and
  {Kilpua}}}]{li2014}
{Li}, Y., {Luhmann}, J.~G., {Lynch}, B.~J., and {Kilpua}, E.~K.~J. (2014),
  {Magnetic clouds and origins in STEREO era}, \textit{\jgra}, \textit{119}(5),
  3237--3246, \doi{10.1002/2013JA019538}.

\bibitem[{\textit{{Li} et~al.}(2018)\textit{{Li}, {Luhmann}, and
  {Lynch}}}]{li2018}
{Li}, Y., {Luhmann}, J.~G., and {Lynch}, B.~J. (2018), {Magnetic Clouds: Solar
  Cycle Dependence, Sources, and Geomagnetic Impacts}, \textit{\solphys},
  \textit{293}, 135, \doi{10.1007/s11207-018-1356-8}.

\bibitem[{\textit{{Liu} et~al.}(2009)\textit{{Liu}, {Alexander}, and
  {Gilbert}}}]{liu2009}
{Liu}, R., {Alexander}, D., and {Gilbert}, H.~R. (2009), {Asymmetric Eruptive
  Filaments}, \textit{\apj}, \textit{691}(2), 1079--1091,
  \doi{10.1088/0004-637X/691/2/1079}.

\bibitem[{\textit{{Liu} et~al.}(2005)\textit{{Liu}, {Richardson}, and
  {Belcher}}}]{liu2005}
{Liu}, Y., {Richardson}, J.~D., and {Belcher}, J.~W. (2005), {A statistical
  study of the properties of interplanetary coronal mass ejections from 0.3 to
  5.4 AU}, \textit{\planss}, \textit{53}(1-3), 3--17,
  \doi{10.1016/j.pss.2004.09.023}.

\bibitem[{\textit{{Liu} et~al.}(2014)\textit{{Liu}, {Luhmann}, {Kajdi{\v{c}}},
  {Kilpua}, {Lugaz}, {Nitta}, {M{\"o}stl}, {Lavraud}, {Bale}, {Farrugia}, and
  {Galvin}}}]{liu2014}
{Liu}, Y.~D., {Luhmann}, J.~G., {Kajdi{\v{c}}}, P., {Kilpua}, E. K.~J.,
  {Lugaz}, N., {Nitta}, N.~V., {M{\"o}stl}, C., {Lavraud}, B., {Bale}, S.~D.,
  {Farrugia}, C.~J., and {Galvin}, A.~B. (2014), {Observations of an extreme
  storm in interplanetary space caused by successive coronal mass ejections},
  \textit{\natco}, \textit{5}, 3481, \doi{10.1038/ncomms4481}.

\bibitem[{\textit{{Liu} et~al.}(2019)\textit{{Liu}, {Zhao}, {Hu}, {Vourlidas},
  and {Zhu}}}]{liu2019}
{Liu}, Y.~D., {Zhao}, X., {Hu}, H., {Vourlidas}, A., and {Zhu}, B. (2019), {A
  Comparative Study of 2017 July and 2012 July Complex Eruptions: Are Solar
  Superstorms {\textquotedblleft}Perfect Storms{\textquotedblright} in
  Nature?}, \textit{\apjs}, \textit{241}(2), 15,
  \doi{10.3847/1538-4365/ab0649}.

\bibitem[{\textit{{Lugaz} and {Farrugia}}(2014)}]{lugaz2014}
{Lugaz}, N., and {Farrugia}, C.~J. (2014), {A new class of complex ejecta
  resulting from the interaction of two CMEs and its expected
  geoeffectiveness}, \textit{\grl}, \textit{41}(3), 769--776,
  \doi{10.1002/2013GL058789}.

\bibitem[{\textit{{Lugaz} et~al.}(2013)\textit{{Lugaz}, {Farrugia},
  {Manchester}, and {Schwadron}}}]{lugaz2013}
{Lugaz}, N., {Farrugia}, C.~J., {Manchester}, I., W.~B., and {Schwadron}, N.
  (2013), {The Interaction of Two Coronal Mass Ejections: Influence of Relative
  Orientation}, \textit{\apj}, \textit{778}(1), 20,
  \doi{10.1088/0004-637X/778/1/20}.

\bibitem[{\textit{{Lugaz} et~al.}(2017{\natexlab{a}})\textit{{Lugaz}, {Temmer},
  {Wang}, and {Farrugia}}}]{lugaz2017a}
{Lugaz}, N., {Temmer}, M., {Wang}, Y., and {Farrugia}, C.~J.
  (2017{\natexlab{a}}), {The Interaction of Successive Coronal Mass Ejections:
  A Review}, \textit{\solphys}, \textit{292}(4), 64,
  \doi{10.1007/s11207-017-1091-6}.

\bibitem[{\textit{{Lugaz} et~al.}(2017{\natexlab{b}})\textit{{Lugaz},
  {Farrugia}, {Winslow}, {Small}, {Manion}, and {Savani}}}]{lugaz2017b}
{Lugaz}, N., {Farrugia}, C.~J., {Winslow}, R.~M., {Small}, C.~R., {Manion}, T.,
  and {Savani}, N.~P. (2017{\natexlab{b}}), {Importance of CME Radial Expansion
  on the Ability of Slow CMEs to Drive Shocks}, \textit{\apj}, \textit{848}(2),
  75, \doi{10.3847/1538-4357/aa8ef9}.

\bibitem[{\textit{{Lugaz} et~al.}(2020)\textit{{Lugaz}, {Winslow}, and
  {Farrugia}}}]{lugaz2020}
{Lugaz}, N., {Winslow}, R.~M., and {Farrugia}, C.~J. (2020), {Evolution of a
  Long-Duration Coronal Mass Ejection and Its Sheath Region Between Mercury and
  Earth on 9-14 July 2013}, \textit{\jgra}, \textit{125}(1), e2019JA027213,
  \doi{10.1029/2019JA027213}.

\bibitem[{\textit{{Luhmann} et~al.}(2020)\textit{{Luhmann}, {Gopalswamy},
  {Jian}, and {Lugaz}}}]{luhmann2020}
{Luhmann}, J.~G., {Gopalswamy}, N., {Jian}, L.~K., and {Lugaz}, N. (2020),
  {ICME Evolution in the Inner Heliosphere}, \textit{\solphys},
  \textit{295}(4), 61, \doi{10.1007/s11207-020-01624-0}.

\bibitem[{\textit{{Lynch} et~al.}(2003)\textit{{Lynch}, {Zurbuchen}, {Fisk},
  and {Antiochos}}}]{lynch2003}
{Lynch}, B.~J., {Zurbuchen}, T.~H., {Fisk}, L.~A., and {Antiochos}, S.~K.
  (2003), {Internal structure of magnetic clouds: Plasma and composition},
  \textit{\jgr}, \textit{108}, 1239, \doi{10.1029/2002JA009591}.

\bibitem[{\textit{{Lynch} et~al.}(2009)\textit{{Lynch}, {Antiochos}, {Li},
  {Luhmann}, and {DeVore}}}]{lynch2009}
{Lynch}, B.~J., {Antiochos}, S.~K., {Li}, Y., {Luhmann}, J.~G., and {DeVore},
  C.~R. (2009), {Rotation of Coronal Mass Ejections during Eruption},
  \textit{\apj}, \textit{697}, 1918--1927, \doi{10.1088/0004-637X/697/2/1918}.

\bibitem[{\textit{{Lynch} et~al.}(2010)\textit{{Lynch}, {Li}, {Thernisien},
  {Robbrecht}, {Fisher}, {Luhmann}, and {Vourlidas}}}]{lynch2010}
{Lynch}, B.~J., {Li}, Y., {Thernisien}, A.~F.~R., {Robbrecht}, E., {Fisher},
  G.~H., {Luhmann}, J.~G., and {Vourlidas}, A. (2010), {Sun to 1 AU propagation
  and evolution of a slow streamer-blowout coronal mass ejection},
  \textit{\jgr}, \textit{115}, A07106, \doi{10.1029/2009JA015099}.

\bibitem[{\textit{{Manchester} et~al.}(2017)\textit{{Manchester}, {Kilpua},
  {Liu}, {Lugaz}, {Riley}, {T{\"o}r{\"o}k}, and {Vr{\v
  s}nak}}}]{manchester2017}
{Manchester}, W., {Kilpua}, E.~K.~J., {Liu}, Y.~D., {Lugaz}, N., {Riley}, P.,
  {T{\"o}r{\"o}k}, T., and {Vr{\v s}nak}, B. (2017), {The Physical Processes of
  CME/ICME Evolution}, \textit{\ssr}, \doi{10.1007/s11214-017-0394-0}.

\bibitem[{\textit{{Manchester} et~al.}(2004)\textit{{Manchester}, {Gombosi},
  {Roussev}, {Ridley}, {de Zeeuw}, {Sokolov}, {Powell}, and
  {T{\'o}th}}}]{manchester2004}
{Manchester}, W.~B., {Gombosi}, T.~I., {Roussev}, I., {Ridley}, A., {de Zeeuw},
  D.~L., {Sokolov}, I.~V., {Powell}, K.~G., and {T{\'o}th}, G. (2004),
  {Modeling a space weather event from the Sun to the Earth: CME generation and
  interplanetary propagation}, \textit{\jgr}, \textit{109}(A2), A02107,
  \doi{10.1029/2003JA010150}.

\bibitem[{\textit{{Martin}}(1998)}]{martin1998a}
{Martin}, S.~F. (1998), {Filament Chirality: A Link Between Fine-Scale and
  Global Patterns (Review)}, in \textit{IAU Colloq. 167: New Perspectives on
  Solar Prominences}, \textit{\aspcs}, vol. 150, edited by D.~F. {Webb},
  B.~{Schmieder}, and D.~M. {Rust}, p. 419.

\bibitem[{\textit{{Martin}}(2003)}]{martin2003}
{Martin}, S.~F. (2003), {Signs of helicity in solar prominences and related
  features}, \textit{\adv}, \textit{32}, 1883--1893,
  \doi{10.1016/S0273-1177(03)90622-3}.

\bibitem[{\textit{{Marubashi} and {Lepping}}(2007)}]{marubashi2007}
{Marubashi}, K., and {Lepping}, R.~P. (2007), {Long-duration magnetic clouds: a
  comparison of analyses using torus- and cylinder-shaped flux rope models},
  \textit{\angeo}, \textit{25}(11), 2453--2477,
  \doi{10.5194/angeo-25-2453-2007}.

\bibitem[{\textit{{Marubashi} et~al.}(2015)\textit{{Marubashi}, {Akiyama},
  {Yashiro}, {Gopalswamy}, {Cho}, and {Park}}}]{marubashi2015}
{Marubashi}, K., {Akiyama}, S., {Yashiro}, S., {Gopalswamy}, N., {Cho}, K.-S.,
  and {Park}, Y.-D. (2015), {Geometrical Relationship Between Interplanetary
  Flux Ropes and Their Solar Sources}, \textit{\solphys}, \textit{290},
  1371--1397, \doi{10.1007/s11207-015-0681-4}.

\bibitem[{\textit{{Matson} et~al.}(2002)\textit{{Matson}, {Spilker}, and
  {Lebreton}}}]{matson2002}
{Matson}, D.~L., {Spilker}, L.~J., and {Lebreton}, J.-P. (2002), {The
  Cassini/Huygens Mission to the Saturnian System}, \textit{\ssr},
  \textit{104}(1), 1--58, \doi{10.1023/A:1023609211620}.

\bibitem[{\textit{{M{\"o}stl} and {Davies}}(2013)}]{mostl2013}
{M{\"o}stl}, C., and {Davies}, J.~A. (2013), {Speeds and Arrival Times of Solar
  Transients Approximated by Self-similar Expanding Circular Fronts},
  \textit{\solphys}, \textit{285}, 411--423, \doi{10.1007/s11207-012-9978-8}.

\bibitem[{\textit{{M{\"o}stl} et~al.}(2008)\textit{{M{\"o}stl}, {Miklenic},
  {Farrugia}, {Temmer}, {Veronig}, {Galvin}, {Vr{\v s}nak}, and
  {Biernat}}}]{mostl2008}
{M{\"o}stl}, C., {Miklenic}, C., {Farrugia}, C.~J., {Temmer}, M., {Veronig},
  A., {Galvin}, A.~B., {Vr{\v s}nak}, B., and {Biernat}, H.~K. (2008),
  {Two-spacecraft reconstruction of a magnetic cloud and comparison to its
  solar source}, \textit{\angeo}, \textit{26}, 3139--3152,
  \doi{10.5194/angeo-26-3139-2008}.

\bibitem[{\textit{{M{\"o}stl} et~al.}(2009{\natexlab{a}})\textit{{M{\"o}stl},
  {Farrugia}, {Miklenic}, {Temmer}, {Galvin}, {Luhmann}, {Kilpua}, {Leitner},
  {Nieves-Chinchilla}, {Veronig}, and {Biernat}}}]{mostl2009a}
{M{\"o}stl}, C., {Farrugia}, C.~J., {Miklenic}, C., {Temmer}, M., {Galvin},
  A.~B., {Luhmann}, J.~G., {Kilpua}, E.~K.~J., {Leitner}, M.,
  {Nieves-Chinchilla}, T., {Veronig}, A., and {Biernat}, H.~K.
  (2009{\natexlab{a}}), {Multispacecraft recovery of a magnetic cloud and its
  origin from magnetic reconnection on the Sun}, \textit{\jgr}, \textit{114},
  A04102, \doi{10.1029/2008JA013657}.

\bibitem[{\textit{{M{\"o}stl} et~al.}(2009{\natexlab{b}})\textit{{M{\"o}stl},
  {Farrugia}, {Temmer}, {Miklenic}, {Veronig}, {Galvin}, {Leitner}, and
  {Biernat}}}]{mostl2009b}
{M{\"o}stl}, C., {Farrugia}, C.~J., {Temmer}, M., {Miklenic}, C., {Veronig},
  A.~M., {Galvin}, A.~B., {Leitner}, M., and {Biernat}, H.~K.
  (2009{\natexlab{b}}), {Linking Remote Imagery of a Coronal Mass Ejection to
  Its In Situ Signatures at 1 AU}, \textit{\apjl}, \textit{705}(2), L180--L185,
  \doi{10.1088/0004-637X/705/2/L180}.

\bibitem[{\textit{{M{\"o}stl} et~al.}(2015)\textit{{M{\"o}stl}, {Rollett},
  {Frahm}, {Liu}, {Long}, {Colaninno}, {Reiss}, {Temmer}, {Farrugia}, {Posner},
  {Dumbovi{\'c}}, {Janvier}, {D{\'e}moulin}, {Boakes}, {Devos}, {Kraaikamp},
  {Mays}, and {Vr{\v s}nak}}}]{mostl2015}
{M{\"o}stl}, C., {Rollett}, T., {Frahm}, R.~A., {Liu}, Y.~D., {Long}, D.~M.,
  {Colaninno}, R.~C., {Reiss}, M.~A., {Temmer}, M., {Farrugia}, C.~J.,
  {Posner}, A., {Dumbovi{\'c}}, M., {Janvier}, M., {D{\'e}moulin}, P.,
  {Boakes}, P., {Devos}, A., {Kraaikamp}, E., {Mays}, M.~L., and {Vr{\v s}nak},
  B. (2015), {Strong coronal channelling and interplanetary evolution of a
  solar storm up to Earth and Mars}, \textit{\natco}, \textit{6}, 7135,
  \doi{10.1038/ncomms8135}.

\bibitem[{\textit{{M{\"o}stl} et~al.}(2017)\textit{{M{\"o}stl}, {Isavnin},
  {Boakes}, {Kilpua}, {Davies}, {Harrison}, {Barnes}, {Krupar}, {Eastwood},
  {Good}, {Forsyth}, {Bothmer}, {Reiss}, {Amerstorfer}, {Winslow}, {Anderson},
  {Philpott}, {Rodriguez}, {Rouillard}, {Gallagher}, {Nieves-Chinchilla}, and
  {Zhang}}}]{mostl2017}
{M{\"o}stl}, C., {Isavnin}, A., {Boakes}, P.~D., {Kilpua}, E.~K.~J., {Davies},
  J.~A., {Harrison}, R.~A., {Barnes}, D., {Krupar}, V., {Eastwood}, J.~P.,
  {Good}, S.~W., {Forsyth}, R.~J., {Bothmer}, V., {Reiss}, M.~A.,
  {Amerstorfer}, T., {Winslow}, R.~M., {Anderson}, B.~J., {Philpott}, L.~C.,
  {Rodriguez}, L., {Rouillard}, A.~P., {Gallagher}, P., {Nieves-Chinchilla},
  T., and {Zhang}, T.~L. (2017), {Modeling observations of solar coronal mass
  ejections with heliospheric imagers verified with the Heliophysics System
  Observatory}, \textit{\spwea}, \textit{15}, 955--970,
  \doi{10.1002/2017SW001614}.

\bibitem[{\textit{{M{\"u}ller} et~al.}(2017)\textit{{M{\"u}ller}, {Nicula},
  {Felix}, {Verstringe}, {Bourgoignie}, {Csillaghy}, {Berghmans}, {Jiggens},
  {Garc{\'\i}a-Ortiz}, {Ireland}, {Zahniy}, and {Fleck}}}]{muller2017}
{M{\"u}ller}, D., {Nicula}, B., {Felix}, S., {Verstringe}, F., {Bourgoignie},
  B., {Csillaghy}, A., {Berghmans}, D., {Jiggens}, P., {Garc{\'\i}a-Ortiz},
  J.~P., {Ireland}, J., {Zahniy}, S., and {Fleck}, B. (2017), {JHelioviewer.
  Time-dependent 3D visualisation of solar and heliospheric data},
  \textit{\aap}, \textit{606}, A10, \doi{10.1051/0004-6361/201730893}.

\bibitem[{\textit{{Mulligan} et~al.}(1998)\textit{{Mulligan}, {Russell}, and
  {Luhmann}}}]{mulligan1998}
{Mulligan}, T., {Russell}, C.~T., and {Luhmann}, J.~G. (1998), {Solar cycle
  evolution of the structure of magnetic clouds in the inner heliosphere},
  \textit{\grl}, \textit{25}, 2959--2962, \doi{10.1029/98GL01302}.

\bibitem[{\textit{{Nieves-Chinchilla} and
  {Vi{\~n}as}}(2008)}]{nieveschinchilla2008}
{Nieves-Chinchilla}, T., and {Vi{\~n}as}, A.~F. (2008), {Solar wind electron
  distribution functions inside magnetic clouds}, \textit{\jgr},
  \textit{113}(A2), A02105, \doi{10.1029/2007JA012703}.

\bibitem[{\textit{{Nieves-Chinchilla} et~al.}(2005)\textit{{Nieves-Chinchilla},
  {Hidalgo}, and {Sequeiros}}}]{nieveschinchilla2005}
{Nieves-Chinchilla}, T., {Hidalgo}, M.~A., and {Sequeiros}, J. (2005),
  {Magnetic Clouds Observed at 1 Au During the Period 2000-2003},
  \textit{\solphys}, \textit{232}(1-2), 105--126,
  \doi{10.1007/s11207-005-1593-5}.

\bibitem[{\textit{{Nieves-Chinchilla} et~al.}(2011)\textit{{Nieves-Chinchilla},
  {G{\'o}mez-Herrero}, {Vi{\~n}as}, {Malandraki}, {Dresing}, {Hidalgo},
  {Opitz}, {Sauvaud}, {Lavraud}, and {Davila}}}]{nieveschinchilla2011}
{Nieves-Chinchilla}, T., {G{\'o}mez-Herrero}, R., {Vi{\~n}as}, A.~F.,
  {Malandraki}, O., {Dresing}, N., {Hidalgo}, M.~A., {Opitz}, A., {Sauvaud},
  J.~A., {Lavraud}, B., and {Davila}, J.~M. (2011), {Analysis and study of the
  in situ observation of the June 1st 2008 CME by STEREO}, \textit{\jastp},
  \textit{73}(11-12), 1348--1360, \doi{10.1016/j.jastp.2010.09.026}.

\bibitem[{\textit{{Nieves-Chinchilla} et~al.}(2012)\textit{{Nieves-Chinchilla},
  {Colaninno}, {Vourlidas}, {Szabo}, {Lepping}, {Boardsen}, {Anderson}, and
  {Korth}}}]{nieveschinchilla2012}
{Nieves-Chinchilla}, T., {Colaninno}, R., {Vourlidas}, A., {Szabo}, A.,
  {Lepping}, R.~P., {Boardsen}, S.~A., {Anderson}, B.~J., and {Korth}, H.
  (2012), {Remote and in situ observations of an unusual Earth-directed coronal
  mass ejection from multiple viewpoints}, \textit{\jgr}, \textit{117}(A6),
  A06106, \doi{10.1029/2011JA017243}.

\bibitem[{\textit{{Nieves-Chinchilla}
  et~al.}(2018{\natexlab{a}})\textit{{Nieves-Chinchilla}, {Vourlidas},
  {Raymond}, {Linton}, {Al-haddad}, {Savani}, {Szabo}, and
  {Hidalgo}}}]{nieveschinchilla2018a}
{Nieves-Chinchilla}, T., {Vourlidas}, A., {Raymond}, J.~C., {Linton}, M.~G.,
  {Al-haddad}, N., {Savani}, N.~P., {Szabo}, A., and {Hidalgo}, M.~A.
  (2018{\natexlab{a}}), {Understanding the Internal Magnetic Field
  Configurations of ICMEs Using More than 20 Years of Wind Observations},
  \textit{\solphys}, \textit{293}, 25, \doi{10.1007/s11207-018-1247-z}.

\bibitem[{\textit{{Nieves-Chinchilla}
  et~al.}(2018{\natexlab{b}})\textit{{Nieves-Chinchilla}, {Linton}, {Hidalgo},
  and {Vourlidas}}}]{nieveschinchilla2018b}
{Nieves-Chinchilla}, T., {Linton}, M.~G., {Hidalgo}, M.~A., and {Vourlidas}, A.
  (2018{\natexlab{b}}), {Elliptic-cylindrical Analytical Flux Rope Model for
  Magnetic Clouds}, \textit{\apj}, \textit{861}(2), 139,
  \doi{10.3847/1538-4357/aac951}.

\bibitem[{\textit{{Nieves-Chinchilla} et~al.}(2019)\textit{{Nieves-Chinchilla},
  {Jian}, {Balmaceda}, {Vourlidas}, {dos Santos}, and
  {Szabo}}}]{nieveschinchilla2019}
{Nieves-Chinchilla}, T., {Jian}, L.~K., {Balmaceda}, L., {Vourlidas}, A., {dos
  Santos}, L. F.~G., and {Szabo}, A. (2019), {Unraveling the Internal Magnetic
  Field Structure of the Earth-directed Interplanetary Coronal Mass Ejections
  During 1995 - 2015}, \textit{\solphys}, \textit{294}(7), 89,
  \doi{10.1007/s11207-019-1477-8}.

\bibitem[{\textit{{Nieves-Chinchilla} et~al.}(2020)\textit{{Nieves-Chinchilla},
  {Szabo}, {Korreck}, {Alzate}, {Balmaceda}, {Lavraud}, {Paulson}, {Narock},
  {Wallace}, {Jian}, {Luhmann}, {Morgan}, {Higginson}, {Arge}, {Bale}, {Case},
  {Wit}, {Giacalone}, {Goetz}, {Harvey}, {Jones-Melosky}, {Kasper}, {Larson},
  {Livi}, {McComas}, {MacDowall}, {Malaspina}, {Pulupa}, {Raouafi},
  {Schwadron}, {Stevens}, and {Whittlesey}}}]{nieveschinchilla2020}
{Nieves-Chinchilla}, T., {Szabo}, A., {Korreck}, K.~E., {Alzate}, N.,
  {Balmaceda}, L.~A., {Lavraud}, B., {Paulson}, K., {Narock}, A.~A., {Wallace},
  S., {Jian}, L.~K., {Luhmann}, J.~G., {Morgan}, H., {Higginson}, A., {Arge},
  C.~N., {Bale}, S.~D., {Case}, A.~W., {Wit}, T. D.~d., {Giacalone}, J.,
  {Goetz}, K., {Harvey}, P.~R., {Jones-Melosky}, S.~I., {Kasper}, J.~C.,
  {Larson}, D.~E., {Livi}, R., {McComas}, D.~J., {MacDowall}, R.~J.,
  {Malaspina}, D.~M., {Pulupa}, M., {Raouafi}, N.~E., {Schwadron}, N.,
  {Stevens}, M.~L., and {Whittlesey}, P.~L. (2020), {Analysis of the Internal
  Structure of the Streamer Blowout Observed by the Parker Solar Probe During
  the First Solar Encounter}, \textit{\apjs}, \textit{246}(2), 63,
  \doi{10.3847/1538-4365/ab61f5}.

\bibitem[{\textit{{Odstrcil}}(2003)}]{odstrcil2003}
{Odstrcil}, D. (2003), {Modeling 3-D solar wind structure}, \textit{\adv},
  \textit{32}, 497--506, \doi{10.1016/S0273-1177(03)00332-6}.

\bibitem[{\textit{{Odstrcil} et~al.}(2004)\textit{{Odstrcil}, {Riley}, and
  {Zhao}}}]{odstrcil2004}
{Odstrcil}, D., {Riley}, P., and {Zhao}, X.~P. (2004), {Numerical simulation of
  the 12 May 1997 interplanetary CME event}, \textit{\jgr}, \textit{109}(A2),
  A02116, \doi{10.1029/2003JA010135}.

\bibitem[{\textit{{Ogilvie} and {Desch}}(1997)}]{ogilvie1997}
{Ogilvie}, K.~W., and {Desch}, M.~D. (1997), {The wind spacecraft and its early
  scientific results}, \textit{\adv}, \textit{20}, 559--568,
  \doi{10.1016/S0273-1177(97)00439-0}.

\bibitem[{\textit{{Ogilvie} et~al.}(1995)\textit{{Ogilvie}, {Chornay},
  {Fritzenreiter}, {Hunsaker}, {Keller}, {Lobell}, {Miller}, {Scudder},
  {Sittler}, {Torbert}, {Bodet}, {Needell}, {Lazarus}, {Steinberg}, {Tappan},
  {Mavretic}, and {Gergin}}}]{ogilvie1995}
{Ogilvie}, K.~W., {Chornay}, D.~J., {Fritzenreiter}, R.~J., {Hunsaker}, F.,
  {Keller}, J., {Lobell}, J., {Miller}, G., {Scudder}, J.~D., {Sittler}, E.~C.,
  Jr., {Torbert}, R.~B., {Bodet}, D., {Needell}, G., {Lazarus}, A.~J.,
  {Steinberg}, J.~T., {Tappan}, J.~H., {Mavretic}, A., and {Gergin}, E. (1995),
  {SWE, A Comprehensive Plasma Instrument for the Wind Spacecraft},
  \textit{\ssr}, \textit{71}, 55--77, \doi{10.1007/BF00751326}.

\bibitem[{\textit{{Owens}}(2008)}]{owens2008}
{Owens}, M.~J. (2008), {Combining remote and in situ observations of coronal
  mass ejections to better constrain magnetic cloud reconstruction},
  \textit{\jgr}, \textit{113}(A12), A12102, \doi{10.1029/2008JA013589}.

\bibitem[{\textit{{Owens}}(2018)}]{owens2018a}
{Owens}, M.~J. (2018), {Solar Wind and Heavy Ion Properties of Interplanetary
  Coronal Mass Ejections}, \textit{\solphys}, \textit{293}(8), 122,
  \doi{10.1007/s11207-018-1343-0}.

\bibitem[{\textit{{Owens} et~al.}(2017)\textit{{Owens}, {Lockwood}, and
  {Barnard}}}]{owens2017}
{Owens}, M.~J., {Lockwood}, M., and {Barnard}, L.~A. (2017), {Coronal mass
  ejections are not coherent magnetohydrodynamic structures}, \textit{\natsr},
  \textit{7}, 4152, \doi{10.1038/s41598-017-04546-3}.

\bibitem[{\textit{{Palmaerts} et~al.}(2018)\textit{{Palmaerts}, {Radioti},
  {Grodent}, {Yao}, {Bradley}, {Roussos}, {Lamy}, {Bunce}, {Cowley}, {Krupp},
  {Kurth}, {G{\'e}rard}, and {Pryor}}}]{palmaerts2018}
{Palmaerts}, B., {Radioti}, A., {Grodent}, D., {Yao}, Z.~H., {Bradley}, T.~J.,
  {Roussos}, E., {Lamy}, L., {Bunce}, E.~J., {Cowley}, S.~W.~H., {Krupp}, N.,
  {Kurth}, W.~S., {G{\'e}rard}, J.~C., and {Pryor}, W.~R. (2018), {Auroral
  Storm and Polar Arcs at Saturn{\textemdash}Final Cassini/UVIS Auroral
  Observations}, \textit{\grl}, \textit{45}(14), 6832--6842,
  \doi{10.1029/2018GL078094}.

\bibitem[{\textit{{Palmerio} et~al.}(2017)\textit{{Palmerio}, {Kilpua},
  {James}, {Green}, {Pomoell}, {Isavnin}, and {Valori}}}]{palmerio2017}
{Palmerio}, E., {Kilpua}, E.~K.~J., {James}, A.~W., {Green}, L.~M., {Pomoell},
  J., {Isavnin}, A., and {Valori}, G. (2017), {Determining the Intrinsic CME
  Flux Rope Type Using Remote-sensing Solar Disk Observations},
  \textit{\solphys}, \textit{292}(2), 39, \doi{10.1007/s11207-017-1063-x}.

\bibitem[{\textit{{Palmerio} et~al.}(2018)\textit{{Palmerio}, {Kilpua},
  {M{\"o}stl}, {Bothmer}, {James}, {Green}, {Isavnin}, {Davies}, and
  {Harrison}}}]{palmerio2018}
{Palmerio}, E., {Kilpua}, E.~K.~J., {M{\"o}stl}, C., {Bothmer}, V., {James},
  A.~W., {Green}, L.~M., {Isavnin}, A., {Davies}, J.~A., and {Harrison}, R.~A.
  (2018), {Coronal Magnetic Structure of Earthbound CMEs and In Situ
  Comparison}, \textit{\spwea}, \textit{16}(5), 442--460,
  \doi{10.1002/2017SW001767}.

\bibitem[{\textit{{Palmerio} et~al.}(2019)\textit{{Palmerio}, {Scolini},
  {Barnes}, {Magdaleni{\'c}}, {West}, {Zhukov}, {Rodriguez}, {Mierla}, {Good},
  {Morosan}, {Kilpua}, {Pomoell}, and {Poedts}}}]{palmerio2019}
{Palmerio}, E., {Scolini}, C., {Barnes}, D., {Magdaleni{\'c}}, J., {West},
  M.~J., {Zhukov}, A.~N., {Rodriguez}, L., {Mierla}, M., {Good}, S.~W.,
  {Morosan}, D.~E., {Kilpua}, E. K.~J., {Pomoell}, J., and {Poedts}, S. (2019),
  {Multipoint Study of Successive Coronal Mass Ejections Driving Moderate
  Disturbances at 1 au}, \textit{\apj}, \textit{878}(1), 37,
  \doi{10.3847/1538-4357/ab1850}.

\bibitem[{\textit{{Palmerio} et~al.}(2021)\textit{{Palmerio}, {Kilpua},
  {Witasse}, {Barnes}, {S{\'a}nchez-Cano}, {Weiss}, {Nieves-Chinchilla},
  {M{\"o}stl}, {Jian}, {Mierla}, {Zhukov}, {Guo}, {Rodriguez}, {Lowrance},
  {Isavnin}, {Turc}, {Futaana}, and {Holmstr{\"o}m}}}]{palmerio2021}
{Palmerio}, E., {Kilpua}, E. K.~J., {Witasse}, O., {Barnes}, D.,
  {S{\'a}nchez-Cano}, B., {Weiss}, A.~J., {Nieves-Chinchilla}, T., {M{\"o}stl},
  C., {Jian}, L.~K., {Mierla}, M., {Zhukov}, A.~N., {Guo}, J., {Rodriguez}, L.,
  {Lowrance}, P.~J., {Isavnin}, A., {Turc}, L., {Futaana}, Y., and
  {Holmstr{\"o}m}, M. (2021), {CME Magnetic Structure and IMF Preconditioning
  Affecting SEP Transport}, \textit{\spwea}, \textit{19}(4), e2020SW002654,
  \doi{10.1029/2020SW002654}.

\bibitem[{\textit{{Panasenco} and {Martin}}(2008)}]{panasenco2008}
{Panasenco}, O., and {Martin}, S.~F. (2008), {Topological Analyses of Symmetric
  Eruptive Prominences}, in \textit{Subsurface and Atmospheric Influences on
  Solar Activity}, \textit{Astronomical Society of the Pacific Conference
  Series}, vol. 383, edited by R.~{Howe}, R.~W. {Komm}, K.~S.
  {Balasubramaniam}, and G.~J.~D. {Petrie}, p. 243.

\bibitem[{\textit{{Papaioannou} et~al.}(2020)\textit{{Papaioannou}, {Belov},
  {Abunina}, {Eroshenko}, {Abunin}, {Anastasiadis}, {Patsourakos}, and
  {Mavromichalaki}}}]{papaioannou2020}
{Papaioannou}, A., {Belov}, A., {Abunina}, M., {Eroshenko}, E., {Abunin}, A.,
  {Anastasiadis}, A., {Patsourakos}, S., and {Mavromichalaki}, H. (2020),
  {Interplanetary Coronal Mass Ejections as the Driver of Non-recurrent Forbush
  Decreases}, \textit{\apj}, \textit{890}(2), 101,
  \doi{10.3847/1538-4357/ab6bd1}.

\bibitem[{\textit{{Patsourakos}
  et~al.}(2010{\natexlab{a}})\textit{{Patsourakos}, {Vourlidas}, and
  {Kliem}}}]{patsourakos2010a}
{Patsourakos}, S., {Vourlidas}, A., and {Kliem}, B. (2010{\natexlab{a}}),
  {Toward understanding the early stages of an impulsively accelerated coronal
  mass ejection. SECCHI observations}, \textit{\aap}, \textit{522}, A100,
  \doi{10.1051/0004-6361/200913599}.

\bibitem[{\textit{{Patsourakos}
  et~al.}(2010{\natexlab{b}})\textit{{Patsourakos}, {Vourlidas}, and
  {Stenborg}}}]{patsourakos2010b}
{Patsourakos}, S., {Vourlidas}, A., and {Stenborg}, G. (2010{\natexlab{b}}),
  {The Genesis of an Impulsive Coronal Mass Ejection Observed at Ultra-high
  Cadence by AIA on SDO}, \textit{\apjl}, \textit{724}(2), L188--L193,
  \doi{10.1088/2041-8205/724/2/L188}.

\bibitem[{\textit{{Pesnell} et~al.}(2012)\textit{{Pesnell}, {Thompson}, and
  {Chamberlin}}}]{pesnell2012}
{Pesnell}, W.~D., {Thompson}, B.~J., and {Chamberlin}, P.~C. (2012), {The Solar
  Dynamics Observatory (SDO)}, \textit{\solphys}, \textit{275}, 3--15,
  \doi{10.1007/s11207-011-9841-3}.

\bibitem[{\textit{{Prang{\'e}} et~al.}(2004)\textit{{Prang{\'e}}, {Pallier},
  {Hansen}, {Howard}, {Vourlidas}, {Courtin}, and {Parkinson}}}]{prange2004}
{Prang{\'e}}, R., {Pallier}, L., {Hansen}, K.~C., {Howard}, R., {Vourlidas},
  A., {Courtin}, R., and {Parkinson}, C. (2004), {An interplanetary shock
  traced by planetary auroral storms from the Sun to Saturn}, \textit{\nat},
  \textit{432}(7013), 78--81, \doi{10.1038/nature02986}.

\bibitem[{\textit{{Prise} et~al.}(2015)\textit{{Prise}, {Harra}, {Matthews},
  {Arridge}, and {Achilleos}}}]{prise2015}
{Prise}, A.~J., {Harra}, L.~K., {Matthews}, S.~A., {Arridge}, C.~S., and
  {Achilleos}, N. (2015), {Analysis of a coronal mass ejection and corotating
  interaction region as they travel from the Sun passing Venus, Earth, Mars,
  and Saturn}, \textit{\jgra}, \textit{120}, 1566--1588,
  \doi{10.1002/2014JA020256}.

\bibitem[{\textit{{Provan} et~al.}(2015)\textit{{Provan}, {Tao}, {Cowley},
  {Dougherty}, and {Coates}}}]{provan2015}
{Provan}, G., {Tao}, C., {Cowley}, S.~W.~H., {Dougherty}, M.~K., and {Coates},
  A.~J. (2015), {Planetary period oscillations in Saturn's magnetosphere:
  Examining the relationship between abrupt changes in behavior and solar
  wind-induced magnetospheric compressions and expansions}, \textit{\jgra},
  \textit{120}(11), 9524--9544, \doi{10.1002/2015JA021642}.

\bibitem[{\textit{{Reed} et~al.}(2018)\textit{{Reed}, {Jackman}, {Lamy},
  {Kurth}, and {Whiter}}}]{reed2018}
{Reed}, J.~J., {Jackman}, C.~M., {Lamy}, L., {Kurth}, W.~S., and {Whiter},
  D.~K. (2018), {Low-Frequency Extensions of the Saturn Kilometric Radiation as
  a Proxy for Magnetospheric Dynamics}, \textit{\jgra}, \textit{123}(1),
  443--463, \doi{10.1002/2017JA024499}.

\bibitem[{\textit{{Regnault} et~al.}(2020)\textit{{Regnault}, {Janvier},
  {D{\'e}moulin}, {Auch{\`e}re}, {Strugarek}, {Dasso}, and
  {No{\^u}s}}}]{regnault2020}
{Regnault}, F., {Janvier}, M., {D{\'e}moulin}, P., {Auch{\`e}re}, F.,
  {Strugarek}, A., {Dasso}, S., and {No{\^u}s}, C. (2020), {20 Years of ACE
  Data: How Superposed Epoch Analyses Reveal Generic Features in Interplanetary
  CME Profiles}, \textit{\jgra}, \textit{125}(11), e28150,
  \doi{10.1029/2020JA028150}.

\bibitem[{\textit{{Richardson}}(2014)}]{richardson2014b}
{Richardson}, I.~G. (2014), {Identification of Interplanetary Coronal Mass
  Ejections at Ulysses Using Multiple Solar Wind Signatures},
  \textit{\solphys}, \textit{289}(10), 3843--3894,
  \doi{10.1007/s11207-014-0540-8}.

\bibitem[{\textit{{Richardson}}(2018)}]{richardson2018}
{Richardson}, I.~G. (2018), {Solar wind stream interaction regions throughout
  the heliosphere}, \textit{\lrsp}, \textit{15}, 1,
  \doi{10.1007/s41116-017-0011-z}.

\bibitem[{\textit{{Richardson} and {Cane}}(1995)}]{richardson1995}
{Richardson}, I.~G., and {Cane}, H.~V. (1995), {Regions of abnormally low
  proton temperature in the solar wind (1965-1991) and their association with
  ejecta}, \textit{\jgr}, \textit{100}(A12), 23,397--23,412,
  \doi{10.1029/95JA02684}.

\bibitem[{\textit{{Richardson} and {Cane}}(2010)}]{richardson2010}
{Richardson}, I.~G., and {Cane}, H.~V. (2010), {Near-Earth Interplanetary
  Coronal Mass Ejections During Solar Cycle 23 (1996 - 2009): Catalog and
  Summary of Properties}, \textit{\solphys}, \textit{264}, 189--237,
  \doi{10.1007/s11207-010-9568-6}.

\bibitem[{\textit{{Richardson} and {Cane}}(2011)}]{richardson2011}
{Richardson}, I.~G., and {Cane}, H.~V. (2011), {Galactic Cosmic Ray Intensity
  Response to Interplanetary Coronal Mass Ejections/Magnetic Clouds in 1995 -
  2009}, \textit{\solphys}, \textit{270}(2), 609--627,
  \doi{10.1007/s11207-011-9774-x}.

\bibitem[{\textit{{Richardson} et~al.}(2002)\textit{{Richardson}, {Paularena},
  {Wang}, and {Burlaga}}}]{richardson2002}
{Richardson}, J.~D., {Paularena}, K.~I., {Wang}, C., and {Burlaga}, L.~F.
  (2002), {The life of a CME and the development of a MIR: From the Sun to 58
  AU}, \textit{Journal of Geophysical Research (Space Physics)}, \textit{107},
  1041, \doi{10.1029/2001JA000175}.

\bibitem[{\textit{{Richardson} et~al.}(2006)\textit{{Richardson}, {Liu},
  {Wang}, and {Burlaga}}}]{richardson2006}
{Richardson}, J.~D., {Liu}, Y., {Wang}, C., and {Burlaga}, L.~F. (2006), {ICMES
  at very large distances}, \textit{\adv}, \textit{38}(3), 528--534,
  \doi{10.1016/j.asr.2005.06.049}.

\bibitem[{\textit{{Riley} et~al.}(2000)\textit{{Riley}, {Gosling}, {McComas},
  and {Forsyth}}}]{riley2000}
{Riley}, P., {Gosling}, J.~T., {McComas}, D.~J., and {Forsyth}, R.~J. (2000),
  {Properties and radial trends of coronal mass ejecta and their associated
  shocks observed by Ulysses in the ecliptic plane}, \textit{\jgr},
  \textit{105}(A6), 12,617--12,626, \doi{10.1029/1999JA000169}.

\bibitem[{\textit{{Rodriguez} et~al.}(2008)\textit{{Rodriguez}, {Zhukov},
  {Dasso}, {Mand rini}, {Cremades}, {Cid}, {Cerrato}, {Saiz}, {Aran},
  {Menvielle}, {Poedts}, and {Schmieder}}}]{rodriguez2008}
{Rodriguez}, L., {Zhukov}, A.~N., {Dasso}, S., {Mand rini}, C.~H., {Cremades},
  H., {Cid}, C., {Cerrato}, Y., {Saiz}, E., {Aran}, A., {Menvielle}, M.,
  {Poedts}, S., and {Schmieder}, B. (2008), {Magnetic clouds seen at different
  locations in the heliosphere}, \textit{\angeo}, \textit{26}(2), 213--229,
  \doi{10.5194/angeo-26-213-2008}.

\bibitem[{\textit{{Rodriguez} et~al.}(2016)\textit{{Rodriguez},
  {Mas{\'{\i}}as-Meza}, {Dasso}, {D{\'e}moulin}, {Zhukov}, {Gulisano},
  {Mierla}, {Kilpua}, {West}, {Lacatus}, {Paraschiv}, and
  {Janvier}}}]{rodriguez2016}
{Rodriguez}, L., {Mas{\'{\i}}as-Meza}, J.~J., {Dasso}, S., {D{\'e}moulin}, P.,
  {Zhukov}, A.~N., {Gulisano}, A.~M., {Mierla}, M., {Kilpua}, E., {West}, M.,
  {Lacatus}, D., {Paraschiv}, A., and {Janvier}, M. (2016), {Typical Profiles
  and Distributions of Plasma and Magnetic Field Parameters in Magnetic Clouds
  at 1 AU}, \textit{\solphys}, \textit{291}, 2145--2163,
  \doi{10.1007/s11207-016-0955-5}.

\bibitem[{\textit{{Rouillard}}(2011)}]{rouillard2011}
{Rouillard}, A.~P. (2011), {Relating white light and in situ observations of
  coronal mass ejections: A review}, \textit{\jastp}, \textit{73}(10),
  1201--1213, \doi{10.1016/j.jastp.2010.08.015}.

\bibitem[{\textit{{Rouillard} et~al.}(2009)\textit{{Rouillard}, {Davies},
  {Forsyth}, {Savani}, {Sheeley}, {Thernisien}, {Zhang}, {Howard}, {Anderson},
  and {Carr}}}]{rouillard2009}
{Rouillard}, A.~P., {Davies}, J.~A., {Forsyth}, R.~J., {Savani}, N.~P.,
  {Sheeley}, N.~R., {Thernisien}, A., {Zhang}, T.~L., {Howard}, R.~A.,
  {Anderson}, B., and {Carr}, C.~M. (2009), {A solar storm observed from the
  Sun to Venus using the STEREO, Venus Express, and MESSENGER spacecraft},
  \textit{\jgr}, \textit{114}(A7), A07106, \doi{10.1029/2008JA014034}.

\bibitem[{\textit{{Rouillard} et~al.}(2010)\textit{{Rouillard}, {Lavraud},
  {Sheeley}, {Davies}, {Burlaga}, {Savani}, {Jacquey}, and
  {Forsyth}}}]{rouillard2010}
{Rouillard}, A.~P., {Lavraud}, B., {Sheeley}, N.~R., {Davies}, J.~A.,
  {Burlaga}, L.~F., {Savani}, N.~P., {Jacquey}, C., and {Forsyth}, R.~J.
  (2010), {White Light and In Situ Comparison of a Forming Merged Interaction
  Region}, \textit{\apj}, \textit{719}(2), 1385--1392,
  \doi{10.1088/0004-637X/719/2/1385}.

\bibitem[{\textit{{Roussos} et~al.}(2008)\textit{{Roussos}, {Krupp},
  {Armstrong}, {Paranicas}, {Mitchell}, {Krimigis}, {Jones}, {Dialynas},
  {Sergis}, and {Hamilton}}}]{roussos2008}
{Roussos}, E., {Krupp}, N., {Armstrong}, T.~P., {Paranicas}, C., {Mitchell},
  D.~G., {Krimigis}, S.~M., {Jones}, G.~H., {Dialynas}, K., {Sergis}, N., and
  {Hamilton}, D.~C. (2008), {Discovery of a transient radiation belt at
  Saturn}, \textit{\grl}, \textit{35}(22), L22106, \doi{10.1029/2008GL035767}.

\bibitem[{\textit{{Roussos} et~al.}(2011)\textit{{Roussos}, {Krupp},
  {Paranicas}, {Kollmann}, {Mitchell}, {Krimigis}, {Armstrong}, {Went},
  {Dougherty}, and {Jones}}}]{roussos2011}
{Roussos}, E., {Krupp}, N., {Paranicas}, C.~P., {Kollmann}, P., {Mitchell},
  D.~G., {Krimigis}, S.~M., {Armstrong}, T.~P., {Went}, D.~R., {Dougherty},
  M.~K., and {Jones}, G.~H. (2011), {Long- and short-term variability of
  Saturn's ionic radiation belts}, \textit{\jgr}, \textit{116}(A2), A02217,
  \doi{10.1029/2010JA015954}.

\bibitem[{\textit{{Roussos} et~al.}(2018)\textit{{Roussos}, {Jackman},
  {Thomsen}, {Kurth}, {Badman}, {Paranicas}, {Kollmann}, {Krupp},
  {Bu{\v{c}}{\'\i}k}, {Mitchell}, {Krimigis}, {Hamilton}, and
  {Radioti}}}]{roussos2018}
{Roussos}, E., {Jackman}, C.~M., {Thomsen}, M.~F., {Kurth}, W.~S., {Badman},
  S.~V., {Paranicas}, C., {Kollmann}, P., {Krupp}, N., {Bu{\v{c}}{\'\i}k}, R.,
  {Mitchell}, D.~G., {Krimigis}, S.~M., {Hamilton}, D.~C., and {Radioti}, A.
  (2018), {Solar Energetic Particles (SEP) and Galactic Cosmic Rays (GCR) as
  tracers of solar wind conditions near Saturn: Event lists and applications},
  \textit{\icarus}, \textit{300}, 47--71, \doi{10.1016/j.icarus.2017.08.040}.

\bibitem[{\textit{{Roussos} et~al.}(2019)\textit{{Roussos}, {Krupp},
  {Dialynas}, {Kollmann}, {Paranicas}, {Echer}, {Mitchell}, and
  {Krimigis}}}]{roussos2019}
{Roussos}, E., {Krupp}, N., {Dialynas}, K., {Kollmann}, P., {Paranicas}, C.,
  {Echer}, E., {Mitchell}, D.~G., and {Krimigis}, S.~M. (2019), {Jovian
  Cosmic-Ray Protons in the Heliosphere: Constraints by Cassini Observations},
  \textit{\apj}, \textit{871}(2), 223, \doi{10.3847/1538-4357/aafb2f}.

\bibitem[{\textit{{Roussos} et~al.}(2020)\textit{{Roussos}, {Dialynas},
  {Krupp}, {Kollmann}, {Paranicas}, {Roelof}, {Yuan}, {Mitchell}, and
  {Krimigis}}}]{roussos2020}
{Roussos}, E., {Dialynas}, K., {Krupp}, N., {Kollmann}, P., {Paranicas}, C.,
  {Roelof}, E.~C., {Yuan}, C., {Mitchell}, D.~G., and {Krimigis}, S.~M. (2020),
  {Long- and Short-term Variability of Galactic Cosmic-Ray Radial Intensity
  Gradients between 1 and 9.5 au: Observations by Cassini, BESS, BESS-Polar,
  PAMELA, and AMS-02}, \textit{\apj}, \textit{904}(2), 165,
  \doi{10.3847/1538-4357/abc346}.

\bibitem[{\textit{{Salman} et~al.}(2020)\textit{{Salman}, {Winslow}, and
  {Lugaz}}}]{salman2020}
{Salman}, T.~M., {Winslow}, R.~M., and {Lugaz}, N. (2020), {Radial Evolution of
  Coronal Mass Ejections Between MESSENGER, Venus Express, STEREO, and L1:
  Catalog and Analysis}, \textit{\jgra}, \textit{125}(1), e2019JA027084,
  \doi{10.1029/2019JA027084}.

\bibitem[{\textit{{Savani} et~al.}(2010)\textit{{Savani}, {Owens}, {Rouillard},
  {Forsyth}, and {Davies}}}]{savani2010}
{Savani}, N.~P., {Owens}, M.~J., {Rouillard}, A.~P., {Forsyth}, R.~J., and
  {Davies}, J.~A. (2010), {Observational Evidence of a Coronal Mass Ejection
  Distortion Directly Attributable to a Structured Solar Wind}, \textit{\apjl},
  \textit{714}, L128--L132, \doi{10.1088/2041-8205/714/1/L128}.

\bibitem[{\textit{{Scherrer} et~al.}(2012)\textit{{Scherrer}, {Schou}, {Bush},
  {Kosovichev}, {Bogart}, {Hoeksema}, {Liu}, {Duvall}, {Zhao}, {Title},
  {Schrijver}, {Tarbell}, and {Tomczyk}}}]{scherrer2012}
{Scherrer}, P.~H., {Schou}, J., {Bush}, R.~I., {Kosovichev}, A.~G., {Bogart},
  R.~S., {Hoeksema}, J.~T., {Liu}, Y., {Duvall}, T.~L., {Zhao}, J., {Title},
  A.~M., {Schrijver}, C.~J., {Tarbell}, T.~D., and {Tomczyk}, S. (2012), {The
  Helioseismic and Magnetic Imager (HMI) Investigation for the Solar Dynamics
  Observatory (SDO)}, \textit{\solphys}, \textit{275}, 207--227,
  \doi{10.1007/s11207-011-9834-2}.

\bibitem[{\textit{{Schmidt} and {Cargill}}(2004)}]{schmidt2004}
{Schmidt}, J., and {Cargill}, P. (2004), {A numerical study of two interacting
  coronal mass ejections}, \textit{\angeo}, \textit{22}(6), 2245--2254,
  \doi{10.5194/angeo-22-2245-2004}.

\bibitem[{\textit{{Schrijver} and {Title}}(2011)}]{schrijver2011}
{Schrijver}, C.~J., and {Title}, A.~M. (2011), {Long-range magnetic couplings
  between solar flares and coronal mass ejections observed by SDO and STEREO},
  \textit{\jgr}, \textit{116}(A4), A04108, \doi{10.1029/2010JA016224}.

\bibitem[{\textit{{Schwenn} et~al.}(2005)\textit{{Schwenn}, {dal Lago},
  {Huttunen}, and {Gonzalez}}}]{schwenn2005}
{Schwenn}, R., {dal Lago}, A., {Huttunen}, E., and {Gonzalez}, W.~D. (2005),
  {The association of coronal mass ejections with their effects near the
  Earth}, \textit{\angeo}, \textit{23}, 1033--1059,
  \doi{10.5194/angeo-23-1033-2005}.

\bibitem[{\textit{{Scolini} et~al.}(2020)\textit{{Scolini}, {Chan{\'e}},
  {Temmer}, {Kilpua}, {Dissauer}, {Veronig}, {Palmerio}, {Pomoell},
  {Dumbovi{\'c}}, {Guo}, {Rodriguez}, and {Poedts}}}]{scolini2020}
{Scolini}, C., {Chan{\'e}}, E., {Temmer}, M., {Kilpua}, E. K.~J., {Dissauer},
  K., {Veronig}, A.~M., {Palmerio}, E., {Pomoell}, J., {Dumbovi{\'c}}, M.,
  {Guo}, J., {Rodriguez}, L., and {Poedts}, S. (2020), {CME--CME Interactions
  as Sources of CME Geoeffectiveness: The Formation of the Complex Ejecta and
  Intense Geomagnetic Storm in 2017 Early September}, \textit{\apjs},
  \textit{247}(1), 21, \doi{10.3847/1538-4365/ab6216}.

\bibitem[{\textit{{Srivastava} et~al.}(2018)\textit{{Srivastava}, {Mishra}, and
  {Chakrabarty}}}]{srivastava2018}
{Srivastava}, N., {Mishra}, W., and {Chakrabarty}, D. (2018), {Interplanetary
  and Geomagnetic Consequences of Interacting CMEs of 13 - 14 June 2012},
  \textit{\solphys}, \textit{293}, 5, \doi{10.1007/s11207-017-1227-8}.

\bibitem[{\textit{{Subramanian} et~al.}(2014)\textit{{Subramanian}, {Arunbabu},
  {Vourlidas}, and {Mauriya}}}]{subramanian2014}
{Subramanian}, P., {Arunbabu}, K.~P., {Vourlidas}, A., and {Mauriya}, A.
  (2014), {Self-similar Expansion of Solar Coronal Mass Ejections: Implications
  for Lorentz Self-force Driving}, \textit{\apj}, \textit{790}(2), 125,
  \doi{10.1088/0004-637X/790/2/125}.

\bibitem[{\textit{{SunPy Community} et~al.}(2015)\textit{{SunPy Community},
  {Mumford}, {Christe}, {P{\'e}rez-Su{\'a}rez}, {Ireland}, {Shih}, {Inglis},
  {Liedtke}, {Hewett}, {Mayer}, {Hughitt}, {Freij}, {Meszaros}, {Bennett},
  {Malocha}, {Evans}, {Agrawal}, {Leonard}, {Robitaille}, {Mampaey},
  {Campos-Rozo}, and {Kirk}}}]{sunpy2015}
{SunPy Community}, {Mumford}, S.~J., {Christe}, S., {P{\'e}rez-Su{\'a}rez}, D.,
  {Ireland}, J., {Shih}, A.~Y., {Inglis}, A.~R., {Liedtke}, S., {Hewett},
  R.~J., {Mayer}, F., {Hughitt}, K., {Freij}, N., {Meszaros}, T., {Bennett},
  S.~M., {Malocha}, M., {Evans}, J., {Agrawal}, A., {Leonard}, A.~J.,
  {Robitaille}, T.~P., {Mampaey}, B., {Campos-Rozo}, J.~I., and {Kirk}, M.~S.
  (2015), {SunPy{\textemdash}Python for solar physics}, \textit{\csd},
  \textit{8}(1), 014009, \doi{10.1088/1749-4699/8/1/014009}.

\bibitem[{\textit{{SunPy Community} et~al.}(2020)\textit{{SunPy Community},
  {Barnes}, {Bobra}, {Christe}, {Freij}, {Hayes}, {Ireland }, {Mumford},
  {Perez-Suarez}, {Ryan}, {Shih}, {Chanda}, {Glogowski}, {Hewett}, {Hughitt},
  {Hill}, {Hiware}, {Inglis}, {Kirk}, {Konge}, {Mason}, {Maloney}, {Murray},
  {Panda}, {Park}, {Pereira}, {Reardon}, {Savage}, {Sip{\H{o}}cz}, {Stansby},
  {Jain}, {Taylor}, {Yadav}, {Rajul}, and {Dang}}}]{sunpy2020}
{SunPy Community}, {Barnes}, W.~T., {Bobra}, M.~G., {Christe}, S.~D., {Freij},
  N., {Hayes}, L.~A., {Ireland }, J., {Mumford}, S., {Perez-Suarez}, D.,
  {Ryan}, D.~F., {Shih}, A.~Y., {Chanda}, P., {Glogowski}, K., {Hewett}, R.,
  {Hughitt}, V.~K., {Hill}, A., {Hiware}, K., {Inglis}, A., {Kirk}, M. S.~F.,
  {Konge}, S., {Mason}, J.~P., {Maloney}, S.~A., {Murray}, S.~A., {Panda}, A.,
  {Park}, J., {Pereira}, T. M.~D., {Reardon}, K., {Savage}, S., {Sip{\H{o}}cz},
  B.~M., {Stansby}, D., {Jain}, Y., {Taylor}, G., {Yadav}, T., {Rajul}, and
  {Dang}, T.~K. (2020), {The SunPy Project: Open Source Development and Status
  of the Version 1.0 Core Package}, \textit{\apj}, \textit{890}(1), 68,
  \doi{10.3847/1538-4357/ab4f7a}.

\bibitem[{\textit{{Svedhem} et~al.}(2007)\textit{{Svedhem}, {Titov}, {McCoy},
  {Lebreton}, {Barabash}, {Bertaux}, {Drossart}, {Formisano}, {H{\"a}usler},
  {Korablev}, {Markiewicz}, {Nevejans}, {P{\"a}tzold}, {Piccioni}, {Zhang},
  {Taylor}, {Lellouch}, {Koschny}, {Witasse}, {Eggel}, {Warhaut}, {Accomazzo},
  {Rodriguez- Canabal}, {Fabrega}, {Schirmann}, {Clochet}, and
  {Coradini}}}]{svedhem2007}
{Svedhem}, H., {Titov}, D.~V., {McCoy}, D., {Lebreton}, J.~P., {Barabash}, S.,
  {Bertaux}, J.~L., {Drossart}, P., {Formisano}, V., {H{\"a}usler}, B.,
  {Korablev}, O., {Markiewicz}, W.~J., {Nevejans}, D., {P{\"a}tzold}, M.,
  {Piccioni}, G., {Zhang}, T.~L., {Taylor}, F.~W., {Lellouch}, E., {Koschny},
  D., {Witasse}, O., {Eggel}, H., {Warhaut}, M., {Accomazzo}, A., {Rodriguez-
  Canabal}, J., {Fabrega}, J., {Schirmann}, T., {Clochet}, A., and {Coradini},
  M. (2007), {Venus Express{\textemdash}The first European mission to Venus},
  \textit{\planss}, \textit{55}, 1636--1652, \doi{10.1016/j.pss.2007.01.013}.

\bibitem[{\textit{{Tao} et~al.}(2005)\textit{{Tao}, {Kataoka}, {Fukunishi},
  {Takahashi}, and {Yokoyama}}}]{tao2005}
{Tao}, C., {Kataoka}, R., {Fukunishi}, H., {Takahashi}, Y., and {Yokoyama}, T.
  (2005), {Magnetic field variations in the Jovian magnetotail induced by solar
  wind dynamic pressure enhancements}, \textit{\jgr}, \textit{110}(A11),
  A11208, \doi{10.1029/2004JA010959}.

\bibitem[{\textit{{Temmer} and {Nitta}}(2015)}]{temmer2015}
{Temmer}, M., and {Nitta}, N.~V. (2015), {Interplanetary Propagation Behavior
  of the Fast Coronal Mass Ejection on 23 July 2012}, \textit{\solphys},
  \textit{290}(3), 919--932, \doi{10.1007/s11207-014-0642-3}.

\bibitem[{\textit{{Temmer} et~al.}(2008)\textit{{Temmer}, {Veronig},
  {Vr{\v{s}}nak}, {Ryb{\'a}k}, {G{\"o}m{\"o}ry}, {Stoiser}, and
  {Mari{\v{c}}i{\'c}}}}]{temmer2008}
{Temmer}, M., {Veronig}, A.~M., {Vr{\v{s}}nak}, B., {Ryb{\'a}k}, J.,
  {G{\"o}m{\"o}ry}, P., {Stoiser}, S., and {Mari{\v{c}}i{\'c}}, D. (2008),
  {Acceleration in Fast Halo CMEs and Synchronized Flare HXR Bursts},
  \textit{\apjl}, \textit{673}(1), L95, \doi{10.1086/527414}.

\bibitem[{\textit{{Temmer} et~al.}(2010)\textit{{Temmer}, {Veronig}, {Kontar},
  {Krucker}, and {Vr{\v{s}}nak}}}]{temmer2010}
{Temmer}, M., {Veronig}, A.~M., {Kontar}, E.~P., {Krucker}, S., and
  {Vr{\v{s}}nak}, B. (2010), {Combined STEREO/RHESSI Study of Coronal Mass
  Ejection Acceleration and Particle Acceleration in Solar Flares},
  \textit{\apj}, \textit{712}(2), 1410--1420,
  \doi{10.1088/0004-637X/712/2/1410}.

\bibitem[{\textit{{Temmer} et~al.}(2017)\textit{{Temmer}, {Reiss}, {Nikolic},
  {Hofmeister}, and {Veronig}}}]{temmer2017}
{Temmer}, M., {Reiss}, M.~A., {Nikolic}, L., {Hofmeister}, S.~J., and
  {Veronig}, A.~M. (2017), {Preconditioning of Interplanetary Space Due to
  Transient CME Disturbances}, \textit{\apj}, \textit{835}(2), 141,
  \doi{10.3847/1538-4357/835/2/141}.

\bibitem[{\textit{{Thernisien}}(2011)}]{thernisien2011}
{Thernisien}, A. (2011), {Implementation of the Graduated Cylindrical Shell
  Model for the Three-dimensional Reconstruction of Coronal Mass Ejections},
  \textit{\apjs}, \textit{194}, 33, \doi{10.1088/0067-0049/194/2/33}.

\bibitem[{\textit{{Thernisien} et~al.}(2009)\textit{{Thernisien}, {Vourlidas},
  and {Howard}}}]{thernisien2009}
{Thernisien}, A., {Vourlidas}, A., and {Howard}, R.~A. (2009), {Forward
  Modeling of Coronal Mass Ejections Using STEREO/SECCHI Data},
  \textit{\solphys}, \textit{256}, 111--130, \doi{10.1007/s11207-009-9346-5}.

\bibitem[{\textit{{Thernisien} et~al.}(2006)\textit{{Thernisien}, {Howard}, and
  {Vourlidas}}}]{thernisien2006}
{Thernisien}, A.~F.~R., {Howard}, R.~A., and {Vourlidas}, A. (2006), {Modeling
  of Flux Rope Coronal Mass Ejections}, \textit{\apj}, \textit{652}, 763--773,
  \doi{10.1086/508254}.

\bibitem[{\textit{{Thompson} et~al.}(2000)\textit{{Thompson}, {Cliver},
  {Nitta}, {Delann{\'e}e}, and {Delaboudini{\`e}re}}}]{thompson2000}
{Thompson}, B.~J., {Cliver}, E.~W., {Nitta}, N., {Delann{\'e}e}, C., and
  {Delaboudini{\`e}re}, J.-P. (2000), {Coronal dimmings and energetic CMEs in
  April-May 1998}, \textit{\grl}, \textit{27}, 1431--1434,
  \doi{10.1029/1999GL003668}.

\bibitem[{\textit{{Thompson}}(2006)}]{thompson2006}
{Thompson}, W.~T. (2006), {Coordinate systems for solar image data},
  \textit{\aap}, \textit{449}(2), 791--803, \doi{10.1051/0004-6361:20054262}.

\bibitem[{\textit{{T{\"o}r{\"o}k} et~al.}(2011)\textit{{T{\"o}r{\"o}k},
  {Panasenco}, {Titov}, {Miki{\'c}}, {Reeves}, {Velli}, {Linker}, and {De
  Toma}}}]{torok2011}
{T{\"o}r{\"o}k}, T., {Panasenco}, O., {Titov}, V.~S., {Miki{\'c}}, Z.,
  {Reeves}, K.~K., {Velli}, M., {Linker}, J.~A., and {De Toma}, G. (2011), {A
  Model for Magnetically Coupled Sympathetic Eruptions}, \textit{\apjl},
  \textit{739}(2), L63, \doi{10.1088/2041-8205/739/2/L63}.

\bibitem[{\textit{{Tripathi} et~al.}(2006)\textit{{Tripathi}, {Isobe}, and
  {Mason}}}]{tripathi2006}
{Tripathi}, D., {Isobe}, H., and {Mason}, H.~E. (2006), {On the propagation of
  brightening after filament/prominence eruptions, as seen by SoHO-EIT},
  \textit{\aap}, \textit{453}(3), 1111--1116, \doi{10.1051/0004-6361:20064993}.

\bibitem[{\textit{{Veronig} et~al.}(2018)\textit{{Veronig}, {Podladchikova},
  {Dissauer}, {Temmer}, {Seaton}, {Long}, {Guo}, {Vr{\v{s}}nak}, {Harra}, and
  {Kliem}}}]{veronig2018}
{Veronig}, A.~M., {Podladchikova}, T., {Dissauer}, K., {Temmer}, M., {Seaton},
  D.~B., {Long}, D., {Guo}, J., {Vr{\v{s}}nak}, B., {Harra}, L., and {Kliem},
  B. (2018), {Genesis and Impulsive Evolution of the 2017 September 10 Coronal
  Mass Ejection}, \textit{\apj}, \textit{868}(2), 107,
  \doi{10.3847/1538-4357/aaeac5}.

\bibitem[{\textit{{von Steiger} and {Richardson}}(2006)}]{vonsteiger2006}
{von Steiger}, R., and {Richardson}, J.~D. (2006), {ICMEs in the Outer
  Heliosphere and at High Latitudes: An Introduction}, \textit{\ssr},
  \textit{123}(1-3), 111--126, \doi{10.1007/s11214-006-9015-z}.

\bibitem[{\textit{{Vourlidas} et~al.}(2011)\textit{{Vourlidas}, {Colaninno},
  {Nieves-Chinchilla}, and {Stenborg}}}]{vourlidas2011}
{Vourlidas}, A., {Colaninno}, R., {Nieves-Chinchilla}, T., and {Stenborg}, G.
  (2011), {The First Observation of a Rapidly Rotating Coronal Mass Ejection in
  the Middle Corona}, \textit{\apjl}, \textit{733}, L23,
  \doi{10.1088/2041-8205/733/2/L23}.

\bibitem[{\textit{{Vr{\v{s}}nak} and {{\v{Z}}ic}}(2007)}]{vrsnak2007}
{Vr{\v{s}}nak}, B., and {{\v{Z}}ic}, T. (2007), {Transit times of
  interplanetary coronal mass ejections and the solar wind speed},
  \textit{\aap}, \textit{472}(3), 937--943, \doi{10.1051/0004-6361:20077499}.

\bibitem[{\textit{{Vr{\v{s}}nak} et~al.}(2019)\textit{{Vr{\v{s}}nak},
  {Amerstorfer}, {Dumbovi{\'c}}, {Leitner}, {Veronig}, {Temmer}, {M{\"o}stl},
  {Amerstorfer}, {Farrugia}, and {Galvin}}}]{vrsnak2019}
{Vr{\v{s}}nak}, B., {Amerstorfer}, T., {Dumbovi{\'c}}, M., {Leitner}, M.,
  {Veronig}, A.~M., {Temmer}, M., {M{\"o}stl}, C., {Amerstorfer}, U.~V.,
  {Farrugia}, C.~J., and {Galvin}, A.~B. (2019), {Heliospheric Evolution of
  Magnetic Clouds}, \textit{\apj}, \textit{877}(2), 77,
  \doi{10.3847/1538-4357/ab190a}.

\bibitem[{\textit{{Wang} and {Richardson}}(2001)}]{wang2001}
{Wang}, C., and {Richardson}, J.~D. (2001), {Voyager 2 observations of helium
  abundance enhancements from 1-60 AU}, \textit{\jgr}, \textit{106}(A4),
  5683--5692, \doi{10.1029/2000JA000134}.

\bibitem[{\textit{{Wang} and {Richardson}}(2004)}]{wang2004a}
{Wang}, C., and {Richardson}, J.~D. (2004), {Interplanetary coronal mass
  ejections observed by Voyager 2 between 1 and 30 AU}, \textit{\jgr},
  \textit{109}(A6), A06104, \doi{10.1029/2004JA010379}.

\bibitem[{\textit{{Wang} et~al.}(2004)\textit{{Wang}, {Shen}, {Wang}, and
  {Ye}}}]{wang2004b}
{Wang}, Y., {Shen}, C., {Wang}, S., and {Ye}, P. (2004), {Deflection of coronal
  mass ejection in the interplanetary medium}, \textit{\solphys},
  \textit{222}(2), 329--343, \doi{10.1023/B:SOLA.0000043576.21942.aa}.

\bibitem[{\textit{{Wang} et~al.}(2003)\textit{{Wang}, {Ye}, and
  {Wang}}}]{wang2003}
{Wang}, Y.~M., {Ye}, P.~Z., and {Wang}, S. (2003), {Multiple magnetic clouds:
  Several examples during March-April 2001}, \textit{\jgr}, \textit{108}(A10),
  1370, \doi{10.1029/2003JA009850}.

\bibitem[{\textit{{Warwick} et~al.}(1981)\textit{{Warwick}, {Pearce}, {Evans},
  {Carr}, {Schauble}, {Alexander}, {Kaiser}, {Desch}, {Pedersen}, {Lecacheux},
  {Daigne}, {Boischot}, and {Barrow}}}]{warwick1981}
{Warwick}, J.~W., {Pearce}, J.~B., {Evans}, D.~R., {Carr}, T.~D., {Schauble},
  J.~J., {Alexander}, J.~K., {Kaiser}, M.~L., {Desch}, M.~D., {Pedersen}, M.,
  {Lecacheux}, A., {Daigne}, G., {Boischot}, A., and {Barrow}, C.~H. (1981),
  {Planetary Radio Astronomy Observations from Voyager 1 near Saturn},
  \textit{\sci}, \textit{212}(4491), 239--243,
  \doi{10.1126/science.212.4491.239}.

\bibitem[{\textit{{Webb} and {Howard}}(2012)}]{webb2012}
{Webb}, D.~F., and {Howard}, T.~A. (2012), {Coronal Mass Ejections:
  Observations}, \textit{\lrsp}, \textit{9}(1), 3, \doi{10.12942/lrsp-2012-3}.

\bibitem[{\textit{{Weiss} et~al.}(2021)\textit{{Weiss}, {Moestl}, {Davies},
  {Amerstorfer}, {Bauer}, {Hinterreiter}, {Reiss}, {Bailey}, {Horbury},
  {O'Brien}, {Evans}, {Angelini}, {Heiner}, {Richter}, {Auster}, {Magnes},
  {Fischer}, and {Baumjohann}}}]{weiss2021}
{Weiss}, A.~J., {Moestl}, C., {Davies}, E.~E., {Amerstorfer}, T., {Bauer}, M.,
  {Hinterreiter}, J., {Reiss}, M., {Bailey}, R.~L., {Horbury}, T.~S.,
  {O'Brien}, H., {Evans}, V., {Angelini}, V., {Heiner}, D., {Richter}, I.,
  {Auster}, H.-U., {Magnes}, W., {Fischer}, D., and {Baumjohann}, W. (2021),
  {Multi point analysis of coronal mass ejection flux ropes using combined data
  from Solar Orbiter, BepiColombo and Wind}, \textit{\aap}, \textit{in press},
  \doi{10.1051/0004-6361/202140919}.

\bibitem[{\textit{{Winslow} et~al.}(2015)\textit{{Winslow}, {Lugaz},
  {Philpott}, {Schwadron}, {Farrugia}, {Anderson}, and {Smith}}}]{winslow2015}
{Winslow}, R.~M., {Lugaz}, N., {Philpott}, L.~C., {Schwadron}, N.~A.,
  {Farrugia}, C.~J., {Anderson}, B.~J., and {Smith}, C.~W. (2015),
  {Interplanetary coronal mass ejections from MESSENGER orbital observations at
  Mercury}, \textit{\jgra}, \textit{120}(8), 6101--6118,
  \doi{10.1002/2015JA021200}.

\bibitem[{\textit{{Winslow} et~al.}(2016)\textit{{Winslow}, {Lugaz},
  {Schwadron}, {Farrugia}, {Yu}, {Raines}, {Mays}, {Galvin}, and
  {Zurbuchen}}}]{winslow2016}
{Winslow}, R.~M., {Lugaz}, N., {Schwadron}, N.~A., {Farrugia}, C.~J., {Yu}, W.,
  {Raines}, J.~M., {Mays}, M.~L., {Galvin}, A.~B., and {Zurbuchen}, T.~H.
  (2016), {Longitudinal conjunction between MESSENGER and STEREO A: Development
  of ICME complexity through stream interactions}, \textit{\jgra},
  \textit{121}(7), 6092--6106, \doi{10.1002/2015JA022307}.

\bibitem[{\textit{{Winslow} et~al.}(2018)\textit{{Winslow}, {Schwadron},
  {Lugaz}, {Guo}, {Joyce}, {Jordan}, {Wilson}, {Spence}, {Lawrence},
  {Wimmer-Schweingruber}, and {Mays}}}]{winslow2018}
{Winslow}, R.~M., {Schwadron}, N.~A., {Lugaz}, N., {Guo}, J., {Joyce}, C.~J.,
  {Jordan}, A.~P., {Wilson}, J.~K., {Spence}, H.~E., {Lawrence}, D.~J.,
  {Wimmer-Schweingruber}, R.~F., and {Mays}, M.~L. (2018), {Opening a Window on
  ICME-driven GCR Modulation in the Inner Solar System}, \textit{\apj},
  \textit{856}(2), 139, \doi{10.3847/1538-4357/aab098}.

\bibitem[{\textit{{Winslow} et~al.}(2021)\textit{{Winslow}, {Scolini}, {Lugaz},
  and {Galvin}}}]{winslow2021}
{Winslow}, R.~M., {Scolini}, C., {Lugaz}, N., and {Galvin}, A.~B. (2021), {The
  Effect of Stream Interaction Regions on ICME Structures Observed in
  Longitudinal Conjunction}, \textit{\apj}, \textit{916}(1), 40,
  \doi{10.3847/1538-4357/ac0439}.

\bibitem[{\textit{{Witasse} et~al.}(2017)\textit{{Witasse}, {S{\'a}nchez-Cano},
  {Mays}, {Kajdi{\v c}}, {Opgenoorth}, {Elliott}, {Richardson}, {Zouganelis},
  {Zender}, {Wimmer-Schweingruber}, {Turc}, {Taylor}, {Roussos}, {Rouillard},
  {Richter}, {Richardson}, {Ramstad}, {Provan}, {Posner}, {Plaut}, {Odstrcil},
  {Nilsson}, {Niemenen}, {Milan}, {Mandt}, {Lohf}, {Lester}, {Lebreton},
  {Kuulkers}, {Krupp}, {Koenders}, {James}, {Intzekara}, {Holmstrom},
  {Hassler}, {Hall}, {Guo}, {Goldstein}, {Goetz}, {Glassmeier}, {G{\'e}not},
  {Evans}, {Espley}, {Edberg}, {Dougherty}, {Cowley}, {Burch}, {Behar},
  {Barabash}, {Andrews}, and {Altobelli}}}]{witasse2017}
{Witasse}, O., {S{\'a}nchez-Cano}, B., {Mays}, M.~L., {Kajdi{\v c}}, P.,
  {Opgenoorth}, H., {Elliott}, H.~A., {Richardson}, I.~G., {Zouganelis}, I.,
  {Zender}, J., {Wimmer-Schweingruber}, R.~F., {Turc}, L., {Taylor},
  M.~G.~G.~T., {Roussos}, E., {Rouillard}, A., {Richter}, I., {Richardson},
  J.~D., {Ramstad}, R., {Provan}, G., {Posner}, A., {Plaut}, J.~J., {Odstrcil},
  D., {Nilsson}, H., {Niemenen}, P., {Milan}, S.~E., {Mandt}, K., {Lohf}, H.,
  {Lester}, M., {Lebreton}, J.-P., {Kuulkers}, E., {Krupp}, N., {Koenders}, C.,
  {James}, M.~K., {Intzekara}, D., {Holmstrom}, M., {Hassler}, D.~M., {Hall},
  B.~E.~S., {Guo}, J., {Goldstein}, R., {Goetz}, C., {Glassmeier}, K.~H.,
  {G{\'e}not}, V., {Evans}, H., {Espley}, J., {Edberg}, N.~J.~T., {Dougherty},
  M., {Cowley}, S.~W.~H., {Burch}, J., {Behar}, E., {Barabash}, S., {Andrews},
  D.~J., and {Altobelli}, N. (2017), {Interplanetary coronal mass ejection
  observed at STEREO-A, Mars, comet 67P/Churyumov-Gerasimenko, Saturn, and New
  Horizons en route to Pluto: Comparison of its Forbush decreases at 1.4, 3.1,
  and 9.9 AU}, \textit{\jgra}, \textit{122}, 7865--7890,
  \doi{10.1002/2017JA023884}.

\bibitem[{\textit{{Wood} et~al.}(2017)\textit{{Wood}, {Wu}, {Lepping},
  {Nieves-Chinchilla}, {Howard}, {Linton}, and {Socker}}}]{wood2017}
{Wood}, B.~E., {Wu}, C.-C., {Lepping}, R.~P., {Nieves-Chinchilla}, T.,
  {Howard}, R.~A., {Linton}, M.~G., and {Socker}, D.~G. (2017), {A STEREO
  Survey of Magnetic Cloud Coronal Mass Ejections Observed at Earth in
  2008-2012}, \textit{\apjs}, \textit{229}(2), 29,
  \doi{10.3847/1538-4365/229/2/29}.

\bibitem[{\textit{{Yang} et~al.}(2012)\textit{{Yang}, {Jiang}, {Yang}, {Zheng},
  {Yang}, {Hong}, {Li}, and {Bi}}}]{yang2012}
{Yang}, J., {Jiang}, Y., {Yang}, B., {Zheng}, R., {Yang}, D., {Hong}, J., {Li},
  H., and {Bi}, Y. (2012), {The Asymmetrical Eruption of a Quiescent Filament
  and Associated Halo CME}, \textit{\solphys}, \textit{279}(1), 115--126,
  \doi{10.1007/s11207-012-0002-0}.

\bibitem[{\textit{{Yurchyshyn}}(2008)}]{yurchyshyn2008}
{Yurchyshyn}, V. (2008), {Relationship between EIT Posteruption Arcades,
  Coronal Mass Ejections, the Coronal Neutral Line, and Magnetic Clouds},
  \textit{\apjl}, \textit{675}(1), L49, \doi{10.1086/533413}.

\bibitem[{\textit{{Zhang} et~al.}(2006)\textit{{Zhang}, {Baumjohann}, {Delva},
  {Auster}, {Balogh}, {Russell}, {Barabash}, {Balikhin}, {Berghofer},
  {Biernat}, {Lammer}, {Lichtenegger}, {Magnes}, {Nakamura}, {Penz},
  {Schwingenschuh}, {V{\"o}r{\"o}s}, {Zambelli}, {Fornacon}, {Glassmeier},
  {Richter}, {Carr}, {Kudela}, {Shi}, {Zhao}, {Motschmann}, and
  {Lebreton}}}]{zhang2006}
{Zhang}, T.~L., {Baumjohann}, W., {Delva}, M., {Auster}, H.-U., {Balogh}, A.,
  {Russell}, C.~T., {Barabash}, S., {Balikhin}, M., {Berghofer}, G., {Biernat},
  H.~K., {Lammer}, H., {Lichtenegger}, H., {Magnes}, W., {Nakamura}, R.,
  {Penz}, T., {Schwingenschuh}, K., {V{\"o}r{\"o}s}, Z., {Zambelli}, W.,
  {Fornacon}, K.-H., {Glassmeier}, K.-H., {Richter}, I., {Carr}, C., {Kudela},
  K., {Shi}, J.~K., {Zhao}, H., {Motschmann}, U., and {Lebreton}, J.-P. (2006),
  {Magnetic field investigation of the Venus plasma environment: Expected new
  results from Venus Express}, \textit{\planss}, \textit{54}, 1336--1343,
  \doi{10.1016/j.pss.2006.04.018}.

\bibitem[{\textit{{Zhao} et~al.}(2020)\textit{{Zhao}, {Zank}, {Adhikari}, {Hu},
  {Kasper}, {Bale}, {Korreck}, {Case}, {Stevens}, {Bonnell}, {Dudok de Wit},
  {Goetz}, {Harvey}, {MacDowall}, {Malaspina}, {Pulupa}, {Larson}, {Livi},
  {Whittlesey}, and {Klein}}}]{zhao2020}
{Zhao}, L.~L., {Zank}, G.~P., {Adhikari}, L., {Hu}, Q., {Kasper}, J.~C.,
  {Bale}, S.~D., {Korreck}, K.~E., {Case}, A.~W., {Stevens}, M., {Bonnell},
  J.~W., {Dudok de Wit}, T., {Goetz}, K., {Harvey}, P.~R., {MacDowall}, R.~J.,
  {Malaspina}, D.~M., {Pulupa}, M., {Larson}, D.~E., {Livi}, R., {Whittlesey},
  P., and {Klein}, K.~G. (2020), {Identification of Magnetic Flux Ropes from
  Parker Solar Probe Observations during the First Encounter}, \textit{\apjs},
  \textit{246}(2), 26, \doi{10.3847/1538-4365/ab4ff1}.

\bibitem[{\textit{{Zhu} et~al.}(2014)\textit{{Zhu}, {Alexander}, {Sun}, and
  {Daou}}}]{zhu2014}
{Zhu}, C., {Alexander}, D., {Sun}, X., and {Daou}, A. (2014), {The Role of
  Interchange Reconnection in Facilitating a Filament Eruption},
  \textit{\solphys}, \textit{289}(12), 4533--4543,
  \doi{10.1007/s11207-014-0592-9}.

\bibitem[{\textit{{Zhukov} and {Auch{\`e}re}}(2004)}]{zhukov2004}
{Zhukov}, A.~N., and {Auch{\`e}re}, F. (2004), {On the nature of EIT waves, EUV
  dimmings and their link to CMEs}, \textit{\aap}, \textit{427}, 705--716,
  \doi{10.1051/0004-6361:20040351}.

\bibitem[{\textit{{Zurbuchen} and {Richardson}}(2006)}]{zurbuchen2006}
{Zurbuchen}, T.~H., and {Richardson}, I.~G. (2006), {In-Situ Solar Wind and
  Magnetic Field Signatures of Interplanetary Coronal Mass Ejections},
  \textit{\ssr}, \textit{123}, 31--43, \doi{10.1007/s11214-006-9010-4}.

\end{thebibliography}

\end{document}